\documentclass[elsart12,eqsecnum,graphics,cite,nofootinbib,aps,prd,10pt,superscriptaddress]
{revtex4-2}
\usepackage{rotating}
\usepackage{booktabs}

\usepackage{placeins}
\usepackage{epsfig,epsf}
\usepackage{graphicx}
\usepackage{grffile}
\usepackage{amsmath,amsfonts,amssymb,amsthm,nccmath,latexsym,mathtools}
\usepackage[mathscr]{euscript}
\usepackage{color}
\usepackage{xcolor} 
\usepackage{paralist}
\usepackage{hyperref}
\newcommand{\beq}{\begin{eqnarray}}
\newcommand{\eeq}{\end{eqnarray}}
\newcommand{\be}{\begin{eqnarray*}}
\newcommand{\ee}{\end{eqnarray*}}

\begin{document}


\title{\huge \bf The Large Hadron electron Collider\\ as a bridge project for CERN}

\author{F.~Ahmadova}
\affiliation{Max-Planck-Institut f\"ur Physik, Boltzmannstr. 8, 85748 Garching, Germany}
\affiliation{Department of Physics, Universit\"at Z\"urich, Winterthurerstrasse 190,
CH-8057 Z\"urich, Switzerland}
\author{K.~Andr\'e}
\affiliation{CERN, Meyrin, Switzerland}
\author{N.~Armesto}
\affiliation{Instituto Galego de F\'{\i}sica de Altas Enerx\'{\i}as IGFAE, Universidade de Santiago de Compostela, 15782 Santiago de Compostela, Galicia-Spain}
\author{G.~Azuelos}
\affiliation{Group of Particle Physics, Université de Montréal, Montréal QC, Canada} \affiliation{TRIUMF, Vancouver BC, Canada}
\author{O.~Behnke}
\affiliation{Deutsches Elektronen-Synchrotron DESY, Notkestr. 85, 22607 Hamburg, Germany}
\author{M.~Boonekamp}
\affiliation{CEA/IRFU, Gif-sur-Yvette, France}
\author{M.~Bonvini}
\affiliation{INFN, Sezione di Roma, Piazzale Aldo Moro 5, 00185 Roma, Italy}
\author{D.~Britzger}
\affiliation{Max-Planck-Institut f\"ur Physik, Boltzmannstr. 8, 85748 Garching, Germany}
\author{O.~Br\"uning}
\affiliation{CERN, Meyrin, Switzerland}
\author{T.A.~Bud}
\affiliation{CERN, Meyrin, Switzerland}
\author{A.M.~Cooper-Sarkar}
\affiliation{Department of Physics, University of Oxford, Denys Wilkinson Building, Keble Road, Oxford. OX1 3RH. United Kingdom}
\author{J.~D'Hondt}
\affiliation{Nikhef, Science Park 105, 1098 XG Amsterdam, The Netherlands}
\author{M.~D'Onofrio}
\affiliation{Department of Physics, University of Liverpool, Oxford Street Liverpool, L69 7ZE, United Kingdom}
\author{O.~Fischer}
\affiliation{Department of Mathematical Sciences, University of Liverpool, Liverpool L69 3BX, United Kingdom}
\author{L.~Forthomme}
\affiliation{AGH University, Faculty of Physics and Applied Computer Science, Al. Mickiewicza 30, 30-055 Kraków, Poland}
\author{F.~Giuli}
\affiliation{Dipartimento di Fisica, Universit\`a degli Studi di Roma Tor Vergata and INFN, Sezione di Roma Tor Vergata, Via della Ricerca Scientifica 1, 00133 Rome, Italy}
\author{C.~Gwenlan}
\affiliation{Department of Physics, University of Oxford, Denys Wilkinson Building, Keble Road, Oxford. OX1 3RH. United Kingdom}
\author{E.~Hammou}
\affiliation{DAMTP, University of Cambridge, Wilberforce Road, Cambridge, CB3 0WA, United Kingdom}
\author{B.~Holzer}
\affiliation{CERN, Meyrin, Switzerland}
\author{H.~Khanpour}
\affiliation{AGH University, Faculty of Physics and Applied Computer Science, Al. Mickiewicza 30, 30-055 Kraków, Poland}
\author{U.~Klein}
\affiliation{Department of Physics, University of Liverpool, Oxford Street Liverpool, L69 7ZE, United Kingdom}
\author{P.~Kostka}
\affiliation{Department of Physics, University of Liverpool, Oxford Street Liverpool, L69 7ZE, United Kingdom}
\author{T.~Lappi}
\affiliation{University of Jyv\"askyl\"a, Department of Physics, P.O. Box 35, FI-40014 University of Jyv\"askyl\"a, Finland}
\affiliation{Helsinki Institute of Physics, P.O. Box 64, FI-00014 University of Helsinki, Finland}
\author{H.~M\"antysaari}
\affiliation{University of Jyv\"askyl\"a, Department of Physics, P.O. Box 35, FI-40014 University of Jyv\"askyl\"a, Finland}
\affiliation{Helsinki Institute of Physics, P.O. Box 64, FI-00014 University of Helsinki, Finland}
\author{B.~Mellado}
\affiliation{Institute of High Energy Physics, 19B, Yuquan Road, Shijing District, 100049, Beijing, China.}
\affiliation{School of Physics and Institute for Collider Particle Physics, University
of the Witwatersrand, Wits, 2050, Johannesburg, South Africa}
\author{P.R.~Newman}
\affiliation{School of Physics and Astronomy, University of Birmingham, B15 2TT, United Kingdom}
\author{F.I.~Olness}
\affiliation{Department of Physics, Southern Methodist University, Dallas, TX 75275, USA}
\author{J.A.~Osborne}
\affiliation{CERN, Meyrin, Switzerland}
\author{Y.~Papaphilippou}
\affiliation{CERN, Meyrin, Switzerland}
\author{H.~Paukkunen}
\affiliation{University of Jyv\"askyl\"a, Department of Physics, P.O. Box 35, FI-40014 University of Jyv\"askyl\"a, Finland}
\affiliation{Helsinki Institute of Physics, P.O. Box 64, FI-00014 University of Helsinki, Finland}
\author{K.~Piotrzkowski}
\affiliation{AGH University, Faculty of Physics and Applied Computer Science, Al. Mickiewicza 30, 30-055 Kraków, Poland}
\author{A.~Polini}
\affiliation{INFN Sezione di Bologna, Via C. Berti Pichat 4/2, Bologna, 40127, Italy}
\author{J.~Rojo}
\affiliation{Nikhef, Science Park 105, 1098 XG Amsterdam, The Netherlands}
\affiliation{Department of Physics and Astronomy, Vrije Universiteit, NL-1081 HV Amsterdam, The Netherlands}
\author{M.~Schott}
\affiliation{Physikalisches Institut, Universit\"at Bonn, Nussallee 12,
53115 Bonn, Germany}
\author{S.~Schumann}
\affiliation{Institut für Theoretische Physik, Universität Göttingen, 
Friedrich-Hund-Platz 1,  37077 Göttingen, Germany}
\author{C.~Schwanenberger}
\affiliation{Deutsches Elektronen-Synchrotron DESY, Notkestr. 85, 22607 Hamburg, Germany}
\affiliation{Universit\"at Hamburg, Luruper Chaussee 149, 22761 Hamburg, Germany}
\author{A.M.~Sta\'sto}
\affiliation{Department of Physics, Penn State University, University Park, PA 16802, U.S.A.}
\author{A.~Stocchi}
\affiliation{Universit\'e Paris-Saclay, CNRS/IN2P3 IJCLab, Orsay, France}
\author{S.~Tentori}
\affiliation{Center for Cosmology, Particle Physics and Phenomenology, Universit\'e Catholique de Louvain, Louvain-la-Neuve, Belgium}
\author{M.~Tevio}
\affiliation{University of Jyv\"askyl\"a, Department of Physics, P.O. Box 35, FI-40014 University of Jyv\"askyl\"a, Finland}
\affiliation{Helsinki Institute of Physics, P.O. Box 64, FI-00014 University of Helsinki, Finland}
\author{C.~Wang}
\affiliation{Johannes Gutenberg Universit\"at, Mainz, Germany}
\author{K. Wichmann}
\affiliation{Deutsches Elektronen-Synchrotron DESY, Notkestr. 85, 22607 Hamburg, Germany}
\author{Y.~Yamazaki}
\affiliation{Graduate School of Science, Kobe University, 657-8501 Rokko-dai, Nada, Kobe, Japan}

\begin{abstract}
\pagebreak
\centerline{\bf Abstract}
\vskip 0.5cm
The LHeC is the project for delivering electron-nucleon collisions at CERN using the HL-LHC beams. An Energy Recovery Linac in racetrack configuration will provide 50 GeV electrons to achieve centre-of-mass energies around 1\,TeV/nucleon and instantaneous luminosities around $10^{34}$\, cm$^{-2}$s$^{-1}$. The LHeC program elaborated in the CDR of 2021 included a phase with concurrent operation of electron-hadron and hadron-hadron collisions, followed by a standalone phase of electron-hadron collisions only. In view of the current HL-LHC schedule, in this paper we have examined the possibilities of a program after the regular HL-LHC program with only electron-proton operation.
In this operation mode, the LHeC would serve as an impactful bridge project between major colliders at CERN. The standalone physics program comprises electroweak, Higgs, top-quark, BSM and strong-interaction physics. In addition, it empowers the
physics analyses at the HL-LHC by retrofitting measurements and searches with significantly more precise knowledge
of the proton structure and $\alpha_s$. 
The accelerator technology deployed in the Energy Recovery Linac for the LHeC is a major stepping-stone for the performance, cost reduction and training for future colliders. The capital investments in the LHeC electron accelerator can be reused in a cost-efficient way as the injector for the FCC-ee. Finally,
data from the LHeC are essential to enable the physics potential of any new high-energy hadron collider. The
operational plan of 6 years easily fits in the period between two major colliders at CERN. Similar to the LHeC empowering the HL-LHC physics program, the FCC-eh would be an impactful addition to the FCC physics program.
\vskip 1cm
\it{This contribution is dedicated to Professor Max Klein, our collaborator, friend, mentor and inspiration, whose tireless guidance and wisdom have been key for the development of the LHeC.}
\end{abstract}

\maketitle

\pagebreak
\tableofcontents


\pagebreak
\section{The \NoCaseChange{LHeC} "bridge" project}
\label{sec:bridge}




The Large Hadron electron Collider (LHeC)~\cite{LHeCStudyGroup:2012zhm,LHeC:2020van}\footnote{\url{https://indico.cern.ch/e/LHeCFCCeh}.} is presented as the ultimate upgrade of the High-Luminosity--Large Hadron Collider (HL-LHC) program. An Energy Recovery Linac (ERL) based accelerator delivers 50\,GeV high-intensity electron beams to collide with the HL-LHC proton or ion beams at centre-of-mass energies $1.2 (0.8)$\,TeV/nucleon, and luminosities around $10^{34(33)}$ cm$^{-2}$s$^{-1}$, for electron-proton ($ep$) and electron-nucleus ($eA$) collisions, respectively. Therefore, it leverages decades of investments to deliver optimal high-energy proton and ion beams.

\begin{figure}[htbp]
\centering
\includegraphics[width=.99\textwidth]{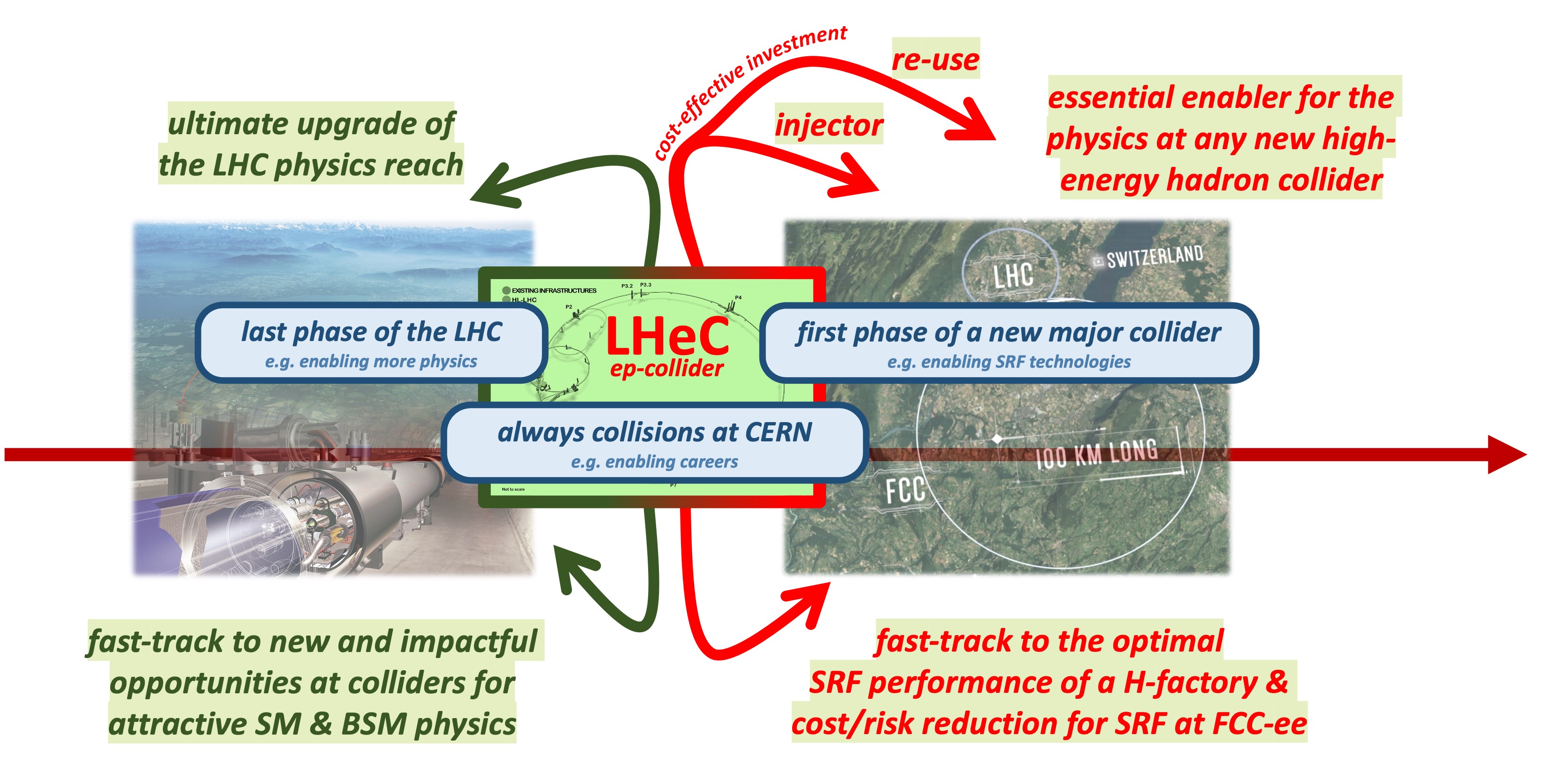}
\caption{The LHeC project as a bridge between current and future major colliders at CERN.\label{fig:bridge-intro}}
\end{figure}

The design presented in~\cite{LHeCStudyGroup:2012zhm,LHeC:2020van} focused on the concurrent operation of $ep/eA$ in one LHC interaction point and $pp/AA$ in the others. In view of the new LHC schedule, here we note that the project fits in the years between the end of the regular HL-LHC program (currently 2041) and the start of a new major collider at CERN. With 6 years of LHeC running, a total integrated luminosity of 1\,ab$^{-1}$ can be collected in $ep$ collision mode, with $eA$ collisions integrated in the $ep$ runs. As illustrated in Fig.~\ref{fig:bridge-intro}, the LHeC is a multi-purpose experiment 
which allows the ultimate exploitation of the HL-LHC results. It also provides a comprehensive program that includes Quantum Chromodynamics (QCD), electroweak, top, Higgs and beyond the standard model (BSM) physics. Finally, it also delivers key physics information required for any future high-energy hadron colliders, and triggers the development of detector and accelerator technology, and deliverables that can be used for the Future Circular Collider~\cite{FCC:2018byv,FCC:2018evy,FCC:2018vvp} in electron-positron mode (FCC-ee). It therefore bridges the HL-LHC to a new flagship project at CERN with impactful physics and innovative technologies, while being fully aligned with the current European Strategy for Particle Physics (ESPP) stating that the full physics potential of the HL-LHC is to be exploited.

Additionally, it will continue to enable the careers of our researchers and engineers through complementing the existing beams at CERN by a new high-energy high-intensity electron beam implemented through a novel accelerator technology, and a new general-purpose detector to be designed, built and operated. The comprehensive LHeC physics program ranges over most fields in HEP. It will ensure a continued development of analysis techniques, and trigger the phenomenological and theoretical efforts required to match the precision of the measurements.

Achieving high-luminosity $ep$ collisions with the LHeC is enabled by the Energy Recovery Linac (ERL) accelerator technology~\cite{Hutton:2023vzv}. This innovative technology addresses the energy sustainability of modern accelerators by recirculating the beam power rather than the particle in the beam. While the ERL technology has been successfully demonstrated for low-energy and low-current accelerators, efforts are underway to demonstrate the energy recovery efficiency with higher beam power adequate for application for the electron accelerator of the LHeC. This accelerator R\&D program promoted in the European Accelerator R\&D Roadmap delivers exciting innovations, and in a cost-effective way the investment in the LHeC electron accelerator can be re-used at future colliders at CERN. At future high-energy hadron colliders exactly the same ERL-based accelerator can be employed to collide with future hadron beams. The completed LHeC accelerator can also be re-used in non-recovery mode and therefore to continuously recirculate electron beams as the injector for a future circular $e^+e^-$ collider. This would allow a major cost efficiency and also to pre-commission a vital component of such a future collider. In addition, there is significant overlap between the Superconducting Radio Frequency (SRF) accelerating systems of the LHeC and those planned for the FCC-ee. Accordingly, the LHeC emerges as a major and unique stepping stone for the development of these SRF components essential for the success of the FCC-ee. Well targeted co-developments on common technologies, with LHeC SRF components as prototypes for the FCC-ee, could enable cost and risk reduction, opportunities for validation in operation, and allow building expertise at CERN by training on SRF components in operation. The LHeC as a bridge project to a future $e^+e^-$ collider is a cost-effective fast-track to the optimal SRF performance of the accelerator and therefore the fastest-track to the best physics.

In the present manuscript, which accompanied the LHeC submission to the 2026 Update of the ESPP~\cite{LHeCSub}, we summarise the physics, detector and accelerator aspects of the LHeC as a bridge project for CERN. Full details can be found in~\cite{LHeCStudyGroup:2012zhm,LHeC:2020van,Andre:2022xeh} and in the references mentioned for the different studies.

Sec.~\ref{sec:standalone} presents a selection of the physics topics that can be studied in deep inelastic scattering (DIS): proton and nuclear partonic structure; Higgs physics enabled by the ample sample, $\sim 200000$, of Higgs bosons that can be produced and studied; top-quark physics; precision SM measurements; BSM searches; physics of high gluon densities and; gamma-gamma physics. 
Sec.~\ref{sec:enabler} shows the impact of the improved determination of proton structure and the strong coupling constant $\alpha_s$ on the extraction of SM parameters in hadron colliders, specifically at the HL-LHC. Then, in Sec.~\ref{sec:global} the impact of LHeC studies is discussed in combination with future machines, such as electron-positron colliders (FCC-ee and CEPC~\cite{CEPCStudyGroup:2018rmc,CEPCStudyGroup:2018ghi}), the FCC in hadron-hadron mode (FCC-hh), or the Electron Ion Collider (EIC)~\cite{AbdulKhalek:2021gbh}.

Sec.~\ref{sec:feasibility} describes the present status of the design of the detector, machine-detector interface and accelerator, with emphasis of the feasibility of the ERL technology for the LHeC requirements. In Sec.~\ref{sec:implementation} the LHeC implementation plan is discussed together with sustainability aspects like power consumption and cost, and environmental impact. Sec.~\ref{sec:perle} presents the demonstrator PERLE~\cite{Angal-Kalinin:2017iup} for the multi-turn high-current ERL technology under construction at IJCLab in Orsay, its current status and the plans to deliver ERL demonstration at the beginning of the 2030ies. Sec.~\ref{sec:technology} discuss the LHeC as a stepping-stone for accelerator technologies and deliverables, and detector design, for a Higgs factory. In Sec.~\ref{sec:cost} estimates for the cost of the accelerator and detector are given. In Sec.~\ref{sec:outlook} we present our summary and outlook for the coming years. Finally, in App.~\ref{sec:lowlumi} we illustrate with some examples the physics that can be achieved with luminosities lower than the nominal one.


\section{The \NoCaseChange{LHeC} at the frontline of particle and nuclear physics}
\label{sec:standalone}


\label{sec:frontier}

\begin{figure}[htb]
\begin{center}
    \includegraphics[width=0.45\textwidth]{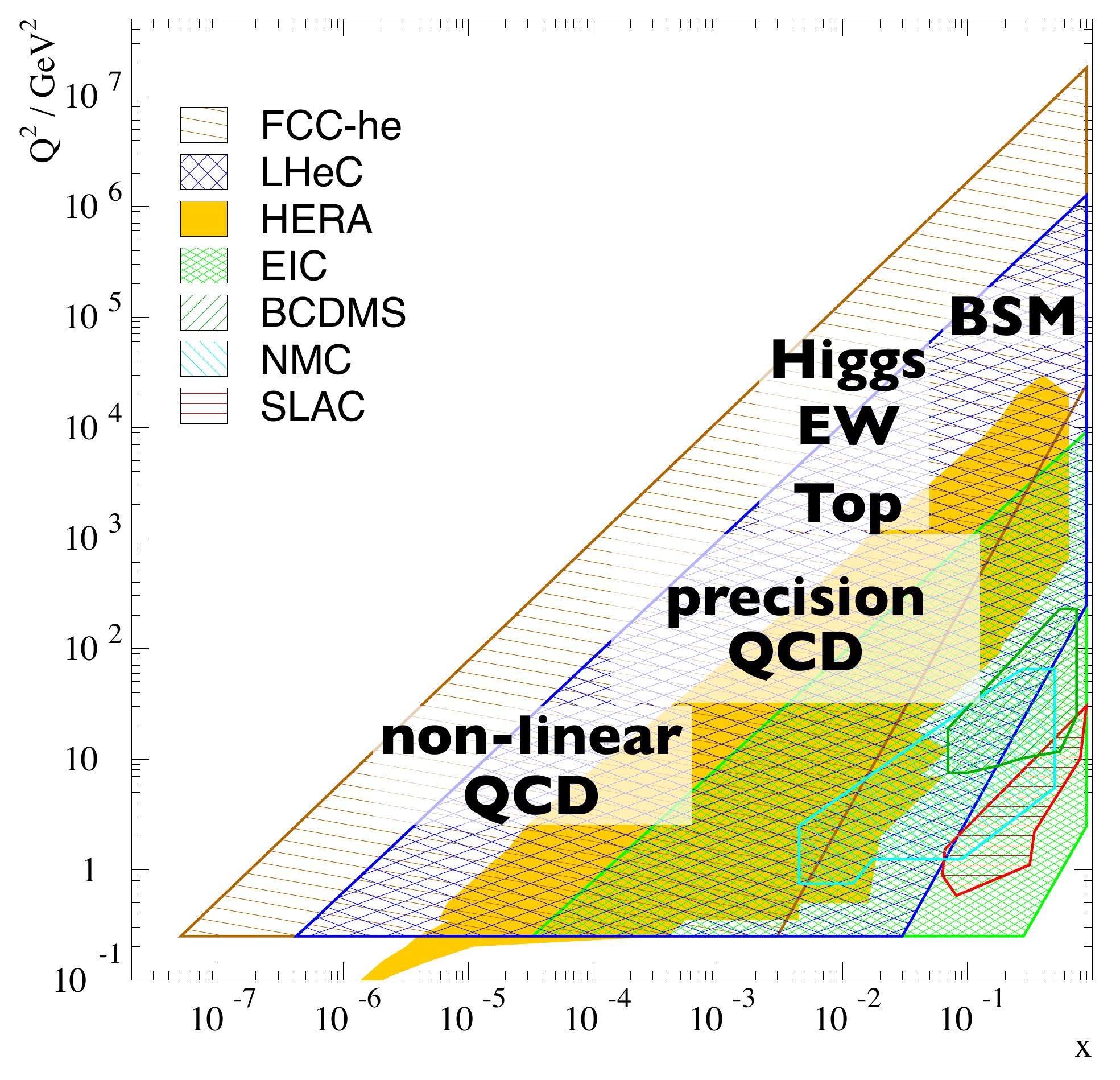}
  \end{center}
\vskip -0.5 cm
\caption{Coverage of the kinematic plane in deep inelastic lepton-proton scattering
by some initial fixed target experiments, with electrons (SLAC~\cite{Whitlow:1991uw}) and
muons (NMC~\cite{NewMuon:1996fwh}, BCDMS~\cite{BCDMS:1989qop}), and by the $ep$ colliders:  the EIC~\cite{AbdulKhalek:2021gbh} (green), HERA~\cite{H1:2015ubc} (yellow), the LHeC~\cite{LHeCStudyGroup:2012zhm,LHeC:2020van} (blue) and 
the FCC-eh~\cite{FCC:2018byv,FCC:2018vvp} (brown). $x$ is the momentum fraction of the proton or nucleus taken by the partons involved in the cross section for a given observable, and $Q^2$ the hard scale of the process squared. The physics topics that the LHeC can cover are illustrated in the kinematic plane. 
Figure modified from~\cite{LHeC:2020van}.
For details regarding specific kinematic ranges, see ``Chapter 18: Structure Functions'' in Ref. \cite{ParticleDataGroup:2024cfk} and the references therein. In particular, Table 18.1 tabulates the primary processes, their corresponding subprocesses, and the relevant $x$ ranges.
}
\label{fig:kinplane}       
\end{figure}

\begin{figure}[htb!]
\begin{center}
    \includegraphics[width=0.52\textwidth]{{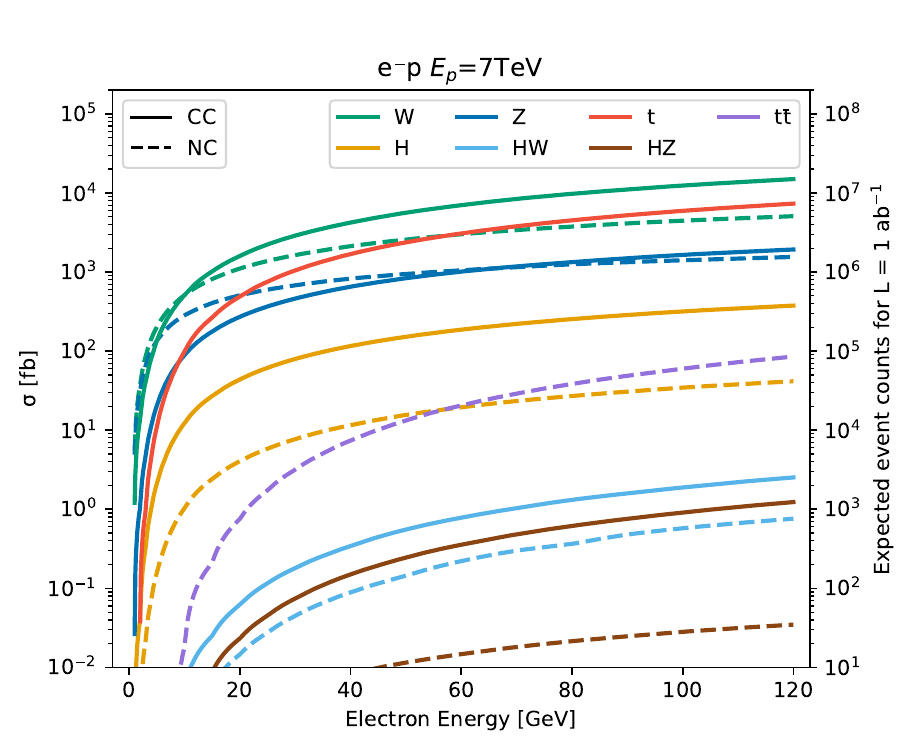}}
  \end{center}
  \vskip -0.5cm
\caption{Selected inclusive production cross sections for an electron--proton collider at the HL-LHC as a function of the electron-beam energy $E_e$. The proton-beam energy is set by the HL-LHC to 7\,TeV, so that the $ep$ centre-of-mass energy and the cross sections depend only on $E_e$ of the to-be-build electron accelerator. Displayed are inclusive production cross sections for prompt $Z$ and $W^\pm$ bosons, the Higgs boson ($H$), single top-quark production ($t$) and top-quark pair production ($t\bar{t}$), each for neutral-current (NC) and charged-current (CC) induced channels, respectively. Furthermore, the cross sections for $HZ$ and $HW$ production are displayed, which are mainly induced by vector-boson-fusion channels. Charged-current mediated cross sections scale as  $(1+P_e)$,  with $P_e$ being the longitudinal electron beam polarization which is set to $P_e=80\%$. For $E_e=50$\,GeV, the total inclusive NC and CC DIS production cross sections,  $6.5\cdot10^8$\,fb and $3.0\cdot10^5$\,fb, respectively, are not displayed, nor more specialized processes like $b\bar{b}$, $c\bar{c}$, jets, prompt photons, and di-boson production ($WW$, $ZZ$), or others like photoproduction or $\gamma\gamma$ induced ones. 
For example, $t\bar{t}$ photoproduction has a cross section of $0.7$\,pb at $E_e=60$\,GeV~\cite{Bouzas:2013jha} which is more than a factor of 30 larger than NC DIS $t\bar{t}$ production displayed here (violet dashed).  
The right-sided axis displays the expected number of events for an integrated luminosity of $\mathcal{L}=1\,\text{ab}^{-1}$. The predictions are calculated with Sherpa3~\cite{Sherpa:2024mfk} at next-to-leading order QCD.}
\label{fig:CrossSections}       
\end{figure}

The physics program at the LHeC~\cite{LHeCStudyGroup:2012zhm,LHeC:2020van,Andre:2022xeh}, based on the extended kinematic plane that it can explore (illustrated in Fig.~\ref{fig:kinplane}), is both comprehensive and distinctive, encompassing a broad spectrum of topics. The LHeC will provide centre-of-mass energies $\sim 1.2 \ (0.8)$\, TeV/nucleon and integrated luminosities of 1\,ab$^{-1}$ (10\,fb$^{-1}$) in $ep$ ($eA$) collisions, respectively, with 50\,fb$^{-1}$ in $ep$ achievable in a first year of running.
The high centre-of-mass energy, together with the large integrated luminosity that is enabled by the high-intense HL-LHC proton beam and the ERL electron beam, enables the measurement of a large variety of processes, as displayed in Fig.~\ref{fig:CrossSections}.

The physics program includes QCD, ranging from precision studies with detailed investigations of hadron structure and parton dynamics -- foundational for the extended physics goals of the LHC and future hadron collider experiments --, determination of the strong coupling constant and measurement of  nuclear parton distributions which provides input essential for heavy-ion physics in general and for Quark-Gluon Plasma characterization in particular, to discovery of a novel non-linear regime of QCD at small values of the momentum fraction $x$. 

The LHeC also enables percent-level precision studies of the Higgs mechanism and unique precision measurements in the electroweak (EWK) sector. It further delivers high-accuracy determinations of key SM parameters, crucial for analyses at the LHC and beyond. Moreover, the LHeC offers strong sensitivity to new physics (BSM) scenarios, including both promptly decaying and long-lived particles. Finally, it also gives access to photon-photon physics. By accessing an unexplored kinematic regime with high luminosity, the LHeC holds the potential to uncover unexpected and groundbreaking discoveries.
Selected highlights of all these aspects are presented in the following Subsections.\footnote{Concerning QCD and EWK studies, much of the simulation work presented here, as well as in the LHeC White Paper~\cite{LHeC:2020van}, was performed using the open-source xFitter analysis framework. xFitter is a versatile, flexible, modular, and comprehensive platform for QCD analyses that provides essential impact studies for both current and future experimental facilities. Further information regarding the xFitter framework and its applications can be found in Refs.~\cite{Alekhin:2014irh,xFitter:2022zjb,xFitterDevelopersTeam:2019ygc,xFitterDevelopersTeam:2018hym}
and at \url{https://gitlab.cern.ch/fitters/xfitter}.}

\subsection{Partons and Proton Structure}
\label{sec:pdf}
Partons, namely quarks and gluons, are confined within the proton. Still a major puzzle in modern physics, they cannot be observed directly. Quarks of any type, $q$ (and $\bar{q}$), and gluons, have a certain probability of carrying a fraction $x$ of the proton's momentum, described by the momentum density function $xq(x),xg(x)$, known as the parton distribution function (PDF).
The characteristics of the proton are defined by its valence quark content, consisting of two up quarks and one down quark. The distribution of the proton's momentum among the quarks varies with $x$, changing as we probe the proton at increasingly small distances in lepton-hadron DIS. This is done through a virtual photon or a $Z$ or $W^{\pm}$ boson, with virtuality $Q^2$, interacting with a quark.

The precision physics era at the HL-LHC will reach its full potential if complemented by precise measurements of PDFs at the LHeC. As detailed in \cite{LHeC:2020van}:  
\begin{itemize}
    \item While charged-current (CC) deep inelastic scattering (DIS) has already been measured at HERA and in fixed-target neutrino experiments, at the LHeC  CC  data will span four orders of magnitude in $x$ and $Q^2$ due to the increased energy. This will mean a breakthrough, compared to the existing situation, in their use as a key foundation for the extraction of proton and nuclear parton structure~\cite{LHeC:2020van,AbdulKhalek:2019mps}.  
    
    \item All PDFs, $xq(x,Q^2)$ and $xg(x,Q^2)$, can be determined over a vast kinematic range, covering $q = u_v, d_v, u, \bar{u}, d, \bar{d},$ $s, c, b$, and even $t$, in a single DIS experiment -- therefore with well-controlled systematics and free of the eventual incompatibilities between data sets of different origin~\cite{LHeC:2020van,Giuli:2026nyh}.
    \footnote{PDFs depend on a factorization scale $\mu_F^2$, which in DIS is typically taken as $Q^2$. We make no distinction between Bjorken $x$ as the kinematic variable in DIS and $x$ as the proton momentum fraction carried by the probed parton, as they coincide in the parton model.}
    
    \item Unlike HERA (Hadron–Electron Ring Accelerator at DESY) or lower-energy fixed-target and $ep/eA$ collider experiments, the LHeC will probe extremely small $x$ values in the DIS region, allowing the study of a new regime of QCD marked by non-linear gluon interactions~\cite{xFitterDevelopersTeam:2018hym,Xie:2022sjm,Bendova:2022xhw}.  
    
\end{itemize}

\begin{figure}[htb]
\begin{center}
    \includegraphics[width=0.49\textwidth]{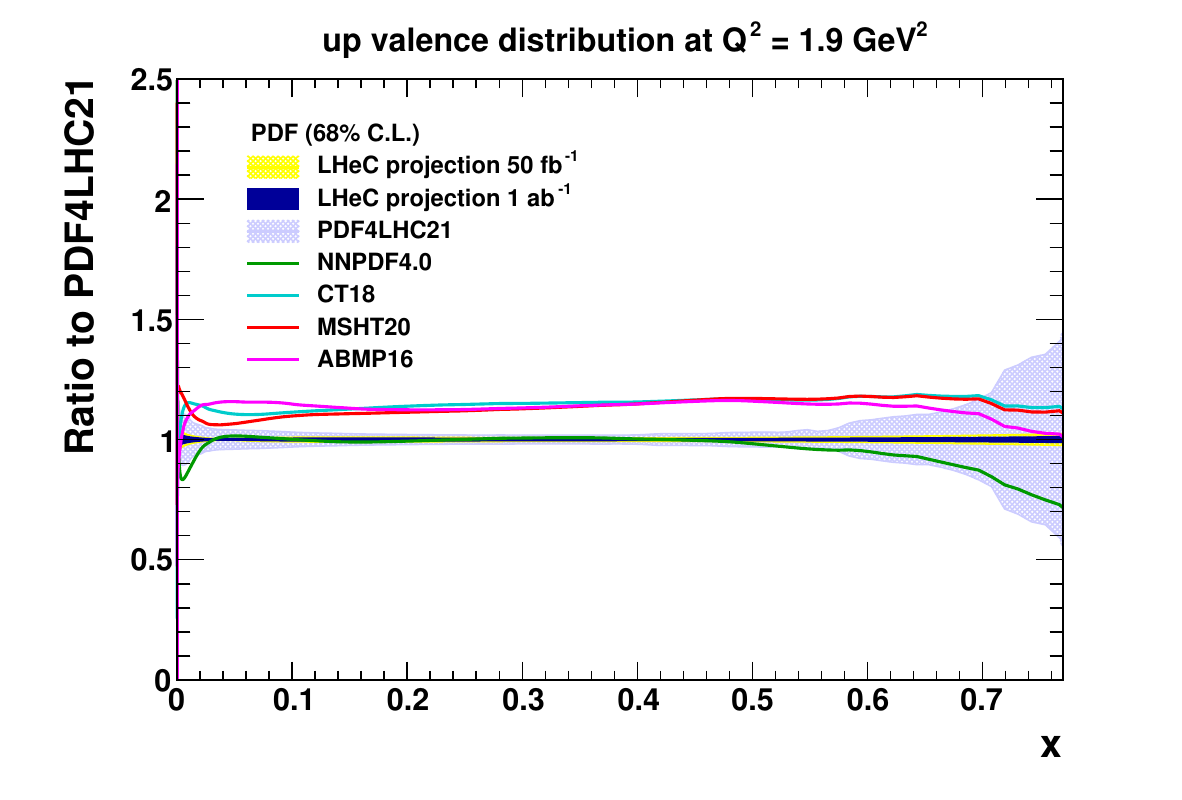}
    \includegraphics[width=0.49\textwidth]{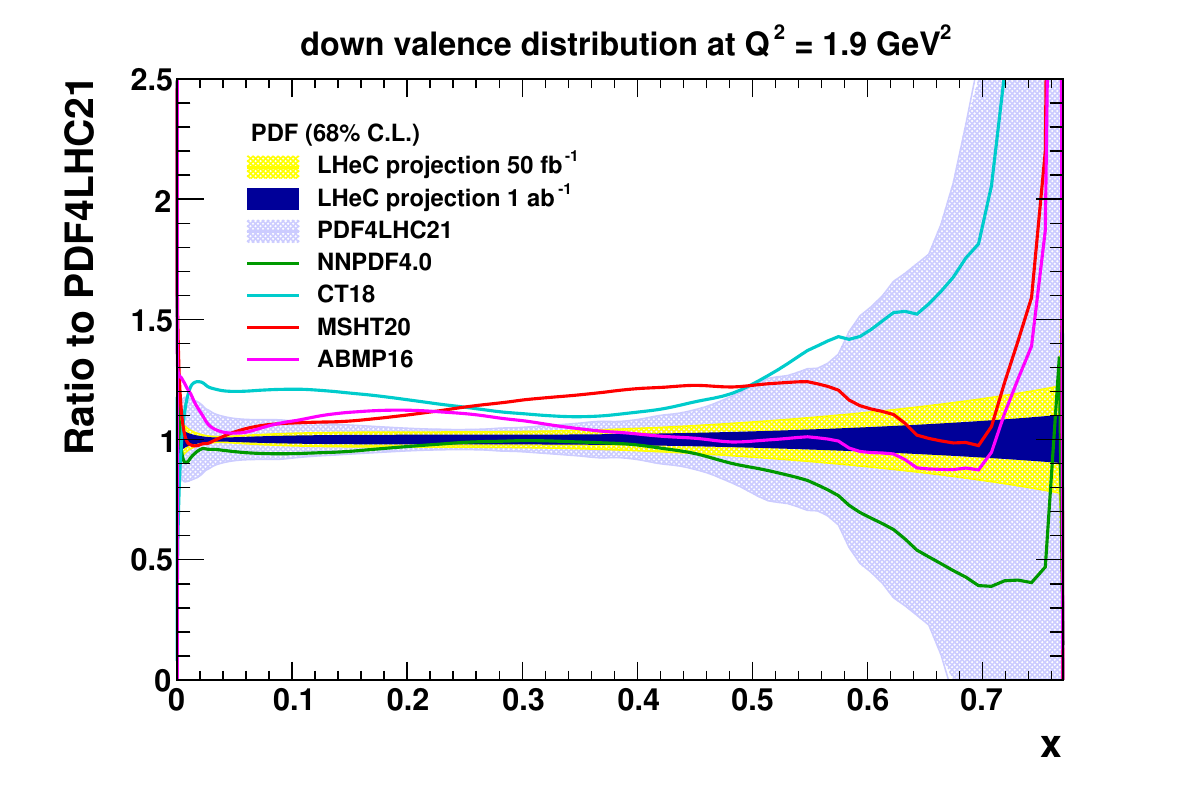}
    \includegraphics[width=0.49\textwidth]{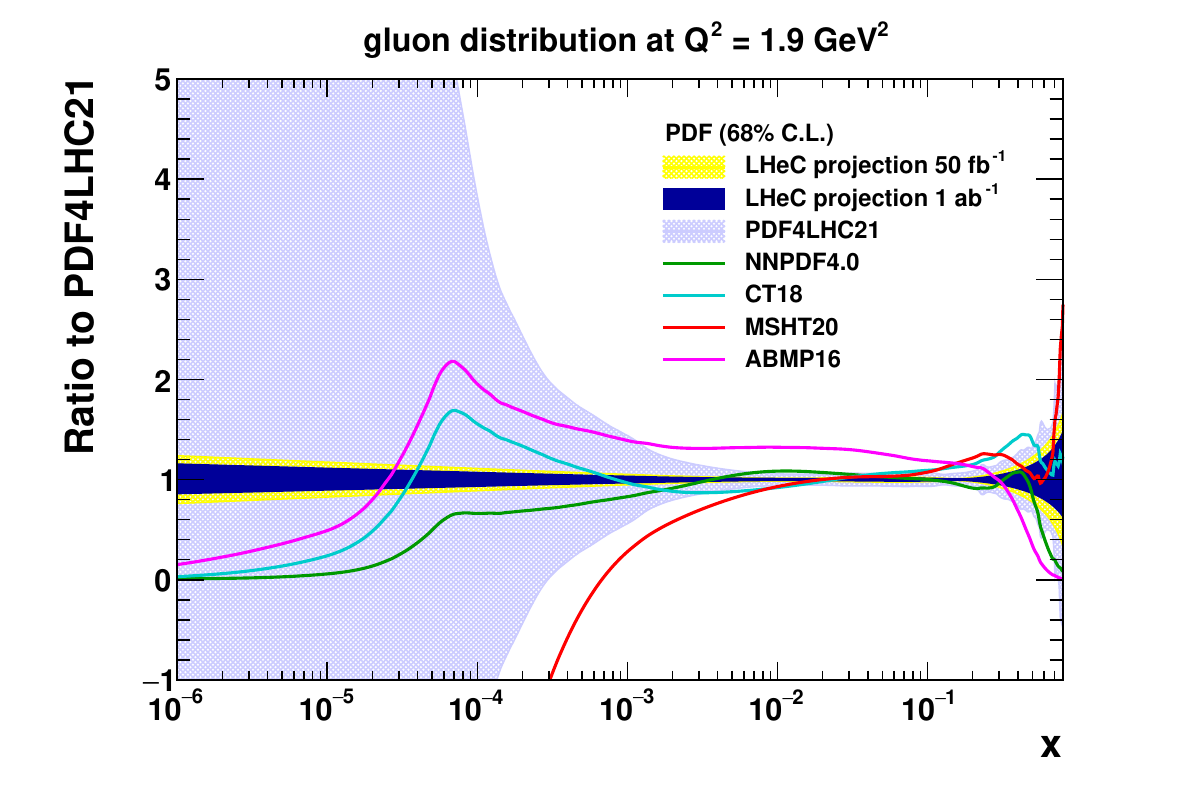} 
    \includegraphics[width=0.49\textwidth]{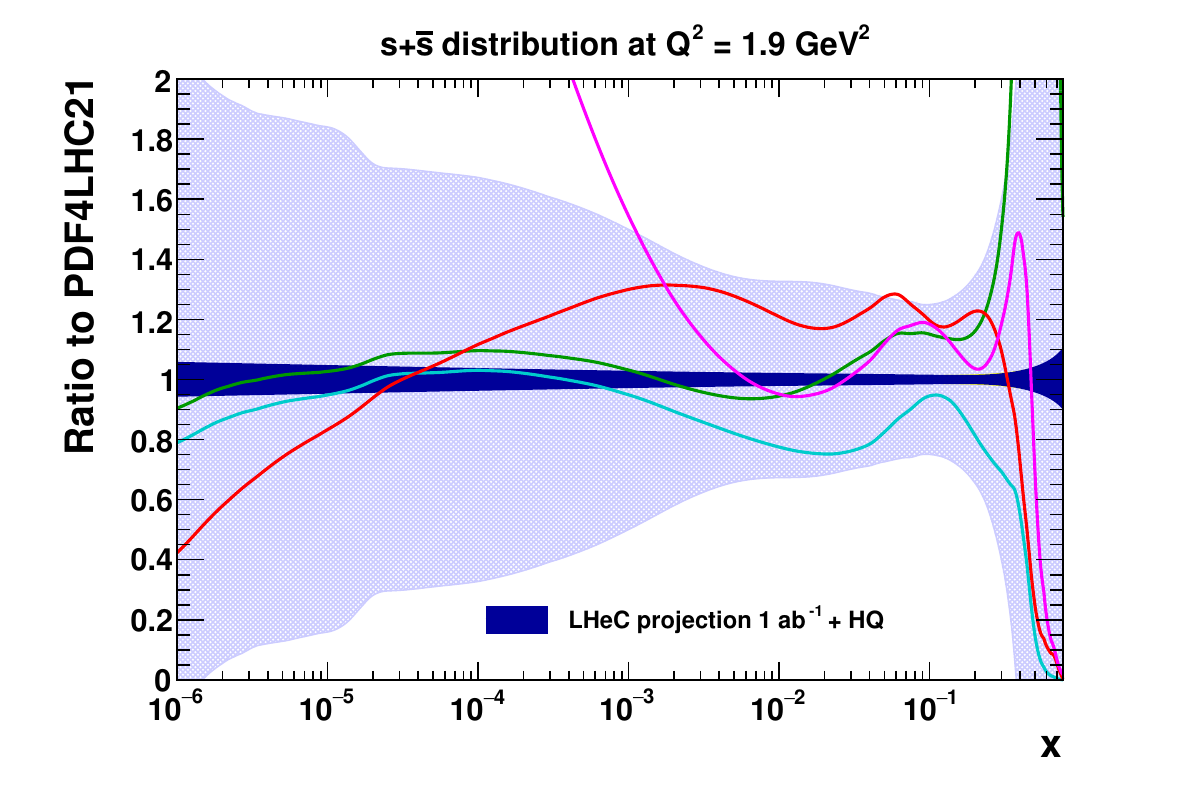}
  \end{center}
\vskip -0.5cm
\caption{Expected precision for the determination of
parton density functions, expressed as a ratio to that of PDF4LHC21~\cite{PDF4LHCWorkingGroup:2022cjn}, as a function of 
$x$ at $Q^2=1.9$\,GeV$^2$: $u_v$ (top left), $d_v$ (top right), $g$ (bottom left) and $s$ (bottom right). LHeC results at next-to-next-to-leading order (NNLO) for integrated luminosities of 50\,fb$^{-1}$ and 1\,ab$^{-1}$  are shown with uncertainty bands together with central values of ABMP16~\cite{Alekhin:2017kpj}, CT18~\cite{Hou:2019efy}, MSHT20~\cite{Bailey:2020ooq} and NNPDF4.0~\cite{NNPDF:2021njg}. Small irregularities, also present in the parton-parton luminosity plots in Sec.~\ref{sec:luminosities}, are due to those in the baseline set PDF4LHC21.}
\label{fig:pplumi}       
\end{figure}

Unprecedented precision in these distributions for protons and nuclei is within reach~\cite{LHeC:2020van}\footnote{In this reference the details on the extraction of PDFs from LHeC data alone can be found, together with the combination with HL-LHC expectations~\cite{AbdulKhalek:2019mps}. The studies were done using xFitter, employing next-to-next-to-leading order (NNLO) evolution equations and a parametrisation for the initial conditions akin to that employed for the HERAPDF2.0  studies~\cite{H1:2015ubc}. LHeC inclusive and heavy-flavour tagged pseudodata were generated in the LHeC $x-Q^2$ region shown in Fig.~\ref{fig:kinplane}, considering systematic uncertainties roughly a factor 2 smaller than those achieved at HERA and the kinematic coverage of the detector as described in Sec.~\ref{sec:detector}.}, including per mille accuracy in $\alpha_s$ and precise measurements of the strange, charm, and bottom quark distributions, as well as the longitudinal structure function $F_L(x,Q^2)$.  
This will provide a comprehensive test of perturbative QCD and parton dynamics, enable potential discoveries at the LHC, such as new physics in high-mass tails from interference effects, and lead to breakthroughs in QCD, including possible factorization breaking beyond diffraction. It will also enhance precision electroweak and Higgs physics at the joint $ep/pp$ LHC facility to an unprecedented level.  
A dedicated study~\cite{LHeC:2020van} has explored LHeC PDF prospects for an initial dataset of $50$\,fb$^{-1}$, compared to the full dataset of 1\,ab$^{-1}$, see Fig.~\ref{fig:pplumi}, where the PDFs and their uncertainties are shown at the starting scale of the QCD evolution (the corresponding uncertainties at larger scales are illustrated in parton-parton luminosity plots in Sec.~\ref{sec:luminosities}).  The LHeC projections shown use the NC and CC inclusive datasets only, with the exception of the extraction of the $s+\bar s$ distribution, which also includes semi-inclusive data on $s$, $c$ and $b$; further results separating the heavy flavour contributions are shown in~\cite{LHeC:2020van}. Heavy flavour contributions at large $x$, not accessible at HERA, will be measured, thus determining the intrinsic heavy flavour content of protons and nuclei.

Beyond the standard collinear PDFs, studies of three-dimensional proton and nuclear structure have been performed~\cite{LHeCStudyGroup:2012zhm,LHeC:2020van,Armesto:2019gxy}, mainly focused on the possibilities through inclusive and exclusive (vector mesons or deeply-virtual Compton scattering) diffraction.

\subsection{Higgs Boson Physics}
The Higgs mechanism is of fundamental importance, as it is linked to the spontaneous breaking of a locally gauge symmetric  theory. It explains why intermediate vector bosons acquire mass while the photon remains massless, and it provides a framework for fermion mass generation via the Higgs field. While the Higgs mechanism also predicted the existence of an elementary particle, the Higgs boson, discovered at the LHC in 2012, a key task in particle physics remains: verifying that the Higgs mechanism functions as predicted by the SM, exploring possible extensions of the Higgs sector and searching for connections to potential exotic particles.

DIS provides a theoretically clean environment for Higgs production in $ep$ collisions, with minimal QED and QCD corrections~\cite{Blumlein:1992eh,Han:2009pe,Jager:2010zm}. At the LHeC, the Higgs boson is mainly produced via CC interactions through the $t$-channel process $WW \to H$, with a smaller but measurable contribution from NC interactions. The Higgs production cross section at the LHeC is about 200\,fb, comparable to the $e^+e^- \to HZ$ cross section at 250\,GeV, see Fig.~\ref{fig:CrossSections}. For the nominal integrated luminosity of 1\,ab$^{-1}$, around $2\cdot 10^5$ Higgs bosons become available for analysis.

The details of the Higgs analyses can be found in~\cite{LHeC:2020van}. Dedicated cut- and BDT-based studies were performed for $H\to b \bar b$, $H\to c \bar c$ and $H \to W^+W^-$ using a DELPHES fast detector simulation~\cite{deFavereau:2013fsa}\footnote{The same DELPHES fast detector simulation was also used in top and BSM studies.}, with the former cross-checked using a more detailed detector simulation~\cite{Behera:2022gnr}. A coarser analysis
was done for other frequent decay channels both in NC and CC, with acceptances
and backgrounds estimated with MadGraph~\cite{Alwall:2014hca} and efficiencies from  Higgs coupling
studies at the LHC. Once the signal strengths were obtained, studies for the couplings in the $\kappa$ framework were performed, see the description in~\cite{LHeC:2020van} and in Sec.~\ref{sec:global}.

The LHeC is sensitive to the six dominant Higgs decay channels: $b \bar{b},~W^+W^-,~gg,~\tau \tau,~c\bar{c}$, and $ZZ$, which together account for 99.6\% of the total SM Higgs decay width. The respective relative uncertainties on the expected measurements of Higgs couplings from the LHeC for the seven most abundant decay channels are summarized in Tab.~\ref{tab:higgskappa}.
High precision measurements with uncertainties at the 1.9 (0.70, 1.2)\% level of Higgs couplings to $b$-quarks  ($W$-bosons, $Z$ bosons), are achievable (the latter particularly clean in $WW \to H \to WW$ (CC) and $ZZ \to H \to ZZ$ (NC) processes), enabling reconstruction of the total SM decay width with percent-level accuracy~\cite{LHeC:2020van}. As discussed in Sect.~\ref{sec:enabler}, combining these measurements with HL-LHC data will further enhance precision, turning the LHeC into a Higgs high precision facility.

\begin{table}[h]
    \begin{tabular}{l|ccccccc}
        \hline \hline
         Coupling $\kappa$ for decay $H\to$ & $b \bar b$ & $WW$ & $gg$ & $\tau \tau$ & $c \bar c$ & $ZZ$ & $\gamma \gamma$ \\ \hline
         Relative uncertainty $\delta\kappa$ (\%) & 1.9 & 0.70 & 3.5 & 3.1 & 3.8 & 1.2 & 6.8 \\
         \hline\hline
    \end{tabular}
    \caption{Summary of uncertainties of Higgs couplings from $ep$ for the seven most abundant decay
channels~\cite{LHeC:2020van}.}
    \label{tab:higgskappa}
\end{table}

Additionally, the LHeC offers a unique opportunity to probe the CP structure of the Higgs Yukawa coupling~\cite{Coleppa:2017rgb}, for instance, via $pe^{-} \to \overline{t}h\nu_e$. Strong constraints can be placed on deviations from the SM-predicted CP phase. Assuming a SM-like top-Yukawa coupling, the coupling magnitude could be measured with 17\% accuracy at the LHeC using 1\,ab$^{-1}$ of integrated luminosity~\cite{Coleppa:2017rgb}. 

It has to be noted that for all these analyses more recent multivariate techniques will further advance the precision of the measurements. In particular, recent improvements of heavy-flavour taggers at the LHC have not been yet considered in the current analyses. Therefore, the quoted values are conservative. 

\subsection{Top-Quark Physics}
\label{sec:top}
\begin{figure} [!h]
\centering
\includegraphics[width=0.49\textwidth]{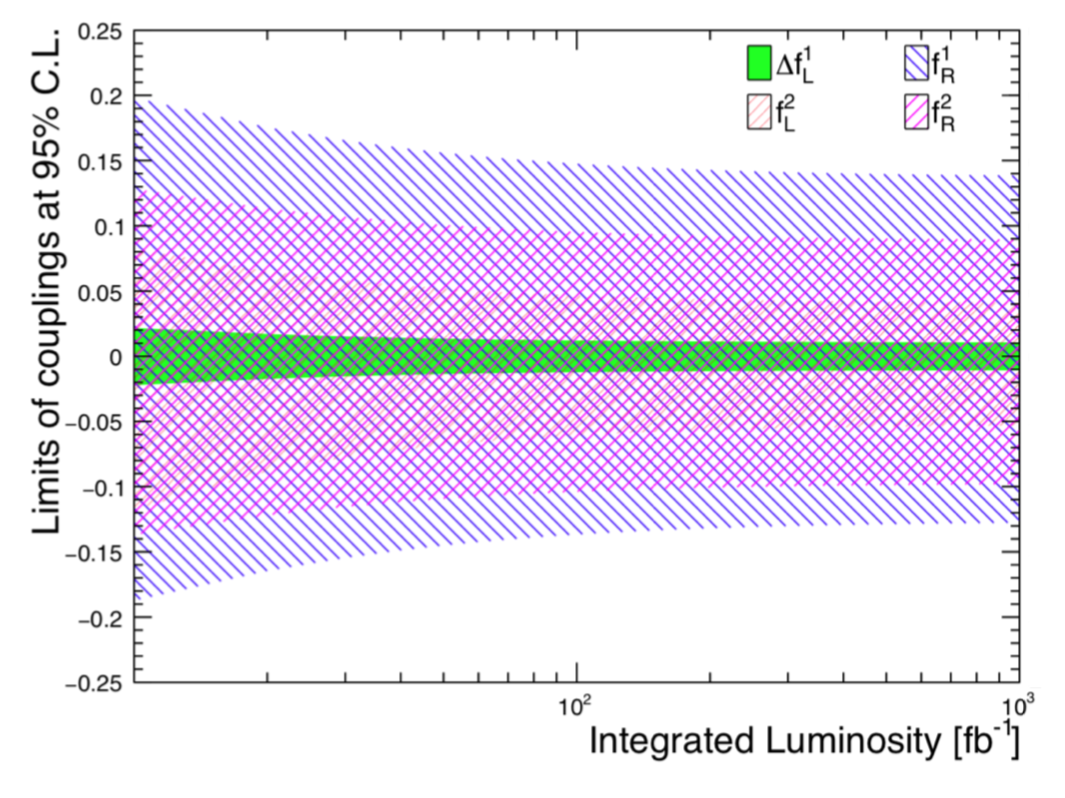}
\includegraphics[width=0.49\textwidth]{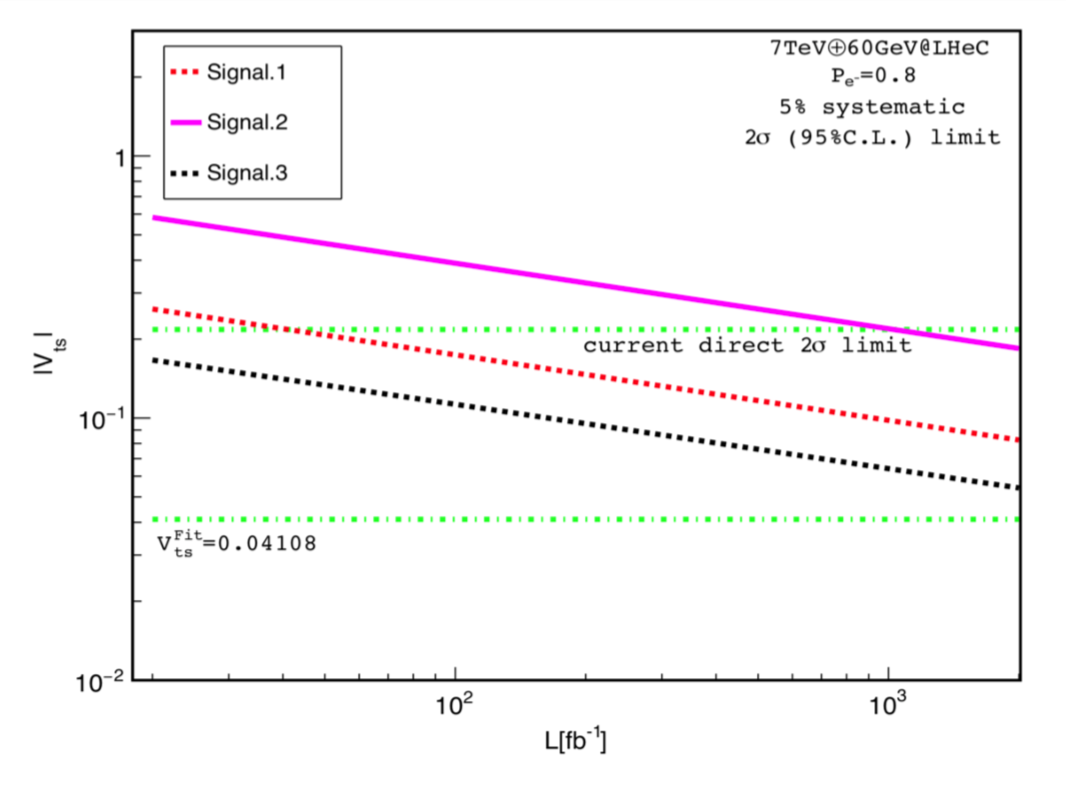}
\vskip -0.3cm
\caption{
Expected sensitivities on
the SM $|V_{tb}|$ ($f^1_L \equiv 1+ \Delta f^1_L$) and on anomalous right-handed vector ($f^1_R$), left-handed tensor ($f^2_L$) and right-handed tensor ($f^2_R$) $Wtb$ couplings at the LHeC~\cite{Dutta:2013mva} (left), and on $|V_{ts}|$ exploring three different signal scenarios (Signal 1: $p e^- \to \nu_e \bar{t} \to \nu_e W^- \bar{b} \to \nu_e \ell^-\nu_\ell \bar{b}$; Signal 2: $p e^- \to \nu_e W^- b \to \nu_e \ell^-\nu_\ell b$; Signal 3: $p e^- \to \nu_e \bar{t} \to \nu_e W^- j \to \nu_e \ell^-\nu_\ell j$)~\cite{Sun:2018gqo} (right), as a function of the integrated luminosity.
}
\label{fig_Wtx_couplings}
\end{figure}

High-energy electron-proton colliders provide a unique environment for probing the electroweak interactions of the top quark, with the LHeC standing out as a powerful single-top production facility(see Sec. 5.3 in~\cite{LHeC:2020van}). At the LHeC, the CC single-top production cross section is 1.9\,pb 
for $E_e=60$\,GeV and a longitudinal electron beam polarization of $80\%$ (see Fig.~\ref{fig:CrossSections}). 
Analyses were done using a DELPHES fast detector simulation. They show that this substantial production rate enables high-precision measurements of the $Wtb$ coupling and sensitive searches for anomalous contributions in the $Wtb$ vertex~\cite{Dutta:2013mva}, as presented in Fig.~\ref{fig_Wtx_couplings} (left). Already with an integrated luminosity of 100\,fb$^{-1}$, relative uncertainties on $Wtb$ measurements can be derived at the per cent level. 
In case of the SM left-handed vector coupling $V_{tb}$, a precision of 1\% can be achieved, improving the currently best measurements at ATLAS~\cite{ATLAS:2024ojr} and CMS~\cite{CMS:2020vac} with 3.1\% and 2.4\% relative uncertainties, respectively. 
With an LHeC dataset of 1000\,fb$^{-1}$ expected uncertainties of the HL-LHC with a dataset of 3000\,fb$^{-1}$~\cite{Deliot:2018jts} can be improved, as presented in Tab.~\ref{tab:anomalous_Wtb}. This provides stringent tests of the SM and potential new physics. Note that even with the small luminosities corresponding to the data collected during the first year of LHeC operation, the measurements are competitive with the results obtained from the full HL-LHC dataset.

\begin{table}[h]
    \begin{tabular}{l||c|c|c}
        \hline \hline
         Anomalous $Wtb$ Coupling & $f^1_R$ & $f^2_L$ & $f^2_R$ \\
             \hline
         HL-LHC, 3000\,fb$^{-1}$ ($\mathcal{R}e$) & [-0.28,0.32] & [-0.17,0.19] & [-0.05,0.02] \\
          \hline
         HL-LHC, 3000\,fb$^{-1}$ ($\mathcal{I}m$) & [-0.30,0.32] & [-0.19,0.18] & [0.11,0.10] \\
          \hline
         LHeC, 1000\,fb$^{-1}$ ($\mathcal{R}e$) & [-0.13,0.14] & [-0.05,0.04] & [-0.10,0.09] \\
         \hline\hline
    \end{tabular}
    \caption{Expected limits at 95\% CL on anomalous right-handed vector ($f^1_R$), left-handed tensor ($f^2_L$) and right-handed tensor ($f^2_R$)  $Wtb$ couplings at the LHeC~\cite{LHeC:2020van} and the HL-LHC~\cite{Deliot:2018jts}. 
    }
    \label{tab:anomalous_Wtb}
\end{table}

Next-to-leading order corrections~\cite{Gao:2021plf} reduce the fiducial single-top production cross section at the LHeC by 14\%, while significantly improving its stability against scale variations. Unlike the LHC, where the large $t\overline{t}$ production cross section makes single-top measurements more challenging, the LHeC offers a much cleaner environment. This distinction underscores the strong complementarity between the two colliders in top-quark physics.

The LHeC’s high precision also enables superior determinations of other CKM matrix elements, such as $V_{td}$ and $V_{ts}$, surpassing the capabilities of the LHC~\cite{Sun:2018gqo,Alvarez:2017ybk}. 
With an integrated luminosity of $\mathcal{O}(1)\,\text{ab}^{-1}$, competitive measurements of these CKM elements could be achieved. In Fig.~\ref{fig_Wtx_couplings} (right) it is shown for three different signal processes that with $\sim 1\,\text{ab}^{-1}$ of integrated luminosity and an electron polarisation of 80\% the $2\sigma$ limits improve on existing limits from the LHC~\cite{CMS:2014mxl} (interpreted by~\cite{Aguilar-Saavedra:2004mfd}) by up to a factor of 5. The study of the most significant signal alone -- but even more the combination of all three different signals -- allows to achieve an accuracy of the order of the actual SM value of
$|V_{ts} | = 0.04108^{+0.0030}_{-0.0057}$ as derived from an indirect global CKM matrix fit~\cite{Charles:2015gya}, providing the first direct high precision measurement of this top-quark coupling. 

Additionally, the LHeC provides an ideal setting for precise studies of top-quark properties, such as spin and polarization, with theoretical and experimental advantages over the LHC~\cite{Atag:2006by}.
Furthermore, the LHeC also provides access to Flavor Changing Neutral Current (FCNC) processes via $\gamma tq$ and $Ztq$ vertices ($q=u,c$)~\cite{TurkCakir:2017rvu,Behera:2018ryv}. Processes such as $e^-p \to e^-\,W\,b+X$ can be precisely measured,
improving current limits on $\gamma tu$ couplings 
and
achieving competitiveness with HL-LHC projections.

Finally, $t\overline{t}$ production at the LHeC 
constitutes a promising area of study due to its relatively large photoproduction cross section of 
$0.7$\,pb at $E_e=60$\,GeV \cite{Bouzas:2013jha} which is much larger than the NC DIS $t\bar{t}$ production of $0.02$\,pb as displayed in Fig.~\ref{fig:CrossSections} (violet dashed).
For example, since in $t\bar{t}$ production energetic photons interact directly with top quarks, the LHeC offers a unique opportunity to probe the $t\overline{t}\gamma$ magnetic and electric dipole moments~\cite{Bouzas:2013jha}. This process provides higher sensitivity than $b\rightarrow s\gamma$ transitions or $t\overline{t}\gamma$ measurements at the LHC.

\subsection{Precision SM Physics}
The SM describes fundamental interactions among elementary particles. Despite its success, key free parameters, such as the strong coupling constant $\alpha_s(M_Z)$ and the weak mixing angle $\sin^2\theta_W$, have only been measured with moderate precision. Any interference with new physics, e.g., from a dark sector, could modify SM predictions at higher scales. Thus, precise measurements of SM parameters, including their scale dependence, are crucial.

\subsubsection*{Strong Coupling Constant}
The colour-neutral electron scattering off a coloured proton parton provides direct access to QCD phenomena, with precision enhanced by advanced vertexing, tracking, and highly granular calorimetry, as well as the unique in-situ calibration techniques in NC DIS. 
At the LHeC, the strong coupling constant $\alpha_s(M_Z)$ can be determined with exceptional precision through scaling violations and jet measurements in the Breit frame, see Sec. 4.1 in~\cite{LHeC:2020van}. 
The precise measurement of inclusive NC and CC DIS cross sections provide precise data for the determination of the PDFs. Through higher-order corrections and the DGLAP formalism as implemented in the xFitter framework, the strong coupling can be determined from these data simultaneously with the PDF parameters, and an uncertainty of $\pm 0.00022$ will be achieved~\cite{LHeC:2020van}, a fourfold improvement over the current world average~\cite{ParticleDataGroup:2024cfk}.

\begin{figure}[hbt!]
    \centering
    \includegraphics[width=0.6\linewidth,clip]{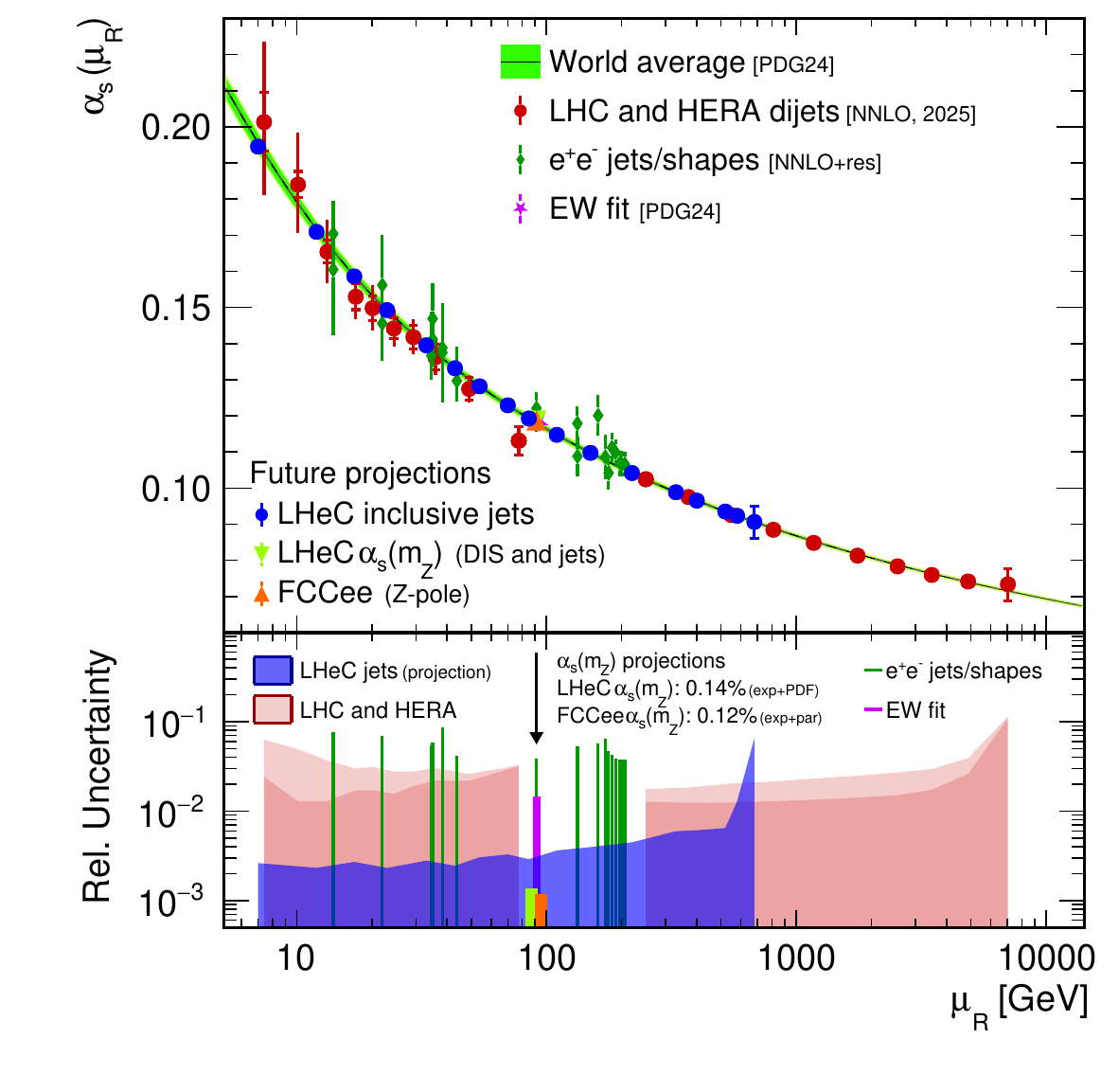}
    \vskip -.7cm
    \caption{Expected measurement of $\alpha_s(\mu_R)$~\cite{LHeC:2020van} and corresponding relative uncertainties at the LHeC 
    compared with presently available measurements~\cite{ParticleDataGroup:2024cfk,Ahmadova:2024emn}, predictions~\cite{dEnterria:2022hzv} and the world average value. The lower panel displays relative uncertainties on $\alpha_s(\mu_R)$, where light-shaded areas show experimental plus theoretical uncertainties and dark shaded areas experimental uncertainties only. 
}
    \label{fig:alphas}
\end{figure}

Inclusive jet cross sections, available at NNLO~\cite{Currie:2016ytq,Currie:2017tpe}, provide further access to $\alpha_s(M_Z)$\footnote{Jet pseudodata were generated in the region $8 < Q^2 < 500 000$\,GeV$^2$ and $4 < p_T < 512$ GeV, using the longitudinally invariant $k_T$ algoritm with $R=1$.}. Leveraging high-resolution calorimetry, in situ calibration, and a jet-energy scale uncertainty of about 0.5\%, $\alpha_s(M_Z)$ can be measured with an uncertainty of $\pm 0.00016$~\cite{LHeC:2020van}, a tenfold improvement over HERA results~\cite{H1:2017bml,Britzger:2019kkb}. These measurements, free of higher-twist and nuclear corrections, challenge lattice results and demand improved higher-order QCD predictions. Moreover, the scale dependence of $\alpha_s$ can be probed over a wide range, as can be seen in Fig.~\ref{fig:alphas}, providing a unique test of the $SU_c(3)$ gauge theory.

\subsubsection*{Electroweak Measurements}
In DIS, space-like momentum transfer occurs via electroweak interactions. At higher $Q^2$,  contributions strongly sensitive to EW physics, as $\gamma Z$ interference and $Z$-boson exchange, and CC DIS mediated exclusively by $W$ bosons, become increasingly significant. The unique kinematic regime of the LHeC allows for high-precision measurements of electroweak parameters 
to be performed through their simultaneous determination in fits together with PDFs within the xFitter framework, see Chapter 5 in~\cite{LHeC:2020van}.

\begin{figure}[hbt!]
    \centering    \includegraphics[width=0.6\linewidth,trim={0 128 0 60},clip]{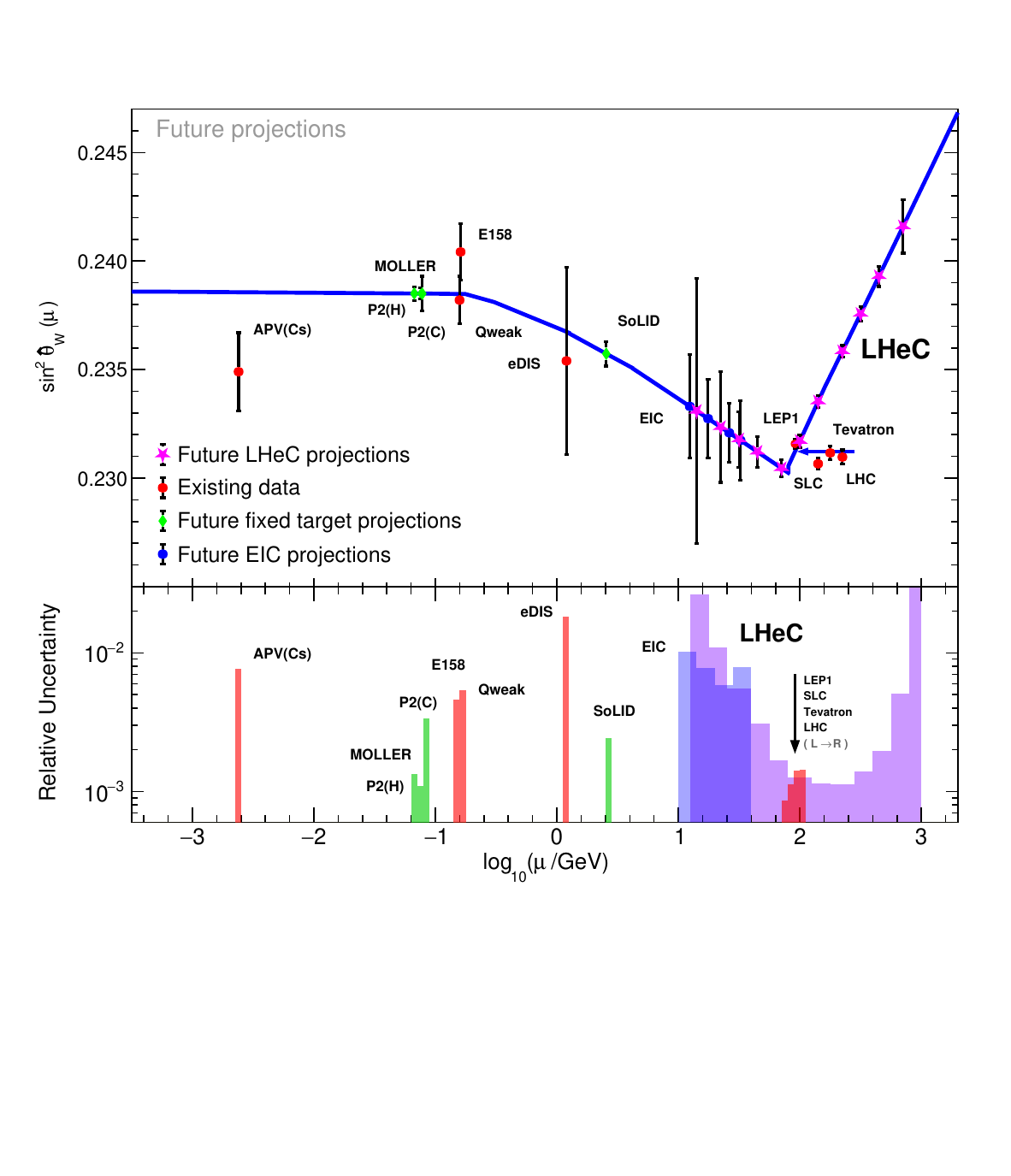}
    \vskip -.7cm
    \caption{Present and future measurements of the running of the weak mixing angle in the $\overline{\textrm{MS}}$ scheme and prospected uncertainties as a function of the scale $\mu$.  (Thanks to: J. Erler, R. Ferro-Hernandez and X. Zheng; updated from~\cite{Erler:2004in,Erler:2017knj}, including recent projections from the Electron Ion Collider (EIC)~\cite{Boughezal:2022pmb}, P2(C) at MESA~\cite{Cadeddu:2024baq} and the LHeC~\cite{Britzger:2020kgg}). The red markers and red uncertainties show present measurements and their relative uncertainties, respectively, and further data points display future projections as indicated.
}
    \label{fig:sin2thevol}
\end{figure}

The highly granular electromagnetic calorimeter enhances measurements by precisely identifying radiated photons. The weak mixing angle $\sin^2\theta_W$ can be determined via polarization asymmetry and neutral-to-charged current (NC/CC) ratios in $e^-p$ scattering, with an uncertainty of $\Delta\sin^2\theta_W = \pm 0.00022$~\cite{Britzger:2020kgg}. While this does not surpass the LEP+SLC $Z$-pole results~\cite{ALEPH:2005ab}, the LHeC uniquely probes the scale dependence of $\sin^2\theta_W$ with percent-level precision over $\sqrt{Q^2}$ from 20\,GeV to nearly 1\,TeV~\cite{LHeC:2020van,Britzger:2022abi}, enabling the first high-scale precision test of the weak gauge structure, as presented in Fig.~\ref{fig:sin2thevol}.

\begin{figure}[htb]
    \centering
    \includegraphics[width=0.34\textwidth,clip]{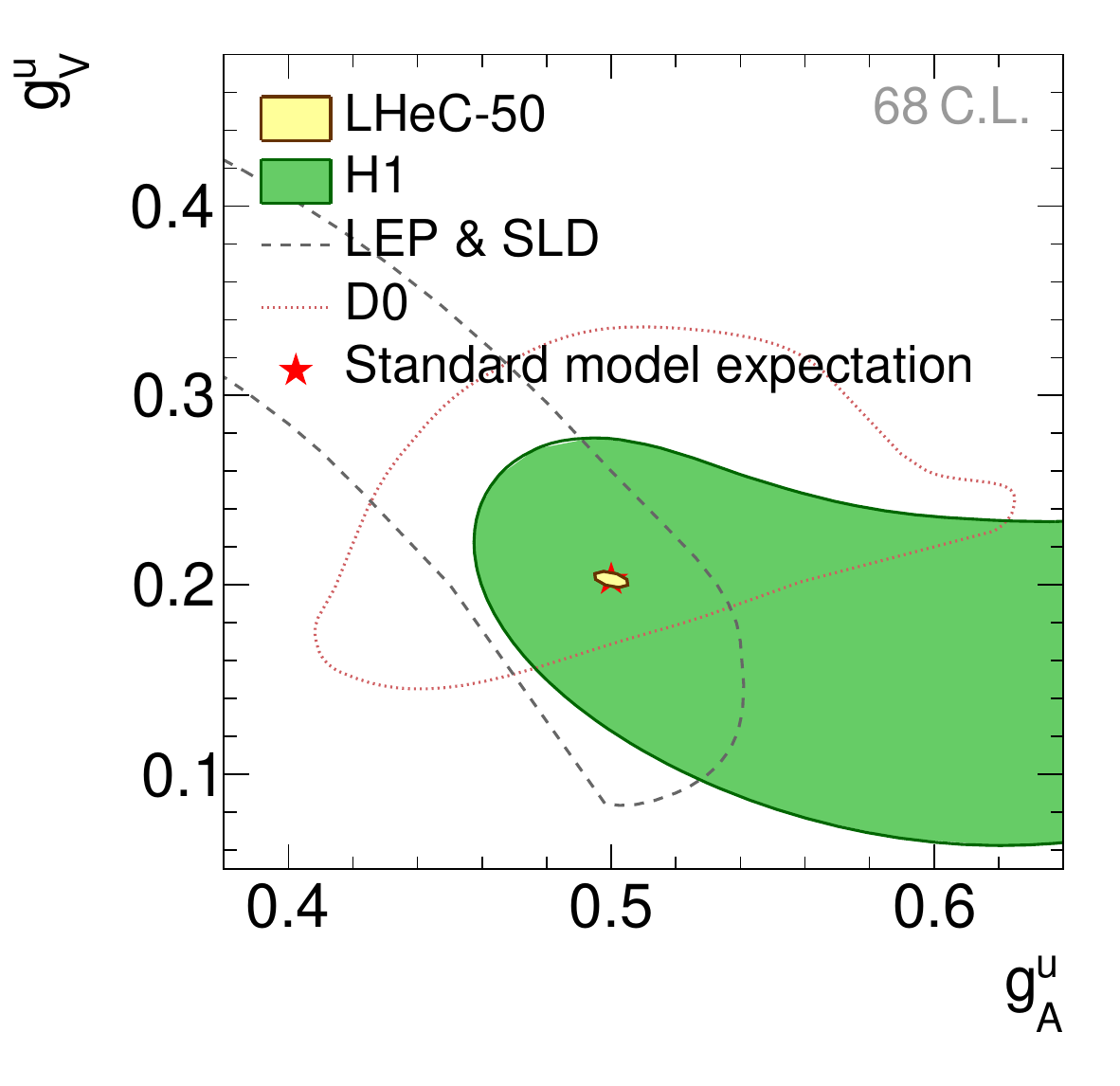}
    \includegraphics[width=0.34\textwidth,clip]{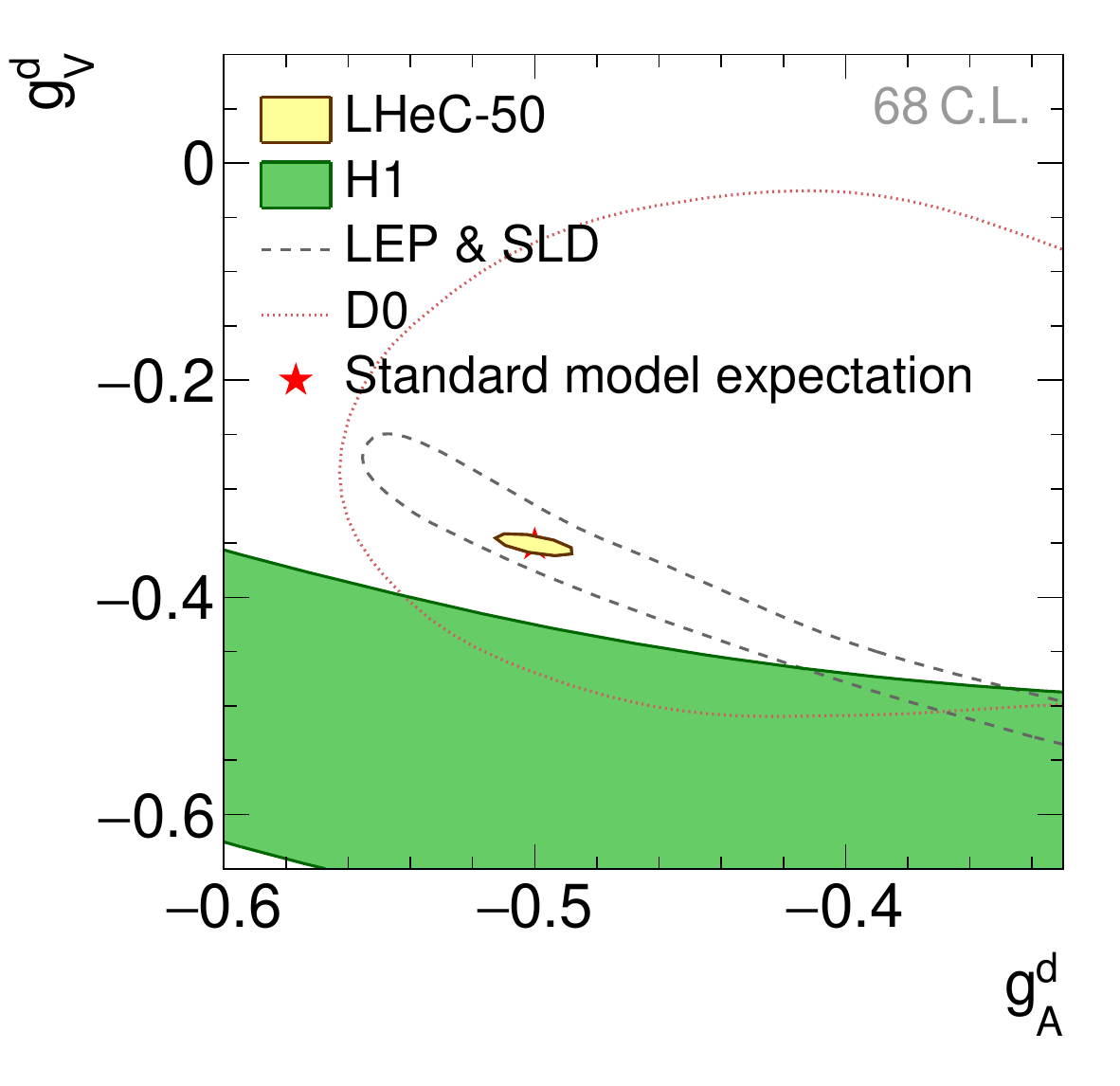}
    \includegraphics[width=0.30\textwidth,trim={0 -41 0 10 }
    ,clip]{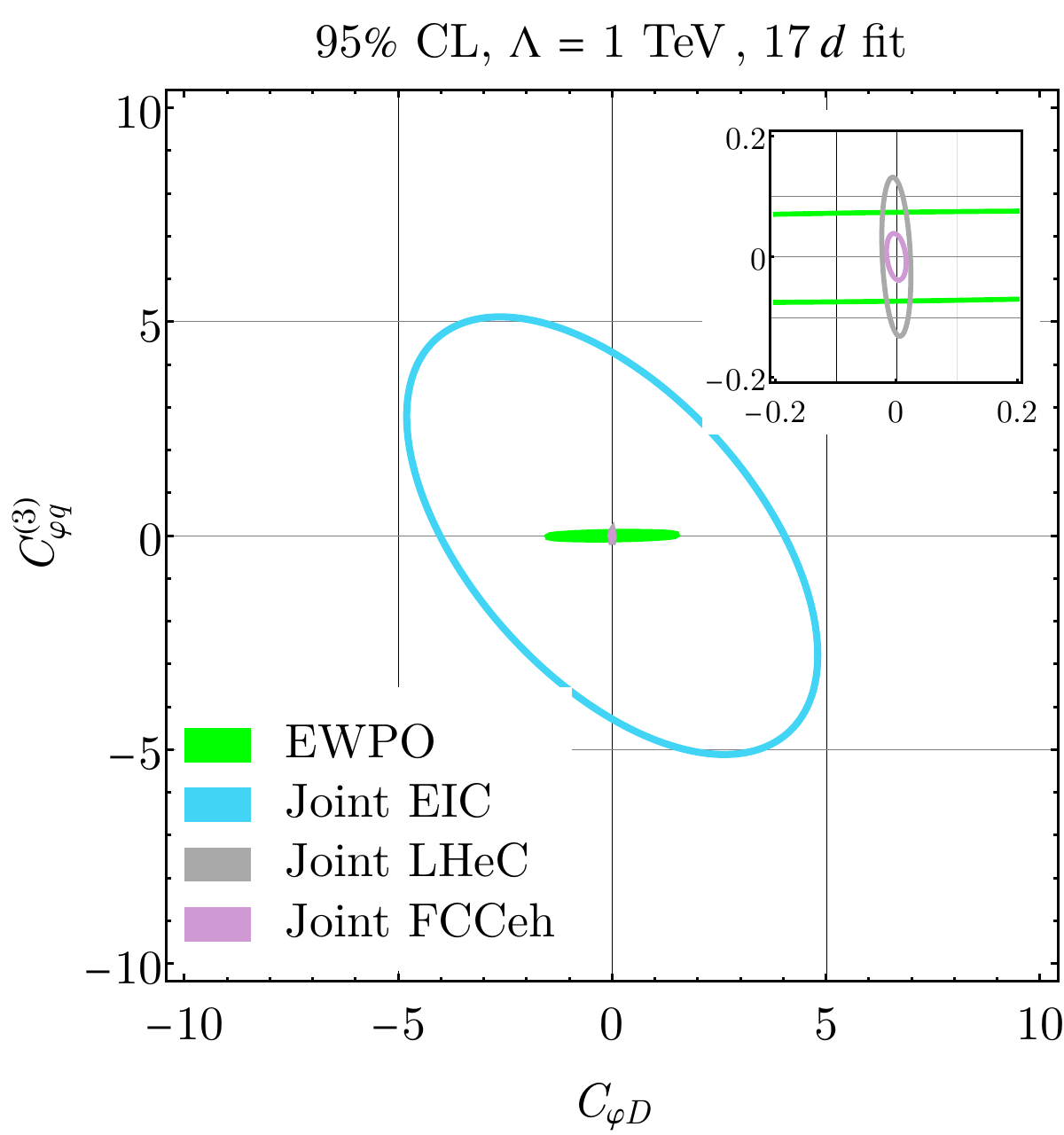}
    \vskip -0.3cm
  \caption{
    Weak NC vector and axial-vector couplings of $u$-type (left)    
    and $d$-type quarks (middle) at 68\,\% confidence level~(CL) 
    for simulated LHeC data with $E_e=50$~GeV.
    The LHeC expectation is compared with results from the 
    combined LEP+SLD experiments~\cite{ALEPH:2005ab}, 
    a single measurement from D0~\cite{D0:2011baz} and one from  H1~\cite{H1:2018mkk}.
    The standard model expectations are diplayed by a red star, partially hidden by the LHeC prospects.
    Right: Marginalized 95\% CL ellipses in the two-parameter fits of SMEFT dimension-6 operators  $C_{\varphi D}$ and $C{\varphi q}$ (right) at UV cut-off scale, $\Lambda=1$~TeV~\cite{Bissolotti:2023vdw}. Shown are joint EIC, LHeC, and FCC-eh fits, as well as the EWPO fit adapted from~\cite{Ellis:2020unq}.
    }
  \label{fig:couplings}
\end{figure}

Beyond the weak mixing angle, the $Z$ exchange also probes weak NC couplings of light quarks ($u, d$) to the $Z$ boson. The two-dimensional uncertainty contours at 68\,\% confidence 
are displayed in Fig.~\ref{fig:couplings} for the two quark families (left and middle) and
compared with available measurements. Reaching a precision at the level of 1\% represents an order-of-magnitude improvement over current most precise measurements~\cite{ParticleDataGroup:2024cfk,H1:2018mkk,Britzger:2022abi,ParticleDataGroup:2018ovx}.

Additionally, in combination with precise $\alpha_s$ and PDF measurements, the LHeC could determine the $W$-boson mass with a precision of just a few MeV~\cite{LHeC:2020van}.

Finally, within the framework of the Standard Model Effective Field Theory (SMEFT), the LHeC is poised to significantly enhance the precision of existing electroweak constraints on the SMEFT parameter space~\cite{Bissolotti:2023vdw}. In many scenarios, it will surpass the accuracy of current bounds while simultaneously disentangling parameter correlations that hinder a comprehensive interpretation of new physics effects, as can be seen for one example in Fig.~\ref{fig:couplings} (right).

This underscores the pivotal role of high-precision DIS measurements in probing physics beyond the Standard Model, further discussed in the next Subsection.

\subsection{Beyond the Standard Model Searches}

Since electrons and quarks interact only via electroweak forces, an electron-proton collider offers a unique environment to probe known phenomena with higher precision and complementary search channels, enabling the potential discovery of weak signals. Despite its lower centre-of-mass energy with respect to the HL-LHC, the LHeC can effectively explore new physics (NP) below the TeV scale. Exotic phenomena accessible at $ep$ colliders cover a broad range of models predicting new (heavy) particles, such as in extensions of the SM Higgs boson sector, supersymmetry, lepto-quark, $Z'$ or vector-like quark models, or modified SM interactions, for instance at the loop level, like flavour-changing neutral or triple-gauge boson couplings.     

An extensive review of the prospects has been given in~\cite{LHeC:2020van} where details and further references can be found, and some results in top physics and in SMEFT analyses have been discussed in previous subsections. Here, the most recent progresses made in the investigation of new physics hidden due to weak couplings are highlighted. Further recent comparisons with other existing or projected experiments can be found in~\cite{deBlas:2025gyz}. We note that most BSM studies consider a nominal electron beam energy of 60\,GeV rather than 50\,GeV. While such reduction impacts on the number of possible new physics signal events produced, good sensitivity is still expected for most models. We also note that, as in the studies presented in previous sections, a DELPHES fast detector simulation has been used in the analyses presented here.

Hidden, or “dark” sectors with feeble interactions mediated by mass mixing with SM particles can provide, depending on the model's implementation, an explanation for the origin of neutrino masses, matter–antimatter asymmetry in the Universe and cosmological inflation, as well as insights into the EWK hierarchy, the strong CP problem and the nature of dark matter. Signatures such as displaced vertices or short tracks may emerge, which can be efficiently identified at the LHeC due to the low level of hadronic background and pile-up with respect to the LHC, and as such offering a considerable window for discovery. 

\begin{figure}[htb]
\centering
\includegraphics[width=0.56\textwidth]{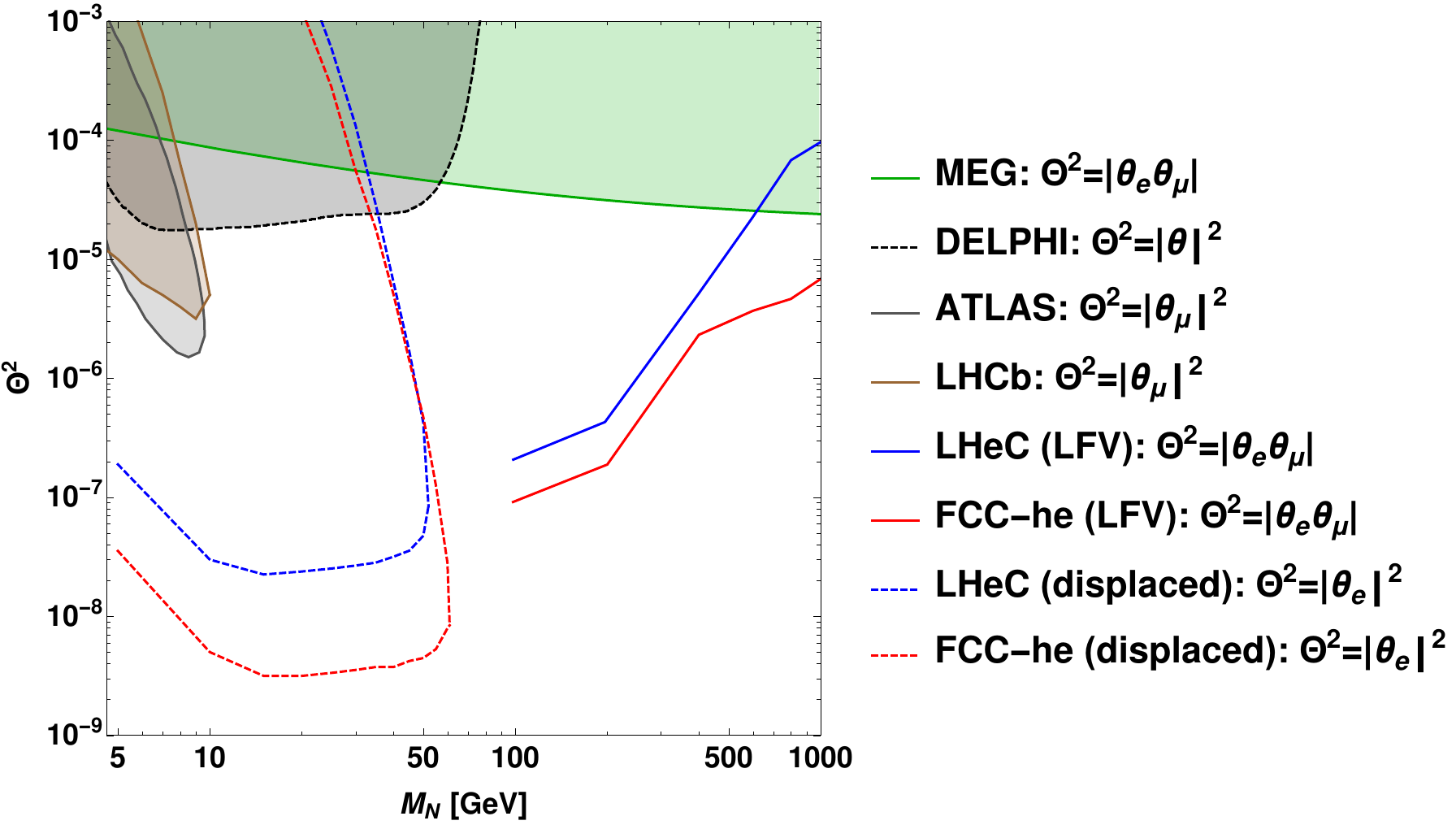}
\includegraphics[width=0.425\textwidth,trim={0 12 0 10 }]{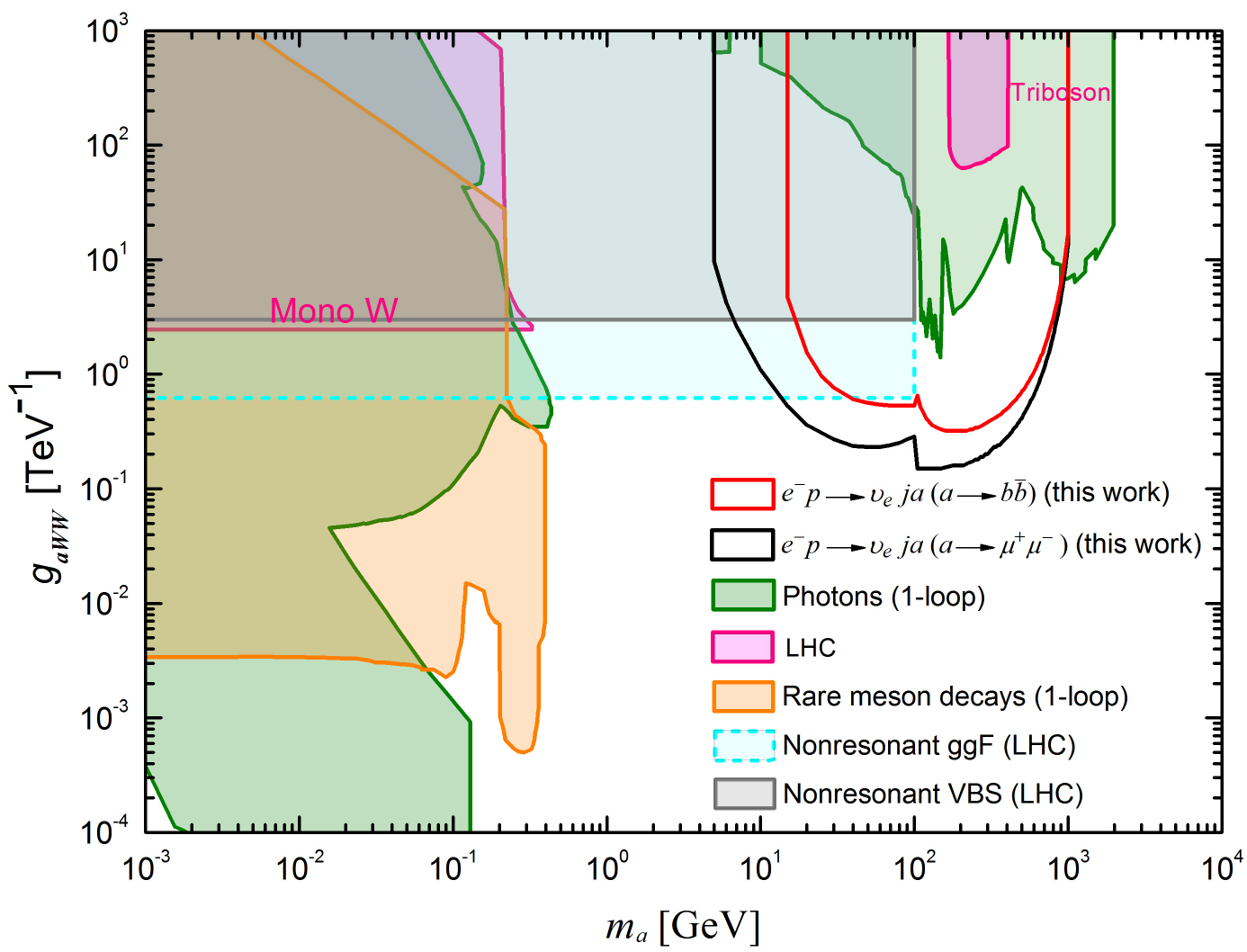}
\caption{Left: Sensitivity of the LFV lepton-trijet searches (at 95\% C.L.) and the displaced vertex searches (at 95\% C.L.)~\cite{Antusch:2019eiz} compared to the exclusion limits from ATLAS~\cite{ATLAS:2019kpx}, LHCb~\cite{Antusch:2017hhu}, LEP~\cite{DELPHI:1996qcc}, and MEG~\cite{MEG:2013oxv}. The sensitivity of the lepton-trijet searches at ep colliders can be generalized to its full $\theta_\alpha$-dependence by replacing $| \theta_e \theta_\mu |$ with $2 |\theta_e|^2 |\theta_\mu|^2/|\theta|^2$.
Right: Projected $2\sigma$ sensitivity limits on the coupling of ALP to $W^{\pm}$ bosons at the LHeC~\cite{Yue:2023mew} in comparison with other current excluded regions.
}
\label{fig:heavy_neutrinos_alps}
\end{figure}

A notable example involves heavy neutrinos (HNs). At the LHeC they can be produced via charged current interactions, with cross sections dependent on the active-sterile mixing parameter $|\theta_e|^2$.
The most promising channels involve lepton flavor violation and displaced vertices, with strong prospect of discovery~\cite{Antusch:2019eiz} as presented in  Fig.~\ref{fig:heavy_neutrinos_alps} (left)\footnote{Recent ATLAS searches for heavy neutrinos, with and without lepton-number violation, constrain sterile neutrinos with masses around 10 GeV to $|U_{\mu/e}^2| \leq 10^{-6}$~\cite{ATLAS:2025lva}. More recent LHCb results set limits at $|U_{\mu N}^2| \lesssim \text{few} \times 10^{-5}$ for masses around 5 GeV~\cite{LHCb:2025ymr}. A direct comparison with the sensitivities presented here is not straightforward due to differing model assumptions. In particular, reinterpreting the ATLAS limits within a more realistic model framework can significantly weaken the constraints~\cite{Tastet:2021vwp}.}
The LHeC sensitivity to heavy neutrinos above 100\,GeV surpasses that of $pp$, due to the lower backgrounds, or $e^+e^-$ colliders. Detection prospects are further improved using jet substructure techniques for boosted $W$ bosons from $N \to e W \to ejj$ decays, which enhances signal discrimination. With 1\,ab$^{-1}$ of data, the LHeC could significantly improve constraints on $|V_{eN}|^2$ over current LHC, $0\nu2\beta$, and EWK limits.
An additional study~\cite{Gu:2022nlj} shows that the lepton number violation signal process $e^-p \to \tau^+jjj$ can be utilized to search for heavy Majorana neutrinos. Results show that the LHeC can probe large additional regions in the parameter space formed by $|V_{\tau N}|^2$ and $|V_{e N}|^2$, with best discovery sensitivity for $m_N \simeq 100$\,GeV. Similar conclusions are extracted in the framework of the $\nu$SMEFT~\cite{Duarte:2025zrg}.

Minimal extensions of the SM often include additional gauge factors, such as $U(1)_X$ symmetries, which can be associated with dark charges linked to dark matter. The corresponding gauge boson, typically called a dark photon ($\gamma'$), gains mass through spontaneous $U(1)_X$ breaking, with QED-like interactions scaled by a small mixing parameter $\epsilon$. They can decay into lepton, hadron, or quark pairs, potentially producing displaced vertices due to their long lifetimes. Search prospects for dark photons at the LHeC focus on processes like $e^- p \to e^- X \gamma'$, where $X$ denotes hadrons and $\gamma'$ decays into charged fermions or mesons~\cite{DOnofrio:2019dcp}. The LHeC’s excellent tracking resolution combined with low backgrounds allows detection of secondary vertices displaced by ${\mathcal O}(0.1)$\,mm. Sensitivity contours in the mass-mixing parameter space show strong reach, complementary to other experiments, particularly for masses below the di-muon threshold where LHC searches face challenges due to pile-up and short-track reconstruction issues.

Axion-like particles (ALPs, $a$), inspired by the QCD axion, are another compelling scenario. Relatively heavy ALPs, i.e., with masses from MeV to TeV, have gained significant interest both as potential dark matter particles and as portals connecting dark and visible matter. At the LHeC, ALPs can be efficiently produced via vector boson fusion processes. Recent studies~\cite{Mosala:2023sse} explore the feasibility of probing ALPs with masses up to $m_a \lesssim 300$\,GeV through channels such as $W^+W^-$, $\gamma\gamma$, $ZZ$, and $Z\gamma$ fusion. The analyses estimate production cross sections and constrain ALP couplings, namely $g_{WW}$, $g_{\gamma\gamma}$, $g_{ZZ}$, and $g_{Z\gamma}$. 
The results show that the LHeC provides stronger limits on $g_{WW}$, $g_{ZZ}$, and $g_{Z\gamma}$ compared to other collider scenarios within the considered mass range, while the limits on $g_{\gamma\gamma}$ are competitive in specific cases. ALP detection prospects via $e^- \gamma \to e^- a$, with $a \to \gamma \gamma$, have been evaluated~\cite{PhysRevD.100.015020}, probing the effective ALP-photon coupling for ALP masses in the range $10\,\text{GeV} < m_a < 3\,\text{TeV}$. 
These studies indicate that the LHeC can significantly improve current LHC bounds, particularly for $m_a < 100$\,GeV. Most recently, studies~\cite{Yue:2023mew} on ALPs through the $e^- p \to \nu_e j a$ process, with $a \to \mu \mu$ or $a \to b\bar{b}$, show that the LHeC might be more sensitive than the LHC in probing ALPs over a range of masses from a few tens of GeV to 900\,GeV. As presented in Fig.~\ref{fig:heavy_neutrinos_alps} (right), the promising sensitivities to the coupling of ALPs with $W^{\pm}$ bosons reach nearly 0.15 and 0.32\,TeV$^{-1}$, respectively.

Additionally, the possibility of BSM physics being absorbed in PDF fits~\cite{Carrazza:2019sec,Greljo:2021kvv,Hammou:2023heg} has been recently addressed in the context of SMEFT. Studies~\cite{Hammou:2024xuj} done for the EIC and the Forward Physics Facility~\cite{Anchordoqui:2021ghd,Adhikary:2024nlv,Feng:2022inv} have been extended to simultaneous fits of PDFs and SMEFT, using the methodology presented in~\cite{Iranipour:2022iak} and released in~\cite{Costantini:2024xae,Cole:2026eex}, {\sc\small SIMUnet}, to fully assess the potential benefits of the LHeC  in disentangling new physics signs from PDFs.
The addition of four-fermion operators to the SM with couplings compatible to present data results~\cite{Hammou} in a negligible effect on NC DIS cross sections at the EIC, while they become relevant for $Q^2>10^4$\,GeV$^2$ and $x>0.5$ accessible at the LHeC. With data at lower $Q^2$ constraining strongly PDFs, the reach for exploration of new physics scenarios, e.g., heavier EW bosons, becomes enlarged.

\subsection{High-energy and high-density measurements of heavy ion collisions}

The LHeC offers a wealth of possibilities for measurements and physics studies in $eA$ collisions~\cite{LHeCStudyGroup:2012zhm,LHeC:2020van,Andre:2022xeh},
with centre-of-mass energies around 0.8\,TeV per nucleon and instantaneous luminosities $\sim 7\cdot 10^{32}$\,cm$^{-2}$s$^{-1}$, to be compared with less than 0.1\,TeV and larger expected luminosities at the EIC~\cite{AbdulKhalek:2021gbh}.
Therefore, the LHeC will extend the kinematic region studied at the EIC by one to two orders of magnitude down in $x$ and up in $Q^2$, roughly corresponding to the same kinematic region accessible in $ep$ shown in Fig.~\ref{fig:kinplane}.

\begin{figure}[h]
    \centering\includegraphics[width=0.5\linewidth]{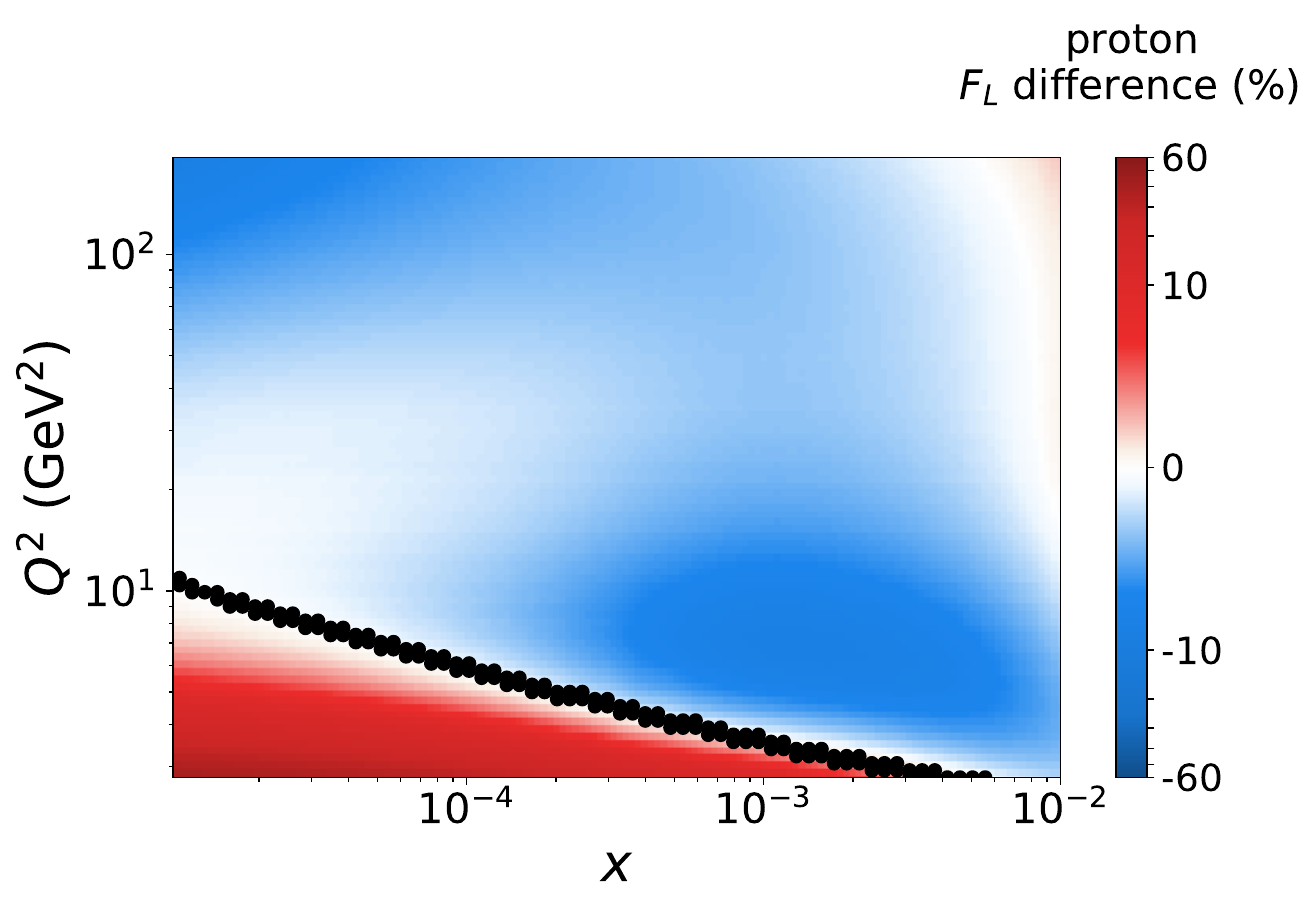}\hfill
    \includegraphics[width=0.5\linewidth]{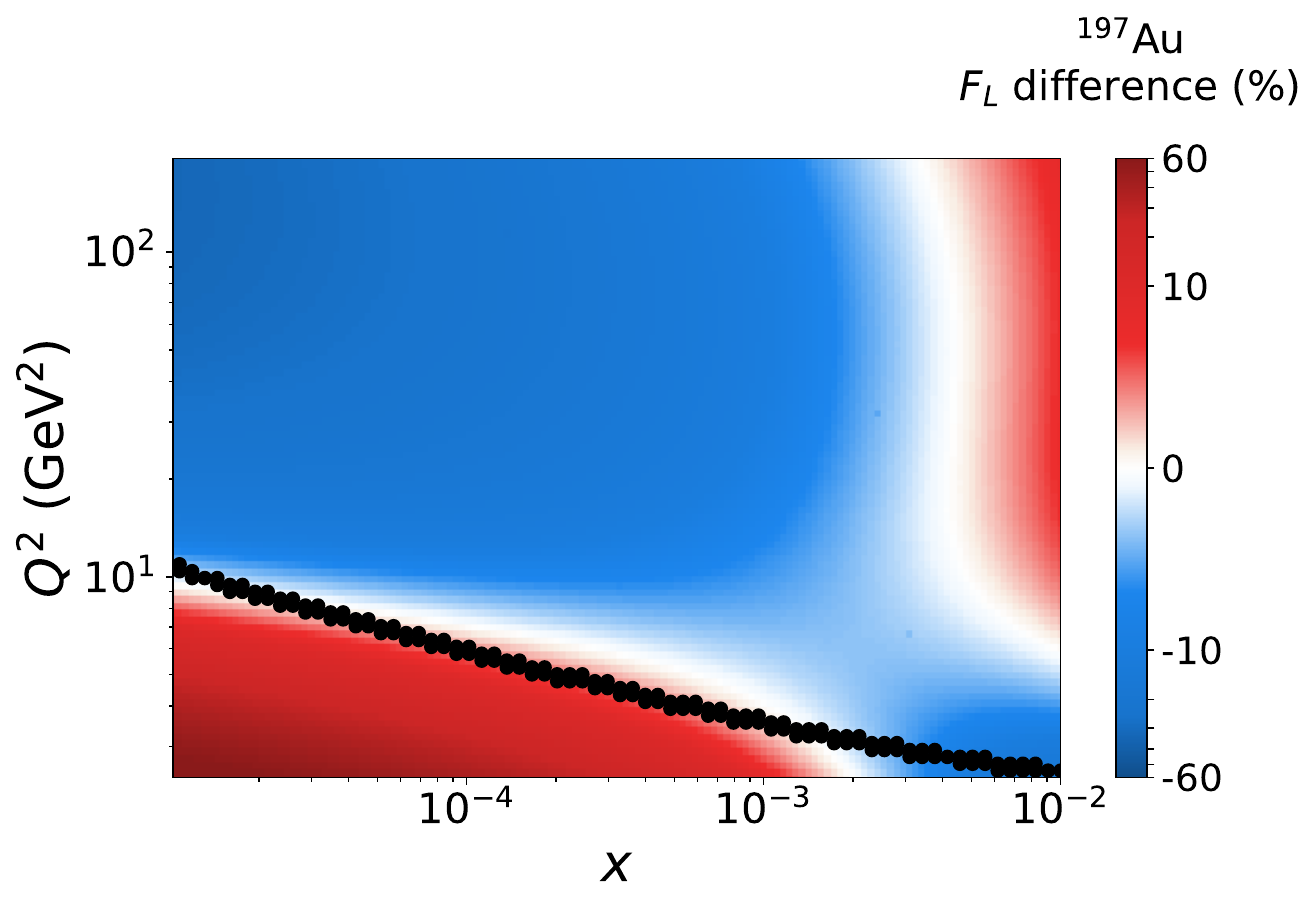}
    \vskip -0.5cm
    \caption{Relative difference, in \%, for the longitudinal structure function $(F_{L}^{\rm BK}-F_{L}^{\rm Rew})/F_{L}^{\rm BK}$ between the saturation predictions (labeled "BK") structure functions and the matched  $F_{\rm L}$ for proton (left) and Au (as example of heavy nuclei, right) obtained from DGLAP fits and named "Rew", as a function of $x$ and $Q^2$. The color scale/axis goes in a linear scale from $-10\, \%$ to $10\, \%$ and in a logarithmic scale outside that range. The black dots indicate the points where results from DGLAP fits are forced to match those of saturation. Plots taken from~\cite{Armesto:2022mxy},where details can be found.}
    \label{fig:smallx}
\end{figure}

Most aspects developed in the mentioned references lie at the core of one of the key questions in nuclear physics: how the nucleon structure is modified when immersed in a nuclear medium. For example, by a precise measurement and complete unfolding of distributions of all parton species in a single nucleus comparable to that in the proton, the LHeC will revolutionize our understanding of nuclear structure in kinematic regions not accessible at the EIC. The longitudinal structure function $F_L$, an observable especially sensitive to the gluon distribution and the dynamics at small-$x$ through its evolution in $Q^2$ at low $x$, will be accessible in a large kinematic region well beyond that at HERA~\cite{LHeC:2020van}. In this way, the LHeC will also clarify the relevance of non-linear (saturation) phenomena by adding the nuclear mass lever arm to $ep$ studies,
see as an example Fig.~\ref{fig:smallx}~\cite{Armesto:2022mxy} where the expected difference in $Q^2$ evolution between linear and saturation approaches is shown to be significant.
This is essential to quantitatively understand saturation as a density effect, which underlies all its existing explanations\footnote{Searches for saturation should also consider the possibility of a different linear evolution, e.g., including small-$x$ resummation~\cite{Ball:2017otu}. But such linear effects would affect the densities of partons in the proton and nuclei equally, while saturation as a density effect would be stronger for the latter.}.

\subsection{$\gamma\gamma$ physics}

The experimental conditions at the LHeC will be particularly favourable
for studies of the exclusive production via photon-photon fusion -- the
event pile-up will be negligible, highly improving the efficiency and
purity of exclusive event selection, and at the same time some sources
of overwhelming backgrounds, as due to the Drell-Yan process for
example, will be absent. Moreover, no event suppression will occur due
to hadronic re-scattering and the associated significant theoretical and
experimental uncertainties will not intervene. Finally, the data
streaming (or the absence of triggering) will allow for detection and
reconstruction of much wider range of final states. As a result, despite
the lower-on-average $\gamma\gamma$ centre-of-mass energy than that at
the HL-LHC, significantly larger statistics of observed
$\gamma\gamma$ events are expected at the LHeC.

The prime example is the two-photon production of $W$-boson pairs:
assuming an integrated $ep$ luminosity of 1\,ab$^{-1}$, about 100000 boson 
pairs will be produced at the LHeC, see~\cite{Forthomme:2024qqi,Forthomme:2024bqq} for studies of the relevant cross sections. As the
semi-leptonic decays will be detected with high efficiency, that will
lead to improved sensitivities to anomalous couplings by an order of
magnitude with respect to the measurements at the HL-LHC. In addition,
the first-ever direct evidence for two-photon production of $Z$-boson pairs
is expected at the LHeC.

A fully exclusive two-photon production of tau-lepton pairs has a large
cross section of almost 50\,pb at the LHeC, for pair invariant masses
above 10\,GeV~\cite{Forthomme:2024qqi,Forthomme:2024bqq}. This leads to very large event
statistics within the acceptance of central detector and excellent
sensitivity to the $\tau$ anomalous magnetic moment $a_\tau$. Already with an integrated $ep$ luminosity of 100\,fb$^{-1}$, the expected LHeC sensitivity is
an order of magnitude better than that achieved at the LHC. As a result,
for the first time the experimental uncertainties will be better than
the higher order corrections to the $\tau$ magnetic moment in the SM.

\section{\NoCaseChange{LHeC} physics enabling HL-LHC \& high-energy proton collider physics}
\label{sec:enabler}

\subsection{LHeC-improved PDFs at hadron colliders}
\label{sec:luminosities}

Deep inelastic scattering data dominate our current understanding of the structure of the proton and form an essential backbone supporting any interpretation of current hadron collider data. At the LHC, searches for new physics and measurement results in the Higgs, electroweak or QCD sectors all reflect the precision of the past HERA data. By the end of the HL-LHC, anticipating a continued improvement in the theoretical calculations, most of the HL-LHC scientific return will be limited by uncertainties in proton structure.

\begin{figure}[htb]
\centering
\includegraphics[width=0.49\textwidth]{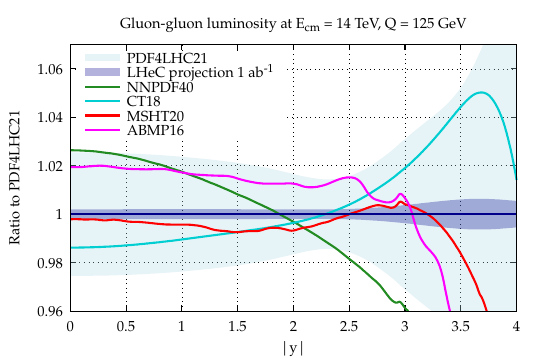}
\includegraphics[width=0.49\textwidth]{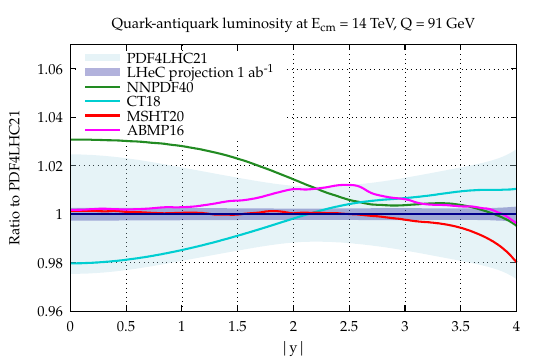}
\includegraphics[width=0.49\textwidth]{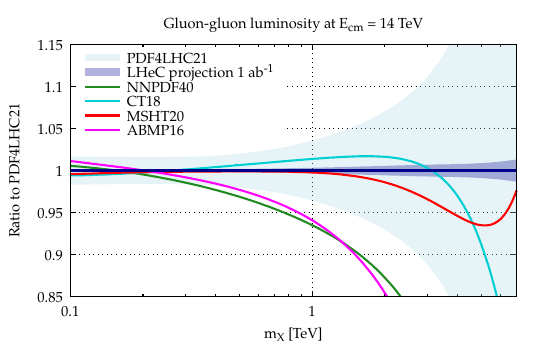}
\includegraphics[width=0.49\textwidth]{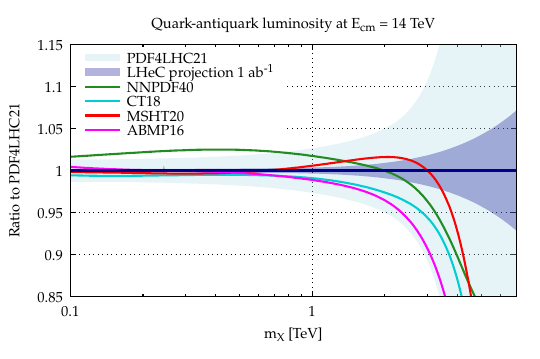}
\includegraphics[width=0.49\textwidth]{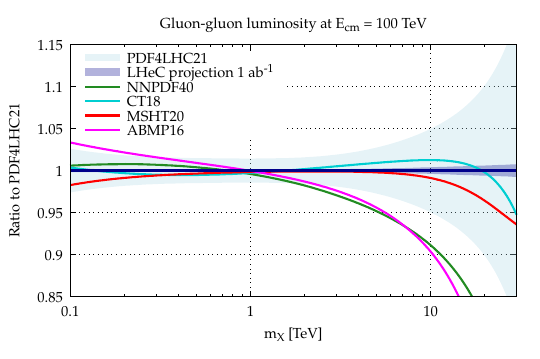}
\includegraphics[width=0.49\textwidth]{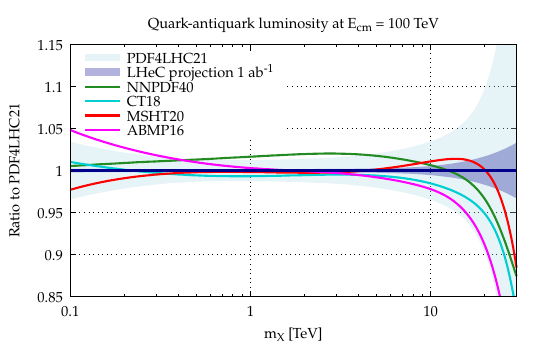}
\vskip -0.3cm
\caption{Parton-parton luminosities as a function of rapidity ($gg$ on the left, $q\bar q$ on the right), at $\sqrt{s}=14$~\,TeV (top), and as a function invariant mass at $\sqrt{s}=14$ (middle) and 100\,TeV (bottom), normalized to PDF4LHC21~\cite{PDF4LHCWorkingGroup:2022cjn}. LHeC results at NNLO are shown with uncertainty bands together with central values of ABMP16~\cite{Alekhin:2017kpj}, CT18~\cite{Hou:2019efy}, MSHT20~\cite{Bailey:2020ooq} and NNPDF4.0~\cite{NNPDF:2021njg}.}
\label{pdfconstraints}
\end{figure}

An optimal exploitation of the huge samples expected at the HL-LHC and, on a longer timescale, the new energy frontier at the High-Energy LHC (HE-LHC)~\cite{FCC:2018bvk} or the FCC-hh, will require an $ep$ scattering dataset of corresponding kinematics and precision. This context forms a major motivation for the LHeC. The impact of LHeC data on parton-parton luminosities relevant at the HL-LHC and FCC-hh, derived from the PDF analyses presented in Sec.~\ref{sec:pdf}, is illustrated in Fig.~\ref{pdfconstraints}. In the context of precision Higgs and Drell-Yan measurements at $\sqrt{s}=14$\,TeV, the uncertainty in the rapidity dependence of the $gg$ and $q\bar{q}$ luminosities is of particular relevance. At $\sqrt{s}=100$\,TeV, the mass dependence of these luminosities is relevant for searches, and illustrates the expected uncertainties in the predictions of processes below the weak scale, where additional uncertainties come from the eventual breaking of standard collinear factorization~\cite{Bonvini:2018iwt}. The anticipaged LHeC uncertainty is a factor of at least 5 better than that of the PDF4LHC21 PDF set. It also clearly discriminates between the different global fits from which PDF4LHC21 is derived. Note that, as shown in Fig.~\ref{fig:pplumi}, the impact of LHeC data will already be very large with just one year of running.

\subsection{Impact on hadron-collider physics}
\label{sec:impact}

Current prospects of the HL-LHC precision in Higgs and electroweak measurements are limited by strong-interaction uncertainties. 
A primary example of the impact of LHeC data on the physics program of the HL-LHC is given by the enhanced precision in the prediction of the Higgs boson cross section in the Standard Model. A breakdown of $gg\to H$ cross section uncertainties, taken from~\cite{Dainese:2019rgk} and updated for recent progress~\cite{Czakon:2023kqm,Czakon:2024ywb,Bonetti:2020hqh,Bonetti:2022lrk}, is given in Tab.~\ref{tab:sigh}. The numbers in Tab.~\ref{tab:sigh} show the present perturbative and PDF$+\alpha_S$ uncertainty; scenario "S2" of~\cite{Dainese:2019rgk} roughly corresponds to a halving of these uncertainties on the timescale of the HL-LHC. The expected impact of LHeC on PDFs and $\alpha_S$ is estimated following~\cite{AbdulKhalek:2019mps}, with an improvement factor of 2.5--3 with respect to S2, reducing the corresponding uncertainty in the $gg\to H$ cross section to about 0.5\% and  making this contribution negligible with respect to the anticipated perturbative uncertainty. This opens up the possibility of a significant decrease in uncertainty when the theoretical errors are further reduced.

\begin{table}[htb]
    \centering
    \begin{tabular}{cc|c|c|c|c|c|c|c|c}
    \hline
    $\sqrt{s}$ [TeV] & $\sigma_{gg\to H}$ [pb] & \multicolumn{2}{c|}{TH uncertainty} & \multicolumn{3}{c|}{PDF+$\alpha_S$ uncertainty} & \multicolumn{3}{c}{Total} \\
    &                         & Ref. & S2 & Ref. & S2 & S2+LHeC & Ref. & S2 & S2+LHeC    \\
    \hline
    14     &  54.7  & 3.9\% & 2.0\% & 3.2\% & 1.6\% & 0.5\% & 5.1\% & 2.6\% & 2.0\% \\
    27     & 146.6  & 4.0\% & 2.0\% & 3.3\% & 1.7\% & 0.6\% & 5.2\% & 2.6\% & 2.1\% \\
    100    & 804.4  & 4.2\% & 2.1\% & 3.7\% & 1.9\% & 0.7\% & 5.6\% & 2.8\% & 2.2\% \\
    \hline
    \end{tabular}
    \caption{Gluon-fusion Higgs cross-sections at $\sqrt{s}$=14, 27 and 100~TeV. The reference values and uncertainties are taken from~\cite{Dainese:2019rgk} and symmetrized. The improved theoretical predictions are from~\cite{Czakon:2023kqm,Czakon:2024ywb,Bonetti:2020hqh,Bonetti:2022lrk}.}
    \label{tab:sigh}
\end{table}

\begin{figure}
\centering
\includegraphics[width=0.6\textwidth]{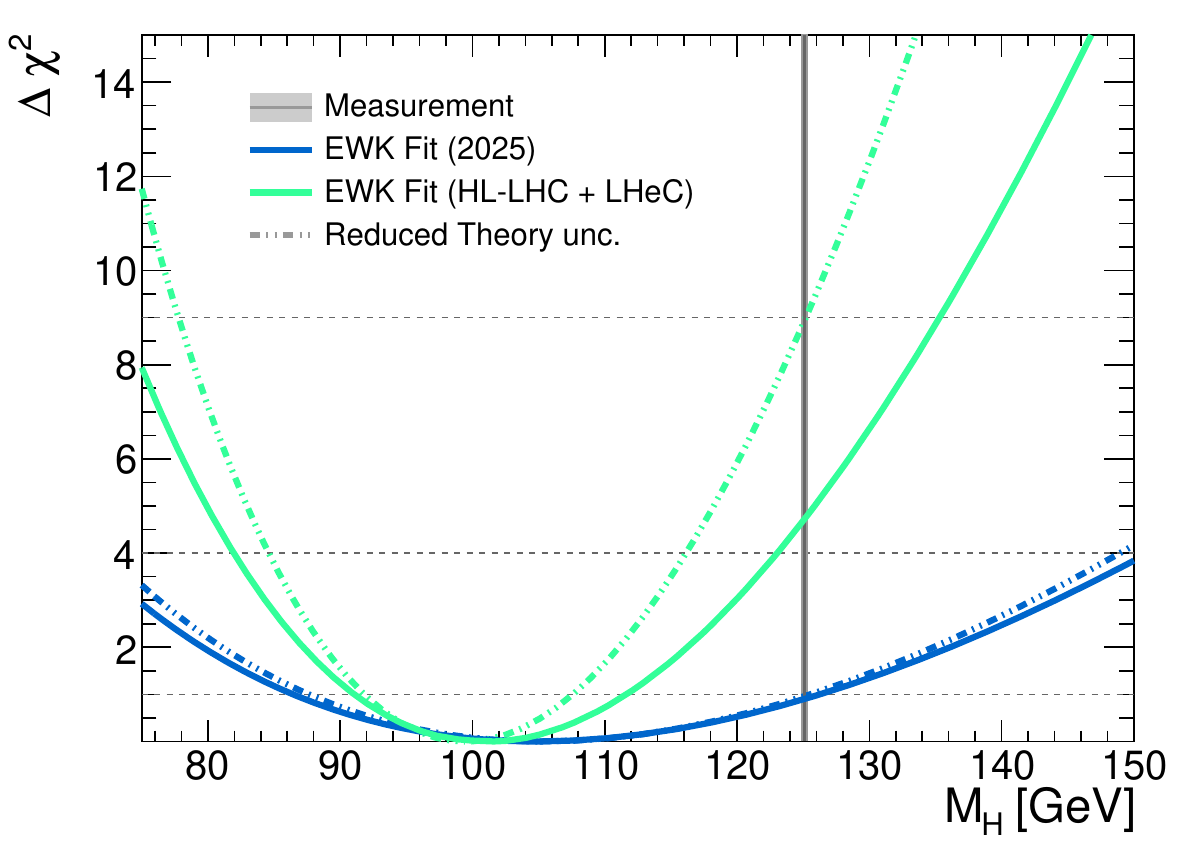}
\vskip -0.3cm
\caption{Present and future precision in the indirect determination of the Higgs-boson mass in the Standard Model.\label{ewpo_fit}}
\end{figure}

The impact of the LHeC on the determination of the fundamental parameters of the electroweak and strong theories has been studied in~\cite{LHeC:2020van}, where the combination of precise LHC and HL-LHC data with LHeC-improved PDFs resulted in expected precisions of 5 MeV in $m_W$, 300~MeV in $m_\text{top}$, $8\times 10^{-5}$ in $\sin^2\theta_\text{eff}^\ell$ and $1.6\times 10^{-4}$ in $\alpha_\text{s}$. Compared to that reference, further improvement in the $W$-boson mass is expected by taking into account the recent measurements and combinations~\cite{ATLAS:2024erm,CMS:2024lrd,LHC-TeVMWWorkingGroup:2023zkn}, with a current combined uncertainty of 7.7 MeV~\cite{ParticleDataGroup:2026aaa}. Improved analysis techniques and the availability of measurements from all LHC collaborations now allow to anticipate an HL-LHC-combined measurement uncertainty of about 5--6\,MeV, which would reduce to 3\,MeV when including LHeC data. Improvements in the top-quark mass are only mildly impacted by the PDFs, but benefit from high statistics at the HL-LHC and from the improved simulation of jet fragmentation and hadronization, as derived from clean jet production studies at the LHeC; an uncertainty below 200 MeV in the top-quark mass is expected. An overview of electroweak precision at the LHeC and its impact on the indirect determination of the Higgs-boson mass in the SM\cite{Baak:2014ora} are summarized in Tab.~\ref{tab:ewpo} and Fig.~\ref{ewpo_fit}. Assuming current theoretical uncertainties, the precision in the indirect determination of $m_H$ reduces from $\delta m_H=25$\,GeV to 10\,GeV, an improvement dominated by the reduced uncertainty in $m_W$. Anticipating a significant reduction in theoretical uncertainties on the timescale of the LHeC, $\delta m_H$ further reduces to 7\,GeV. In this scenario, and assuming measurements stay at their current values, the direct and indirect determinations of $m_H$ can potentially exclude the SM at the level of three standard deviations. The uncertainty in $\Delta\alpha_\text{had}$, the hadronic contribution to the evolution of $\alpha_\text{QED}$, is assumed unchanged, while significant improvement is expected, e.g., from lattice calculations~\cite{Conigli:2025tbb}.

\begin{table}[]
    \centering
    \begin{tabular}{cccccc}
    \hline
     Parameter & Unit & Value & \multicolumn{3}{c}{Uncertainty} \\
               &      &       & Present  & HL-LHC & HL-LHC+LHeC \\
    \hline
    $m_Z$ & MeV & 91187.6 & 2.1 & $<2$ & $<2$ \\
    $m_W$ & MeV & 80369.2 & 13.3 & 5--6 & 3 \\
    $\sin^2\theta^\ell_\textrm{eff}$ & & 0.23152 & 0.00016 & 0.00016 & 0.00008\\
    $m_\textrm{top}$ & GeV & 172.57 & 0.29 & $<0.2$ & $<0.2$ \\
    $\alpha_S$ & & 0.1179 & 0.0010 & 0.0008 & 0.00016\\
    \hline
    \end{tabular}
    \caption{Present and future precision in the determination of selected parameters of the electroweak and strong interactions.}
    \label{tab:ewpo}
\end{table}

Measurements at the LHeC will have strong implications on the heavy ion (HI) program at the HL-LHC~\cite{Citron:2018lsq}. First, they will provide information on nuclear partonic structure, both collinear nuclear PDFs and more differential information via nuclear transverse momentum and generalized parton distributions. Second, they will clarify the dynamics for particle production in nuclear collisions (from soft to hard: non-linear dynamics, nuclear shadowing and cold nuclear matter effects on hard probes like jets and quarkonia) in the region of relevance for the LHC, a region that the EIC cannot access. Such information is crucial for the characterization of the quark-gluon plasma and of the processes leading to the fast applicability of relativistic hydrodynamics in high-energy HI collisions, directly linked to the small system problem.
As to the implications on future colliders, the LHeC will provide details on partonic structure that are essential for any future HI program, like that at the FCC-hh~\cite{FCC:2018vvp}. Together with data taken at the EIC, it will enable a comprehensive picture of nuclear partonic structure from small to large $x$, making extrapolations towards even smaller $x$ more reliable as a basis for studies at future colliders. Such details are key for establishing the benchmark on top of which a precise characterisation of the quark-gluon plasma as a new phase of matter is performed.

\section{\NoCaseChange{LHeC} in the global context}
\label{sec:global}

In terms of new physics searches, the LHeC and HL-LHC provide complementary approaches to probing anomalous interactions and deviations from the SM, particularly in scenarios involving Long Lived Particles (LLP) and new physics frameworks such as Axion Like Particles (ALP),  dark photons, sterile neutrinos or additional scalar fields.

In the context of LLP searches, the LHeC’s clean environment is particularly advantageous for studying displaced vertices or non-prompt decays with shorter lifetimes. The LHeC’s ability to detect such decays has been extensively reviewed in~\cite{LHeC:2020van}. For example, good sensitivity can be achieved for Higgs boson decays to ALPs, dark photons, or other scalars that can produce displaced signatures more challenging to identify at the HL-LHC due to pile-up. Short-lived charged particles decaying within the tracking volume can be detected for $c\tau$ in the range $10^{-3} - 10^{-5}$~m, as opposed to the HL-LHC reach for $c\tau$ around $10^{-2} - 10^{-3}$~m or above. 

While the LHeC is limited in its reach for high-energy phenomena, it has great potential in precision studies of couplings, parton distributions, and certain effective operators that manifest as deviations in DIS cross-sections or angular distributions. For example, the LHeC is uniquely suited to constraining modifications to structure functions or probing new states coupled to electrons via suppressed or rare processes, making it complementary to the HL-LHC’s greater center-of-mass energy reach. 

Other than HERA, the EIC~\cite{AbdulKhalek:2021gbh} is the only other
high energy DIS facility that is expected to be operational
before the LHeC, with first collisions scheduled for 2034.
The EIC aims for comparable luminosities to those of the LHeC, 
in a centre-of-mass energy range 
($30 < \surd s < 140 \ {\rm GeV}$) that is intermediate between
fixed target facilities and HERA. 
The scientific priorities of the EIC lie in the 3-dimensional mapping of proton structure, leading to an understanding of 
nucleon spin and mass generation. Being enabled to a large extent
by polarising the protons, these aims are complementary to those
of the LHeC. 
Both projects aim to improve our knowledge of proton and nuclear
collinear PDFs, though with the emphasis on different kinematic
regions. 
The combination of EIC and LHeC thus leads to a comprehensive
understanding all aspects of the structure and 
dynamics of strongly interacting particles, with the LHeC 
extending the kinematic range to considerably lower $x$
values and also offering a BSM and electroweak program that
is not possible at the EIC's lower $\surd s$ values. 
There are also considerable 
synergies between the two projects in terms of
detectors technologies, as discussed in more detail in 
Sec.~\ref{detector:steppingstone}. 

\begin{figure}[htb!]
\begin{center}
\includegraphics[width=0.4\textwidth]{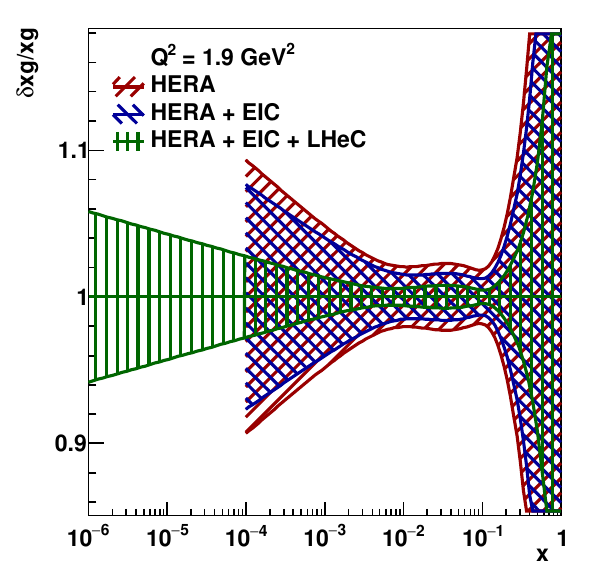}\hskip 1 cm\includegraphics[width=0.4\textwidth]{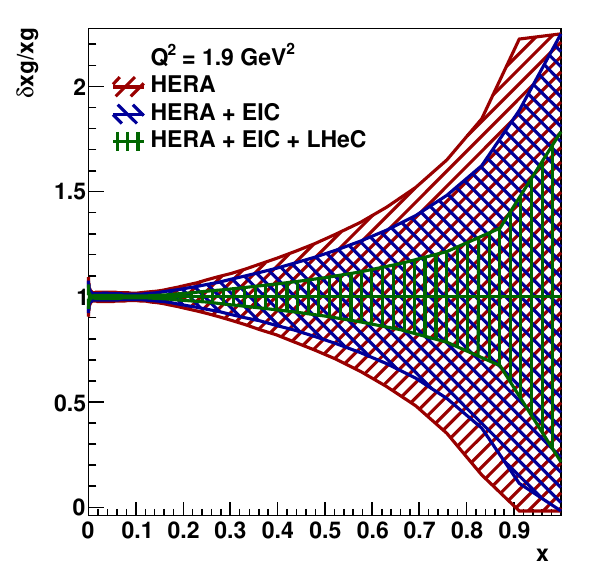}
\end{center}
\vskip -0.7cm
\caption{Uncertainty in the determination of the gluon distribution at $Q^2=1.9$\,GeV$^2$ in logarithmic (left) and linear (right) scale in fits to HERA data plus EIC projections plus LHeC~\cite{Wichmann}.}
\label{fig:glueEIC}
\end{figure}

Studies of complementarities between the EIC and LHeC
arising from their differing kinematic coverages
have been carried out in the context of proton PDFs and the determination of $\alpha_s$ through inclusive DIS measurements~\cite{Wichmann}.
The EIC in combination with existing HERA data will provide improvements in proton and nuclear PDFs~\cite{Armesto:2023hnw,Khalek:2021ulf} and $\alpha_s$~\cite{Cerci:2023uhu}. On top of this, see Fig.~\ref{fig:glueEIC}, the LHeC will further improve the PDF determination in a much larger kinematic region and decrease the uncertainty in $\alpha_s$ by more than a factor 2 compared to HERA+EIC.
It should be noted that addition of FCC-eh projected pseudodata to these fits improves the PDF determination only marginally in the LHeC kinematic region~\cite{Wichmann}. Further complementarities can be observed for the weak mixing angle in Fig.~\ref{fig:sin2thevol}, where the EIC explores the evolution with comparable precision at  scales small compared to those achievable at the LHeC.

The role of LHeC in the physics of the Higgs boson is considered in two alternate scenarios. In the first scenario, CERN proceeds with the FCC program, and the LHeC acts as the ultimate upgrade of the LHC and as a technological bridge between the HL-LHC and FCC-ee. In the second scenario, a circular $e^+e^-$ collider is built elsewhere (e.g., CEPC in China), and CERN proceeds with a fast-track hadron-collider; in this case, LHeC optimizes the scientific reach of both its predecessor and its successor. 

\begin{figure}[h]
\begin{center}
\includegraphics[width=0.49\textwidth]{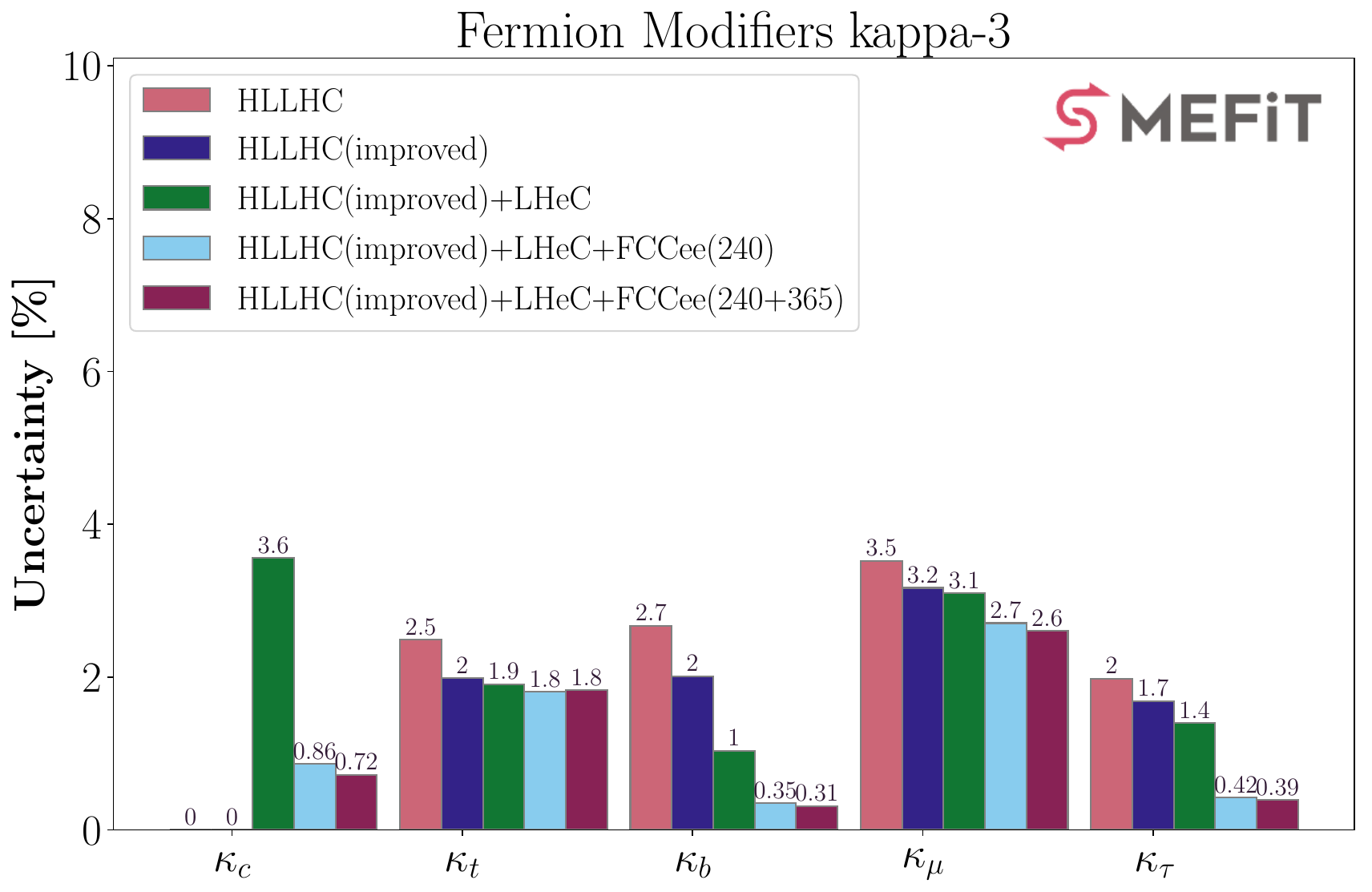}
\includegraphics[width=0.49\textwidth]{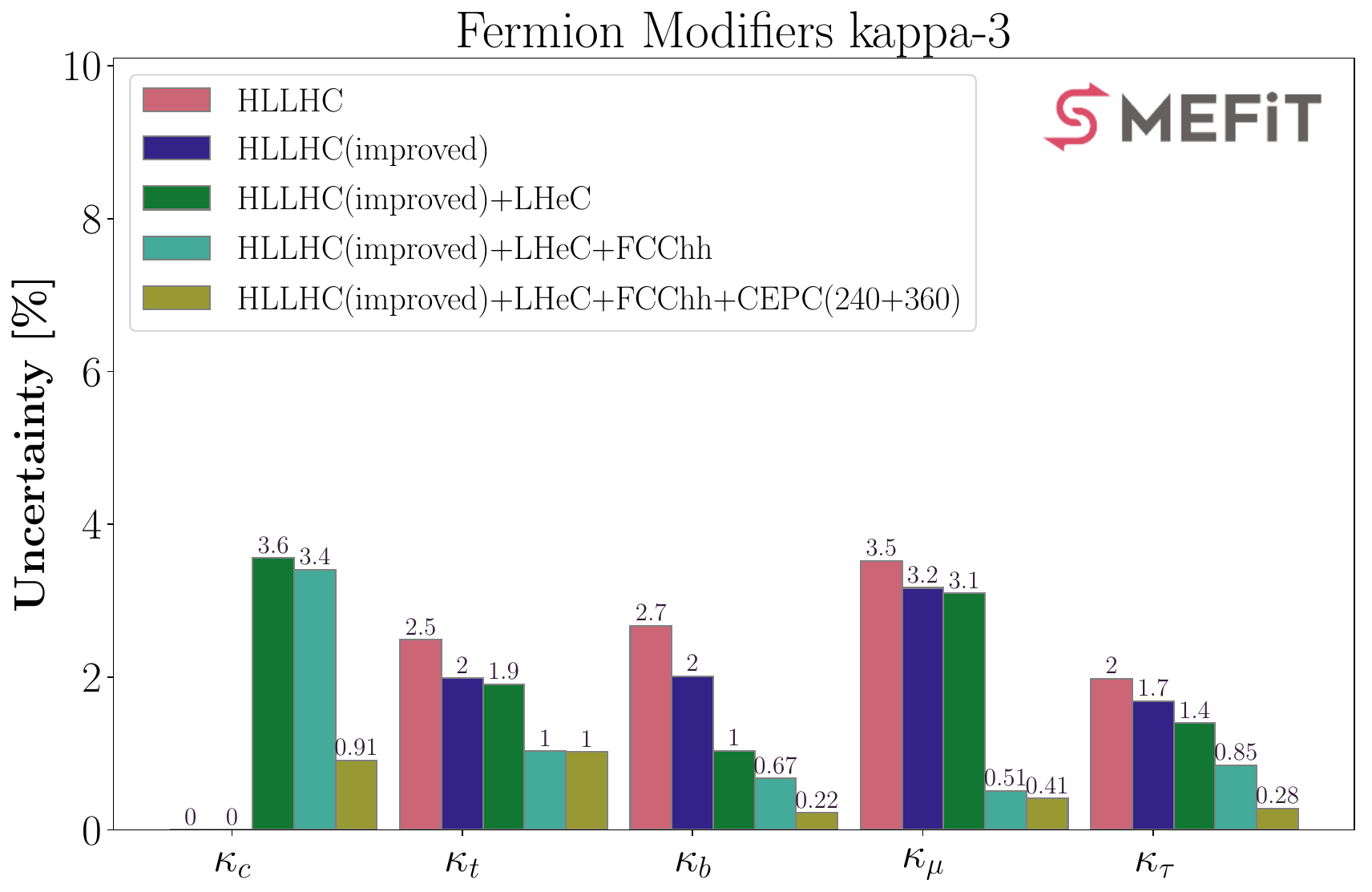}
\includegraphics[width=0.49\textwidth]{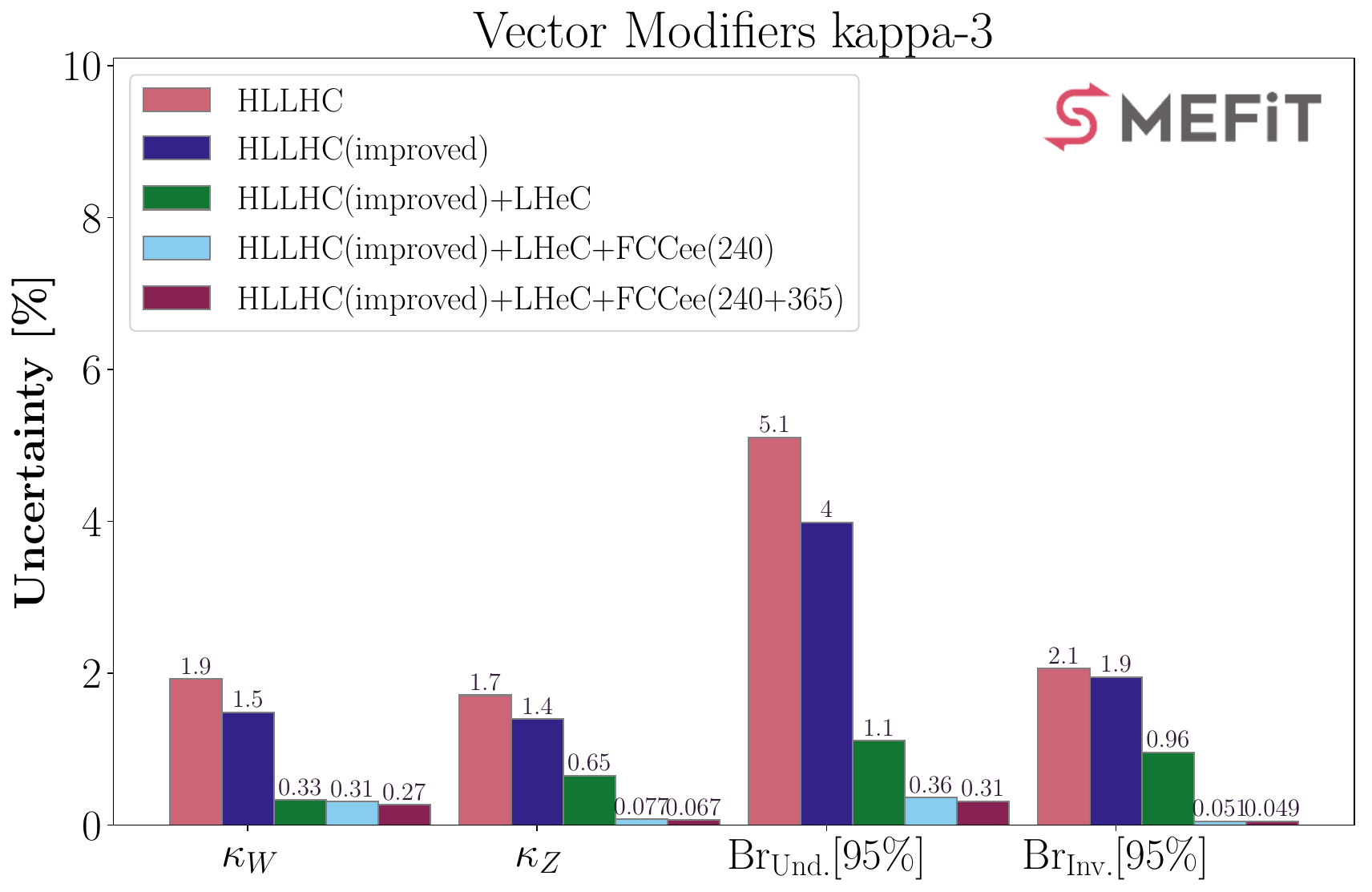}
\includegraphics[width=0.49\textwidth]{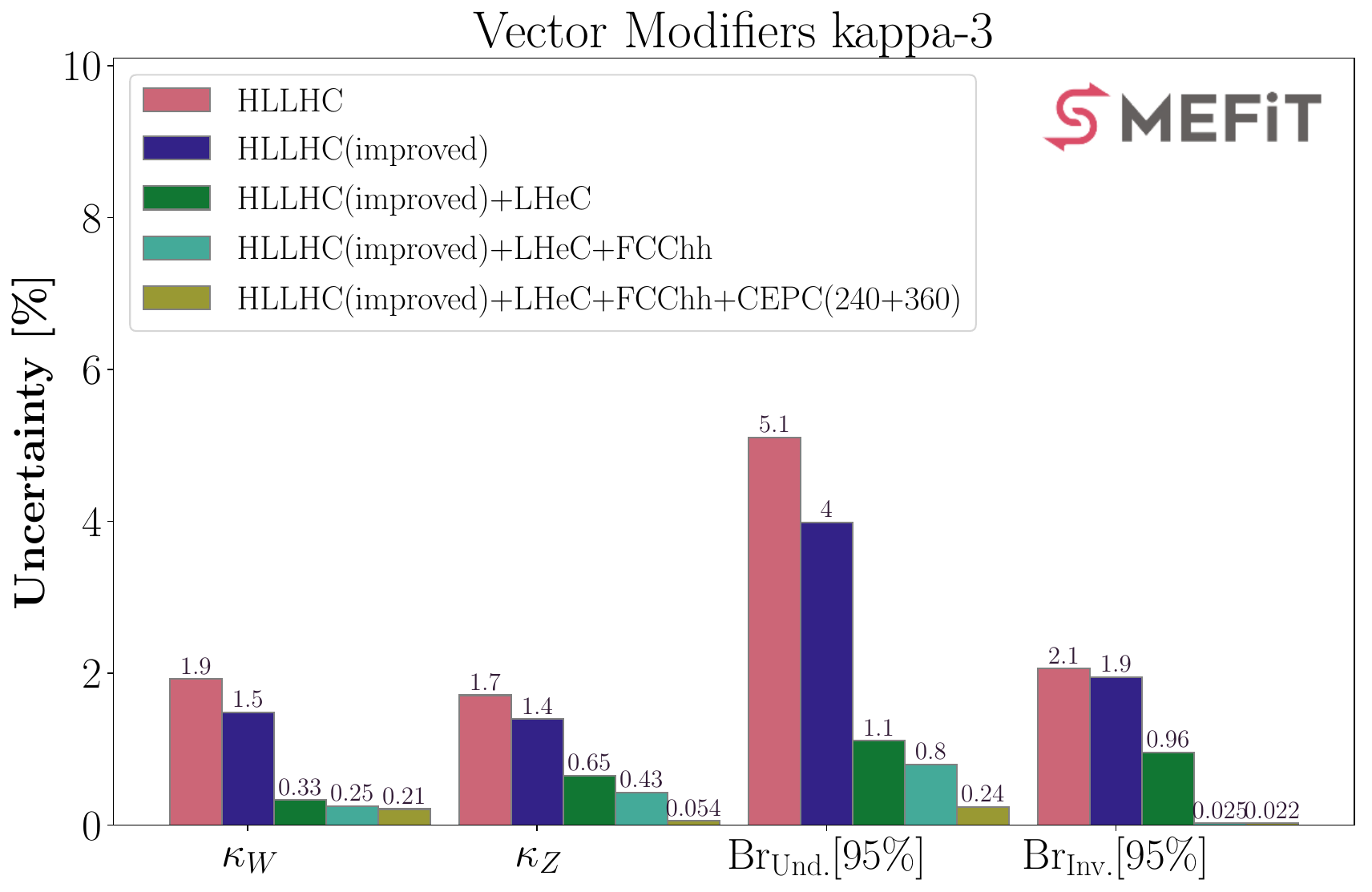}
\includegraphics[width=0.49\textwidth]{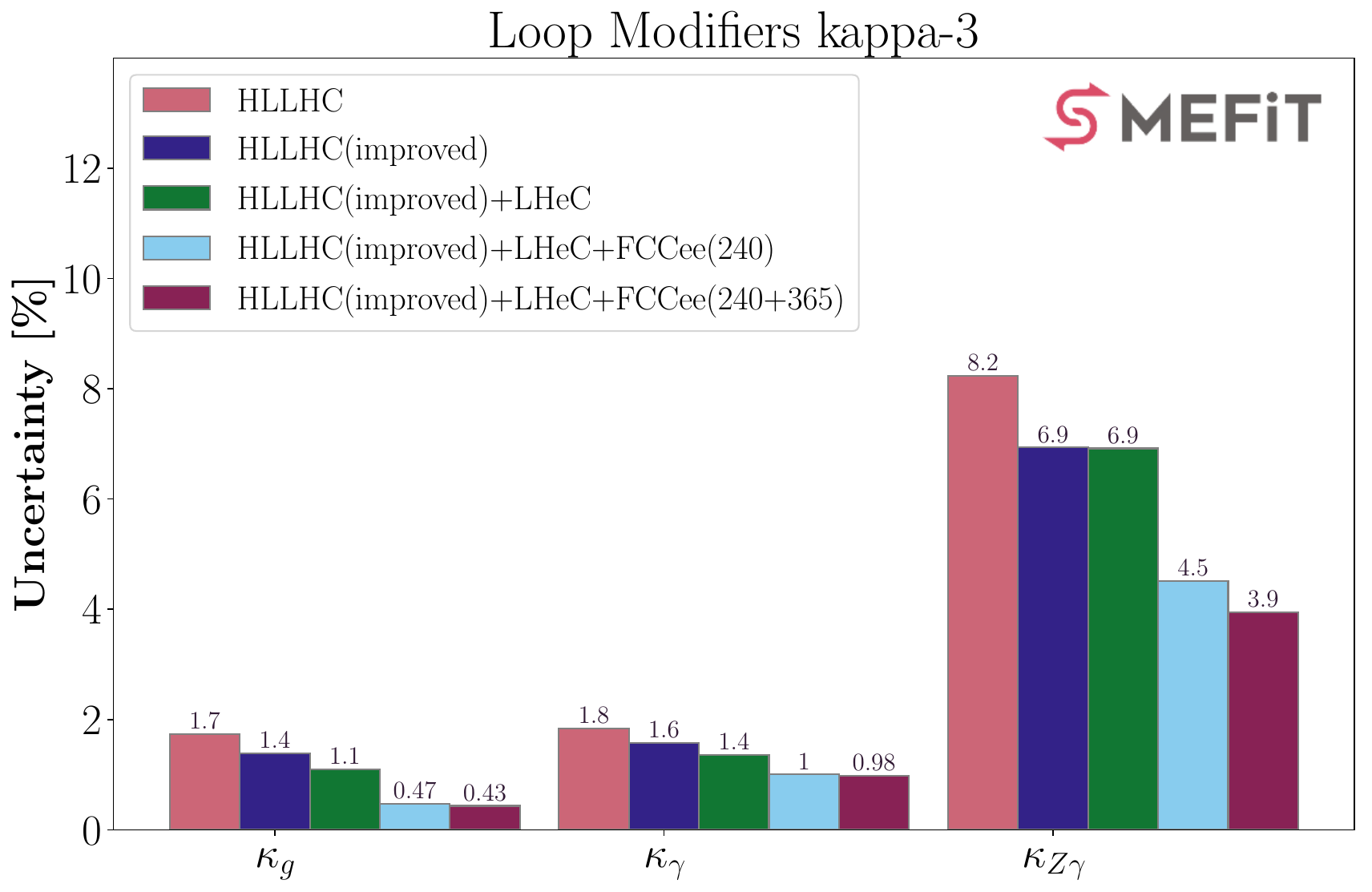}
\includegraphics[width=0.49\textwidth]{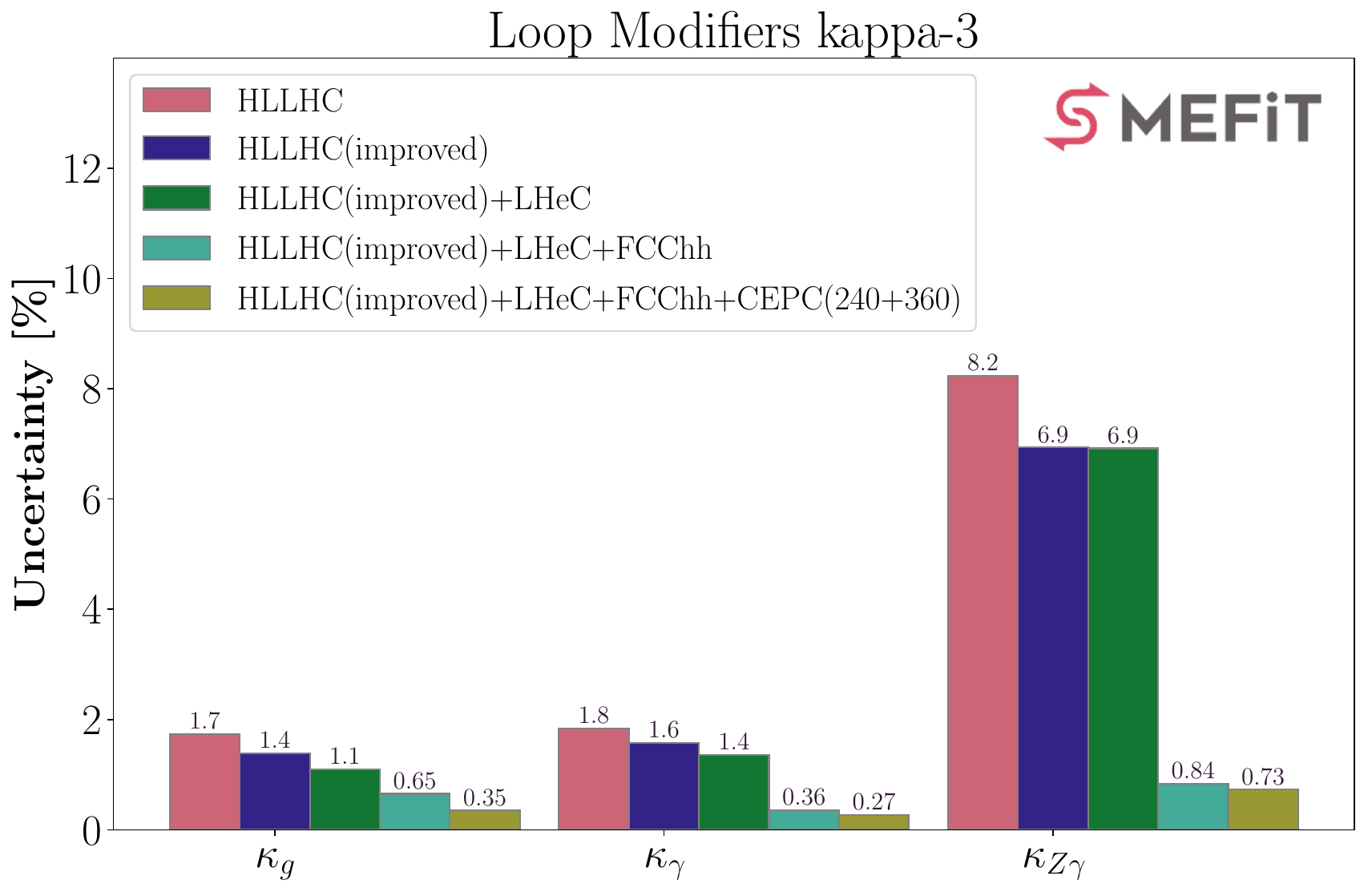}
\end{center}
\vskip -0.6cm
\caption{Relative uncertainty in the coupling modifiers obtained in the kappa-3 framework~\cite{deBlas:2019rxi,deBlas:2022ofj}, for different combinations of the HL-LHC, LHeC, and other future collider datasets. "HL-LHC(improved)" refers to the impact of reduced PDF+$\alpha_s$ uncertainties in the HL-LHC measurements. These data are further combined with LHeC Higgs-boson measurements, and with either FCC-ee data at $\sqrt{s}=240$ and $365$~GeV (left), or with FCC-hh and CEPC data (right). The fermion, vector, and loop modifiers are shown from top to bottom respectively. Since neither HL-LHC nor LHeC give direct access to the Higgs width, the condition $\kappa_V(\kappa_W, \kappa_Z)\le 1$ is imposed on the fit and the uncertainty on these modifiers is defined as $1-\kappa_V(68\%)$. Results have been obtained with the {\sc\small SMEFiT} framework~\cite{Celada:2024mcf,terHoeve:2023pvs,Giani:2023gfq,Ethier:2021bye,Armadillo:2026mvp}.
}
\label{fig:smefit}
\end{figure}

The expected evolution of the uncertainty in the Higgs boson couplings is illustrated in Fig.~\ref{fig:smefit}, assuming the kappa-3 framework~\cite{deBlas:2019rxi,deBlas:2022ofj}
and using the most recent version of the {{\sc\small SMEFiT}} fitting tool \cite{Armadillo:2026mvp}. The baseline LHeC inputs have been taken from \cite{deBlas:2019rxi,LHeC:2020van}. The HL-LHC inputs were built starting from the Snowmass projection \cite{deBlas:2022ofj}, breaking down the total uncertainties presented there into statistical and systematic components~\cite{Cepeda:2019klc} and adding the theory uncertainties from \cite{LHCHiggsCrossSectionWorkingGroup:2016ypw}. A reduction of these uncertainties by a factor of two was applied to fit the S2 scenario of~\cite{LHCHiggsCrossSectionWorkingGroup:2016ypw}. The experimental uncertainties in the $\mu\mu$ and $Z\gamma$ decay channels have been reduced using the updated prospects presented in~\cite{ATLAS:2025eii}. The LHeC will improve the Higgs coupling studies not only by providing specific measurements, but also by reducing the PDF and $\alpha_s$ uncertainties of the LHC measurements, of particular relevance in gluon fusion and $t\bar{t}h$ production. The "HLLHC (improved)" scenario reflects a reduction of the corresponding uncertainties by a further factor of three to reflect these improvements. Starting from the nominal HL-LHC expectations~\cite{ATLAS:2025eii}, Scenario 1 tracks the successive improvements obtained from the LHeC constraints on HL-LHC uncertainties, from Higgs measurements at the LHeC itself, and from FCC-ee data at the $HZ$ cross-section peak ($\sqrt{s}=240$~GeV) and later at the $t\bar{t}$ threshold ($\sqrt{s}=365$~GeV). In the fermion sector, the LHeC provides the first, percent-level measurement of the $H\to c\bar{c}$ decay, and dominates the determination of the Higgs coupling to the top quark. For other fermionic decay modes, the LHeC will reach the 1\% level before the FCC-ee further reduces the uncertainties by typical factors of two to three. While the FCC-ee dominates the determination of $\kappa_Z$, $\kappa_W$ is fully dominated by the LHeC, and only the $t\bar{t}$ run of the FCC-ee will provide a comparable measurement, reducing the overall uncertainty by about 30\%. Finally, the loop-induced coupling modifiers $\kappa_g$, $\kappa_\gamma$ and $\kappa_{Z\gamma}$ are reaching quasi-asymptotic values already after the LHeC era. The scarce sample of observed $H\to Z\gamma$ decays limits the precision in $\kappa_{Z\gamma}$ to about 10\%. In all cases, the LHeC allows improvements of the HL-LHC measurements by about 25--30\%.

\begin{figure}[h]
\begin{center}
\includegraphics[width=0.49\textwidth]{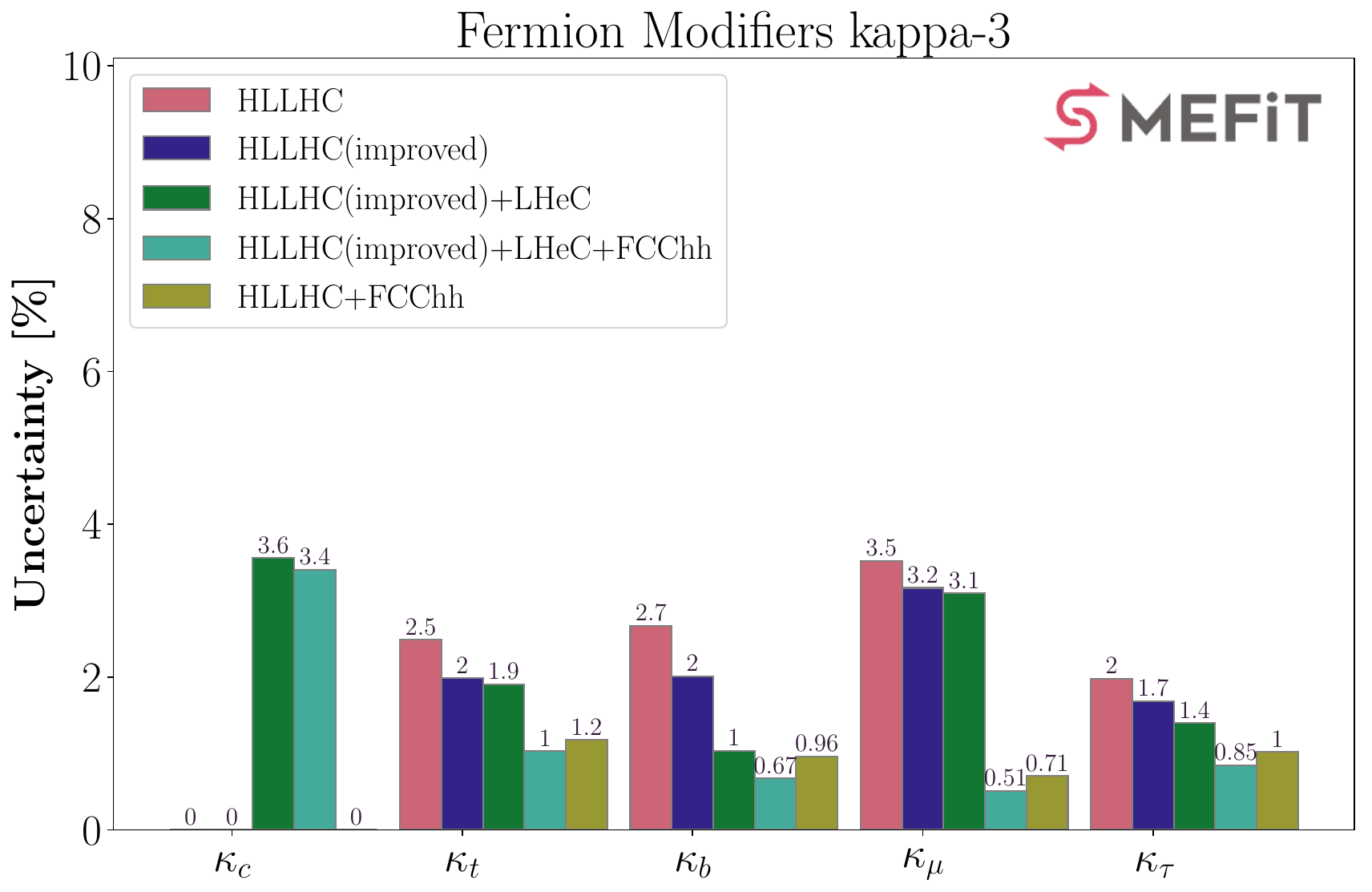}
\includegraphics[width=0.49\textwidth]{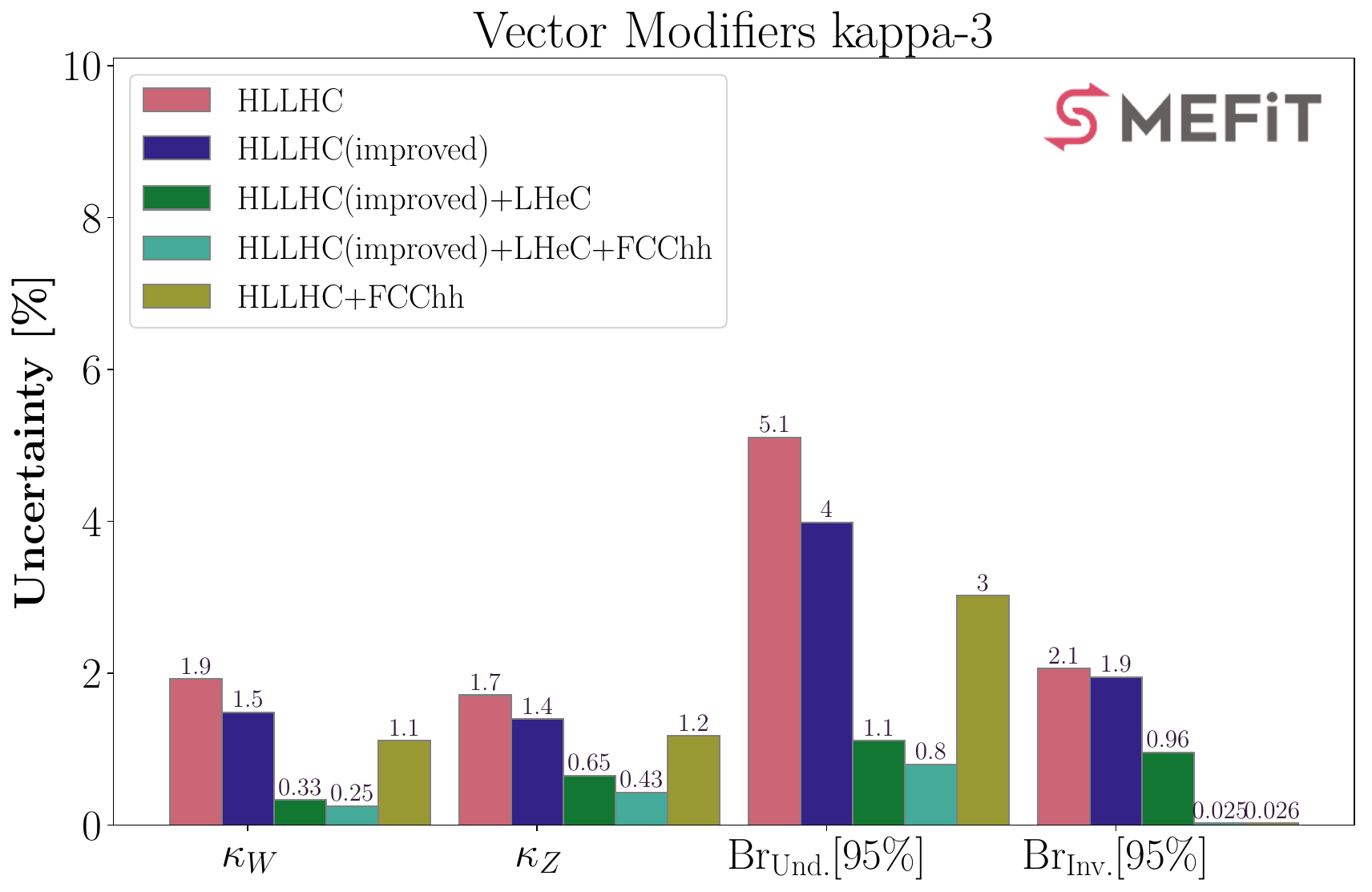}
\includegraphics[width=0.49\textwidth]{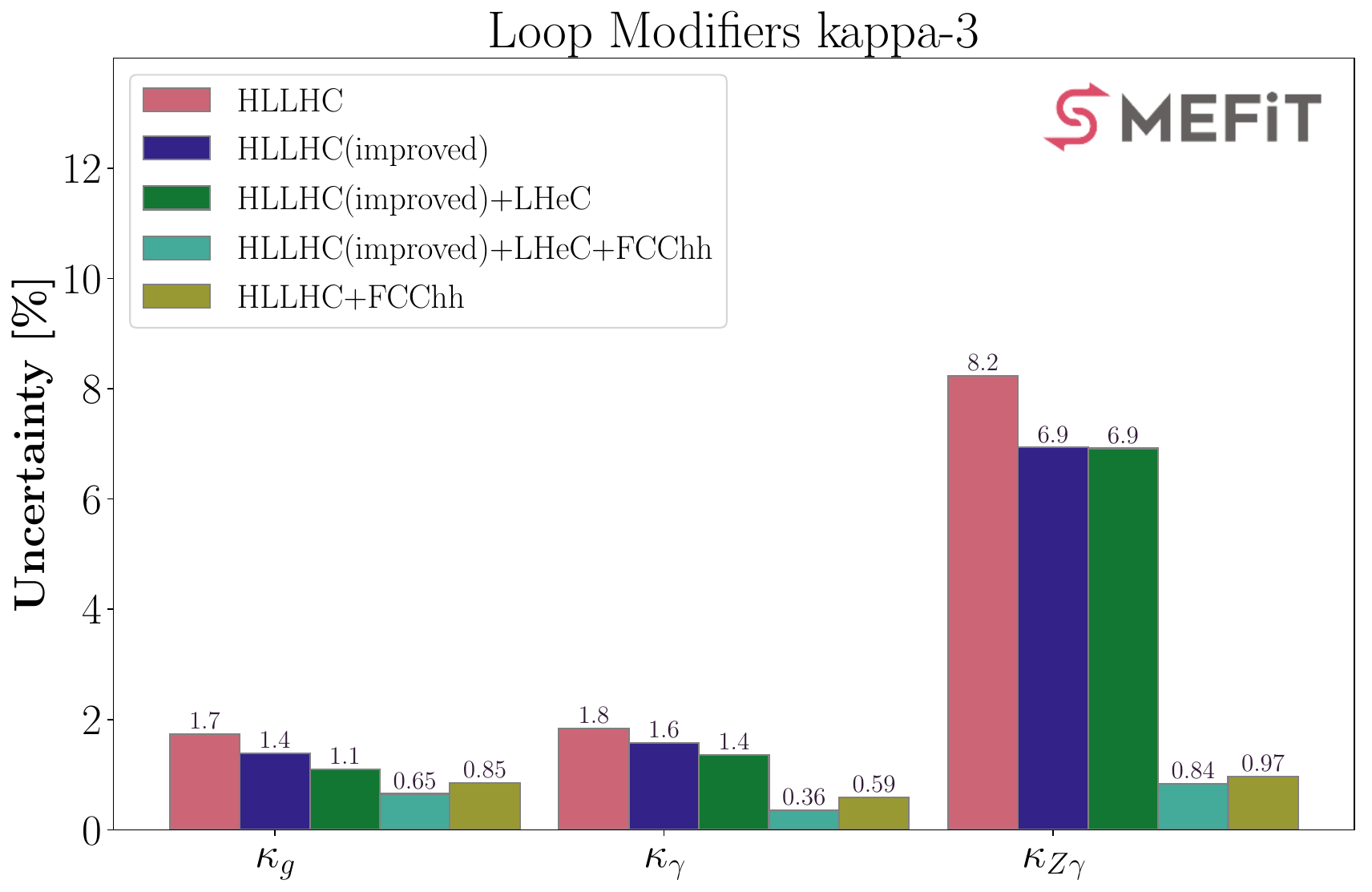}
\end{center}

\vskip -0.6cm
\caption{Id. to  Fig.~\ref{fig:smefit} (right column) but showing  the combination HL-LHC+FCC-hh in  absence of LHeC improvements and data instead of the combination with the CEPC. The fermion, vector, and loop modifiers are shown on the top left, top right and bottom plots, respectively. Results have been obtained with the {\sc\small SMEFiT} framework~\cite{Celada:2024mcf,terHoeve:2023pvs,Giani:2023gfq,Ethier:2021bye,Armadillo:2026mvp}.
}
\label{fig:smefit2}
\end{figure}

Scenario 2 yields comparable observations regarding the role of LHeC in synergy with the FCC-hh and CEPC. An obvious feature of this scenario is the significant improvement obtained from much larger samples of rare Higgs-boson decays ($H\to\mu\mu$, $H\to t\bar{t}$, $H\to Z\gamma$) on the corresponding coupling modifiers. For all couplings, the simultaneous operation of an $e^+e^-$ collider and an $ep/pp$ collider complex provides slightly better uncertainties than those achieved in Scenario 1.
We note that the uncertainties in the determination of $\kappa_W$, $\kappa_Z$, $\kappa_c$, and $\mathrm{Br}_\mathrm{Inv.}$ are smaller for the HL-LHC+LHeC combination than for HL-LHC+FCC-hh -- particularly for the vector couplings --, while both combinations give comparable uncertainties for $\kappa_b$, $\kappa_\tau$ and $\kappa_g$, as shown in Fig.~\ref{fig:smefit2}.

In order to illustrate the situation of the LHeC in the landscape of projects for future colliders, in Tab.~\ref{tab:compcoll} we present the different proposals of lepton-hadron, hadron-hadron and lepton-lepton colliders with some of their characteristics and estimated timeline, taken from~\cite{LHeC:2020van,Caldwell:2025fjg,Arduini:2025hor,Arduini:2957413} where full details can be found. Comparisons of the physics cases, performed in the framework of the Physics Preparatory Group for the 2026 Update of the ESPP, can be found in~\cite{deBlas:2025gyz,Jorge:2957414}.

\begin{table}[htb!]
\centering
\resizebox{\textwidth}{!}{%
\begin{tabular}{@{}l l c c l l l@{}}
\toprule
\textbf{Facility} & \textbf{Years} & \textbf{$E_{cm}$ (GeV)} & \textbf{Luminosity ($10^{33}$ cm$^{-2}$s$^{-1}$)} & \textbf{Ions} & \textbf{Polarization} & \textbf{Status} \\
\midrule \hline
\multicolumn{7}{l}{\textbf{$\mathbf{eh}$ Facilities}} \\
\addlinespace
JLab 11 / JLab 22 & Now / from late 2030s & 4.5 -- 6.5 & $10^2$ -- $10^6$ & p $\to$ Pb & e, p, light nuclei & Running / Concept \\
EIC & $>$ 2034 & 30 -- 140 & 1 -- 10 & p $\to$ U & e, p, $^3$He, Li & Approved \\
LHeC & From early 2040s & 1200 & 23 & *LHC & e possible & Advanced \\
Plasma-based schemes & $>$ 2035 & 530 -- 9100 & $10^{-5}$ -- $10^{-1}$ & *SPS/LHC & e possible & Concept \\
FCC-eh & $>$ 2050 (*FCC-hh) & 3500 & 15 & *FCC-hh & e possible & Intermediate \\
FASER / FPF / AdvSND & 2030--2041 & 30 -- 90 & 0.3 -- 10 & W, Ar & no & Approved / Advanced \\
\addlinespace \hline
\multicolumn{7}{l}{\textbf{hh Facilities}} \\
\addlinespace
HL-LHC & 2030--2041 & 14000 & 50 & p $\to$ Pb & no & Approved \\
FCC-hh & $>$ 2050 & 84600 & 300 & p $\to$ Pb & no & Intermediate \\
\addlinespace \hline
\multicolumn{7}{l}{\textbf{$\mathbf{l^+l^-}$ Facilities}} \\
\addlinespace
LCF & $>$ 2045 & 91.2 -- 550 & 2.8 -- 38.5 & n/a & yes & Advanced \\
FCC-ee & 2046--2065 & 91.2 -- 365 & 1400 -- 14 & n/a & no & Advanced \\
CLIC & $>$ 2045 & 380 -- 1500 & 22 -- 37 & n/a & yes, e$^-$ & Advanced \\
CEPC & $>$ 2040 & 91 -- 360 & 1920 -- 8 & n/a & no & Advanced \\
ILC250 & $>$ 2040 & 250 & 13.5 & n/a & yes & Advanced \\
LEP3 & $>$ 2040 & 91.2 -- 230 & 440 -- 18 & n/a & no & Advanced \\
MuCol & $>$ 2050 & 3200 -- 7600 & 9 -- 101 & n/a & no & Concept \\
Plasma-based schemes & $>$ 2050 & 250 & 7.6 (2.5) & n/a & no & Concept \\
\bottomrule \hline \hline
\end{tabular}%
}
\caption{Illustration of how the proposed LHeC integrates into the broader roadmap for future high-energy physics facilities: electron-hadron ($eh$), lepton-antilepton ($l^+l^-$, all of them $e^+e^-$ except MuCol), and hadron-hadron ($hh$). Numbers taken from~\cite{LHeC:2020van,Caldwell:2025fjg,Arduini:2025hor,Arduini:2957413}, where further details and full references can be found.}
\label{tab:compcoll}
\end{table}


\section{Technical feasibility of the \NoCaseChange{LHeC}}
\label{sec:feasibility}

\subsection{Baseline Detector for the LHeC}
\label{sec:detector}


The baseline detector for the LHeC is designed to fulfill the following requirements from physics: (a) high-purity identification and precise measurement of the scattered electrons for NC processes;
(b) high-resolution measurement of the hadronic final states, for reconstructing NC and CC kinematic variables and jet momenta and; (c) tagging heavy flavour (HF) jets for Higgs boson and top quark decays, BSM searches and semi-inclusive DIS measurements for disentangling the flavour and gluonic contributions to the interaction. The experimental environment is moderate: the average number of interactions per bunch crossing is 0.1, which is three orders of magnitude less particle flow than that for LHC $pp$ collisions, imposing much less constraint to radiation hardness and occupancy.

The baseline detector
is based on advanced but already available technologies, either in use or in preparation for a collider experiment. An 
overview is shown in Fig.~\ref{fig:detector},
corresponding to the version from~\cite{Andre:2022xeh}, which evolved slightly
from the more detailed description in~\cite{LHeC:2020van}.
The separate detector components are briefly described below:

\begin{figure}[ht]
\centering
\includegraphics[width=0.9\textwidth]{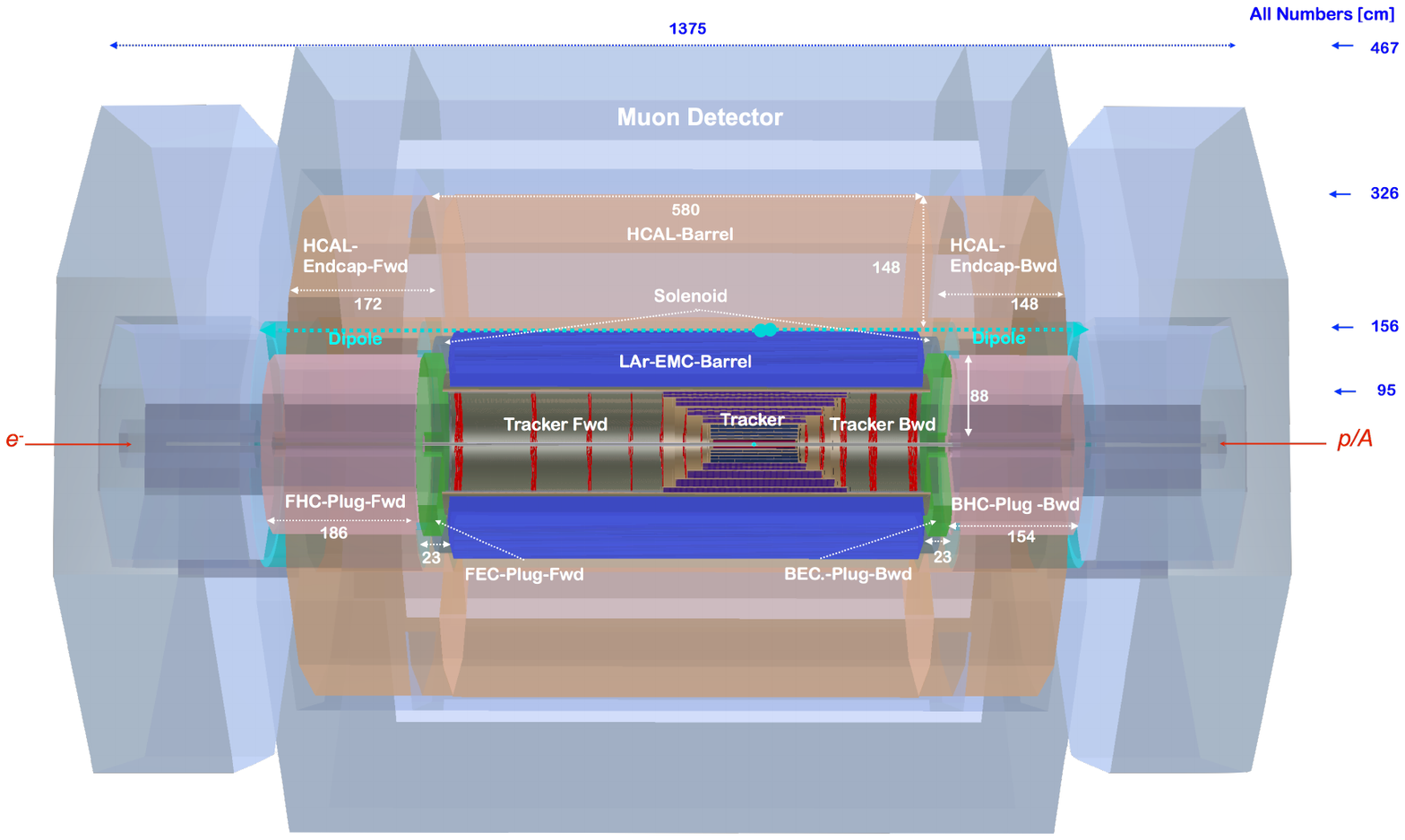}
\vskip -1cm
\caption{Overview of the baseline LHeC detector design and
corresponding dimensions.}
\label{fig:detector}
\end{figure}

\begin{itemize}

\item Thin-layer all-silicon central tracker: 
CMOS-based technologies recently available minimise the degradation of track-parameter resolution thanks to its low material budget ($\lesssim 0.2 X_0$ in total). The sensors can be bent to fit the shape of the beam pipe, allowing to place them as close as possible to it. They also provide enough radiation tolerance. The technology has been adopted for ALICE ITS3 and EIC-ePIC silicon tracker. The LHeC should provide at least two hits up to a very high rapidity range, $\eta < 5.2$.

The challenge here is to place the sensors as close to the beam line to maximise the angular coverage and HF tagging ability, while avoiding the irradiation from synchrotron radiation induced by the electron beam (see Sec.~\ref{subsec:MDI}). Layouts using a secondary vacuum layer around the interaction point may be considered to minimise the acceptance loss in azimuthal angle due to the synchrotron radiation fan.

\item Hermetic EM and hadronic calorimetry: the balance of the energy resolution by high sampling fraction and granularity by fine segmentation is the most important factor for the EM calorimeters to achieve high-purity identification of scattered electrons at reasonable cost. 
For hadron calorimetry (HCAL), various technologies providing high energy resolution, developed for Higgs factories, are also adequate for the LHeC. An exception is the modules around the beam pipe in the forward direction, which should stand for a high level of radiation. The baseline design consists of a LAr EM calorimeter for barrel (similar to EM part of the ATLAS LAr calorimeter), and a plastic scintillator/iron sampling calorimeter for HCAL in barrel and endcap, except for the very forward and backward rapidity ranges where silicon sampling layers are used for granularity and radiation tolerance (similar to CMS HGCAL\,\cite{CMS-HGCAL-TDR} but with higher numbers of sampling layers).

\item Muon system: 
the background for muons is mainly meson decays in jets of the same event. Technologies currently available for muon chambers (e.g., drift tubes) and timing layers with the second coordinate measurement (e.g., RPC layers) will serve for this purpose. No dedicated magnet system is assumed.

\item Backward and Forward detectors: the photon rate from the process $ep \to ep\gamma$, which has been used at HERA for luminosity measurement, reaches above 1\,GHz for high luminosity operation. A development after the CDR update~\cite{LHeC:2020van} is to use a two-stage absorber, where the first absorber is to reduce the rate by absorbing photons, while the second layer works as a converter. 
In the forward direction, forward proton spectrometer 
stations could be based on Roman pot insertions to the beampipe in similar
locations to those at the LHC with minor changes. There is also the additional
possibility of more distant stations from the interaction point, accessing
lower fractional proton momentum losses, as was explored previously
in the LHC context in~\cite{FP420RD:2008jqg}. There are also opportunities to
benefit from the experience at the EIC, which plans an unprecedented level
of instrumentation for forward proton production. 
The radiation dose at zero-degree calorimeter is expected at the order of $10^{15}-10^{16}\ {\rm n_{eq}/cm^2}$,  allowing for operating the calorimeter for the entire period of physics runs.

\end{itemize}

Further improvements over the baseline detector are foreseen, both in terms of cost and performance, especially for tracking and calorimetry; see Sec.~\ref{sec:technology}.

An implementation of the baseline detector is available~\cite{lhecsw} for simulation studies. The simulation code is based on the DD4Hep~\cite{frank_markus_2018_1464634} framework. Interface to several generators is prepared. The output data are formatted such that they can be reconstructed using packages based on the Gaudi software framework widely used for LHC experiments and FCC studies.

\subsection{Machine-Detector Interface}
\label{subsec:MDI}

In order to avoid magnets around the beam pipe inside the main detector, the two beams are brought to collisions with a weak dipole magnet (about 0.2\,T) integrated with the detector solenoid. 
Protecting detectors from the synchrotron radiation (SR) from the 50\,GeV electron beam, both from the bending upstream and from the dipole, is to be investigated carefully. The experience from HERA-II high-luminosity upgrade has shown that at least the secondary scattering of the SR photons should be simulated when designing masks and absorbers. Efforts~\cite{Andre:2022xeh} were made after~\cite{LHeC:2020van} to decrease the critical energy of the SR to soften the spectrum for efficient absorption. 
Taking full advantage of the length of the LHC straight section, $L^*=23$\,m, a re-optimization of the separation scheme had been performed for lowest critical energy of the emitted synchrotron light. A reduction from $E_c=255$\, keV to  $E_c=114$\, keV is obtained, which is close  to the maximum photon absorption coefficient of the foreseen collimator and absorber schemes. The corresponding power of the emitted light in the interaction region reduces accordingly from $E_p=18$\,kW to a comfortable level of  $6$\,kW.
A study to optimise the masks for the SR fans is being conducted.

\subsection{Baseline Accelerator Technology}

The ERL design including several options and various optimizations was developed~\cite{Andre:2022zqx}. A careful design of the magnetic lattice and beam optics was undertaken for emittance preservation during the ERL operation including the vertical deflection system and detector bypass, providing a minimised transverse emittance growth, for all three complete ERL versions corresponding to $1/3^{rd}$, $1/4^{th}$ and $1/5^{th}$ of the LHC circumference. A full front-to-end simulation including synchrotron radiation, longitudinal wakefields and beam-beam disruption at the interaction point (IP) showed excellent energy recovery performance that reduces with the ERL circumference. 
The particle tracking simulations constitute a solid basis for future LHeC studies involving magnet errors and alignment tolerances. Besides, the tools developed for the design and optimisation of the ERL layout, as well as the front-to-end, tracking simulations including synchrotron radiation and beam disruption at the IP are easy to adapt and re-use for similar collider studies such as the FCC-eh.

Some further design options and optimization to be pursued include a separate FFAG arc design~\cite{Trbojevic:2024reo} and impact on emittance preservation, lattice and robustness studies with respect to imperfections with improved simulation framework (X-suite) but also impact of Coherent Synchrotron Radiation and micro-bunching instability handling. It would be indeed important to continue building ERL expertise through collaboration with PERLE and bERLinPRO~\cite{Kamps:2019pgl} for gaining operational experience and as stepping stone for proving several accelerator physics concepts and prototyping/testing technologies.

An electron interaction region has been optimized to minimise the synchrotron radiation
power. The local impact of the electron magnets on the proton beam orbit and optics can be
corrected in the HL-LHC. Several modular proton optics have been developed by using the arcs and additional insertions to
match optics, the tune and the chromaticity. Tracking simulations were undertaken to investigate the impact of the proton beams on each other interaction region
(IR) optics design, including asymmetric magnets using HL-LHC short model coils.
The local impact of an optimized electron IR on the proton beam orbit and optics can be corrected
in the HL-LHC, enabling $ep$ collisions with a luminosity
above $10^{34}~\text{cm}^{-2}~\text{s}^{-1}$~\cite{LHeC:2020van}, and above $10^{33}$\,cm$^{-2}$s$^{-1}$ in conventional recirculating linac mode when scaling the maximum electron beam power to 60\,MW.

\begin{figure}[h]
\centering
\includegraphics[width=0.5\textwidth]{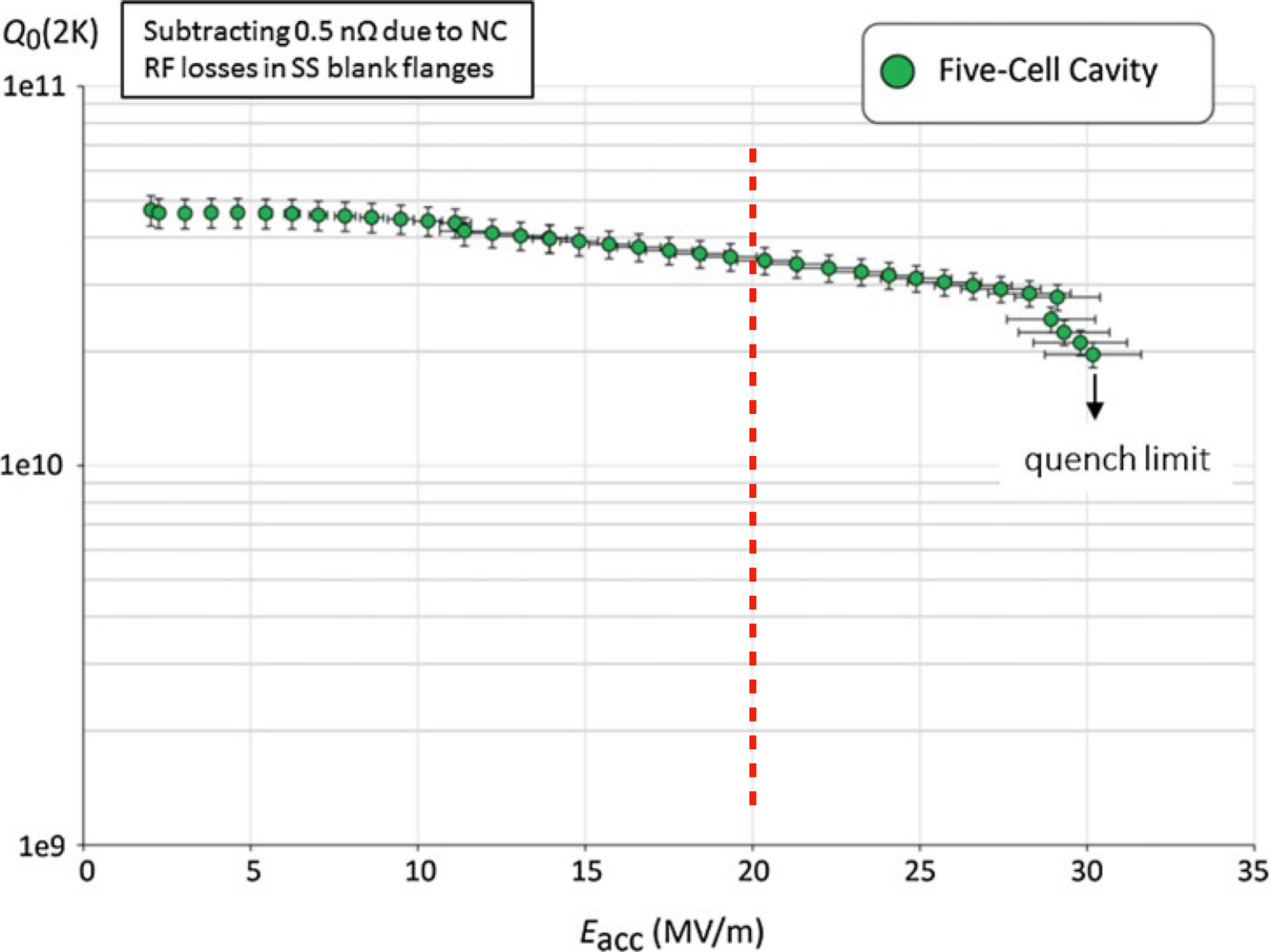}
\caption{Vertical test results for the five-cell 802 MHz niobium cavity prototype. Taken from~\cite{LHeC:Marhau}.}
 \label{fig:rf_cavity}
\end{figure}

The LHeC conceptual design is based on the assumption of 800\,MHz superconducting radio frequency (SRF) technology with a gradient of around 20 MV/m in CW mode. Design studies at Jefferson Lab have demonstrated the technical feasibility of such a 5-cell superconducting resonator: in vertical test stands field gradients of up to $E_{acc}=30$\,MV/m have been reached, well above the required values of the LHeC operation. Fig.~\ref{fig:rf_cavity} shows the cavity quality $Q_0$-factor as a function of the acceleration field. The ultimate quench limit is reached far above the required LHeC operation parameters, marked as red dotted line. 

The key technical challenge is the development of SRF technology including High-Temperature SRF at 4.5~K,
for cavities with high $Q_0$ for CW operation.  In this context, the development of Fast Reactive Tuner (FRT) is also crucial for microphonics mitigation. 
In particular, a Ferro-Electric FRT is a novel type of RF cavity tuner containing a low loss ferroelectric material. 
They are of particular interest for SRF cavities as they can be placed outside the liquid helium environment and allow for
quick tuning. Initial studies~\cite{Shipman:2807269} show that 
microphonics suppression for low or zero beam loading machines and transient detuning for high-beam
loading machines can be achieved with reduction of RF power consumption by large factors. Improved compensation 
of frequency variations relaxes specifications for cavities and cryomodules leading to reduced cost and enhanced 
feasibility. The RF system could also profit from developments of High-efficiency klystrons~\cite{Catalan-Lasheras:2024xxw}
in particular the very compact (2.8 m) two-stage klystron design for FCC, which targets very high-efficiency of above 80 \%.
All these developments should be coupled with further optimisation studies of the cryogenics, whose system is presently optimised 
for feeding 2~K superconducting cavities. Studies for high-temperature and/or high $Q_0$ SRF would certainly lead to a reduction of power and cost.

\subsection{ERL feasibility for high-power beams}

The feasibility of high current ERL accelerators was successfully demonstrated at Jefferson Lab in the early 2000s. The J-Lab Infra Red Free Electron Laser (FEL) delivered over 2\,kW FEL radiation using a single turn ERL operating with beams between 5 mA and 10 mA and up to 1\,MW beam power. The LHeC concept as described in~\cite{LHeC:2020van} requires an ERL operation with multi-turn [3 acceleration and 3 deceleration passages through the linear accelerator] and with a beam power above 500\,MW at the top energy. Extrapolating the ERL concept from single turn operation with 1 MW beam power to a multi-turn operation with over 500 MW beam power clearly requires a stepwise demonstration of the scalability of the ERL concept.

As discussed in the next Section, the current LHC schedule makes installation after 2041 the most realistic option for the LHeC, requiring start of the construction by 2037 and project approval by 2034. The scalability of the ERL concept clearly needs to be demonstrated prior to the project approval and aiming for a high current multi-turn ERL demonstration by the late 2020ies / early 2030ies is a key requirement for the LHeC proposal. However, one should observe that the LHeC could still reach luminosity performance levels of $10^{33}$\,cm$^{-2}$s$^{-1}$ without ERL configuration, when operating in a conventional recirculating linac mode and while still satisfying the request for a less than 100\,MW additional total power footprint\footnote{The physics case with smaller integrated luminosities is discussed in App.~\ref{sec:lowlumi}, where it is shown that a large fraction of the physics program remains with integrated luminosities down to 1/20th the nominal value.}.

The ERL feasibility at high beam currents and beam power with the required beam characteristics will be checked in PERLE at IJCLab in Orsay, see Sec.~\ref{sec:perle}. Specifically, the current plan shown in Sec.~\ref{sec:perlestatus} considers 1-turn and 3-turn operation in 2029 and 2030 respectively, in time for LHeC consideration in the next ESPP and for the start of its installation in 2041.

\section{The \NoCaseChange{LHeC} implementation plan}
\label{sec:implementation}
The 2021 operation planning foresaw an LHC operation until end of 2025, followed by a ca. 3 year Long Shutdown (LS3) for the installation and commissioning of the new HL-LHC equipment and then three HL-LHC running periods, Run4, Run5 and Run6, separated by two Long Shutdown (LS) periods, LS4 and LS5.
The LHeC planning laid out in the 2021 CDR revision of the LHeC study~\cite{LHeC:2020van} assumed the ERL machine installation partially in parallel of the Run4 operation period and the detector installation during LS4 of the HL-LHC operation period.
The design of the ERL and of the interaction region~\cite{Andre:2022xeh} allows concurrent $ep$ operation at the LHeC interaction point (IP) and $pp$ operation at the other LHC IPs, with detailed studies on beam-beam interactions and the configuration to host a third, non-colliding proton beam. Given that a LS5 is no longer foreseen in the present schedule due to the delays of the HL-LHC, the scenario where the LHeC runs after the nominal end of the program is considered as we discuss in the following. An alternative scenario has been considered elsewhere~\cite{Andre:2025iar}, in which the interaction region in IP2 is modified during
LS4, allowing collisions of 20\,GeV electrons from a single-pass ERL
during Run5. However, the implementation of this scenario requires a
very quick decision regarding its implementation, as well as the
adaptation of the planned ALICE3~\cite{ALICE:2022wwr}.
   
\begin{figure}
\centering
\includegraphics[width=1.0\textwidth]{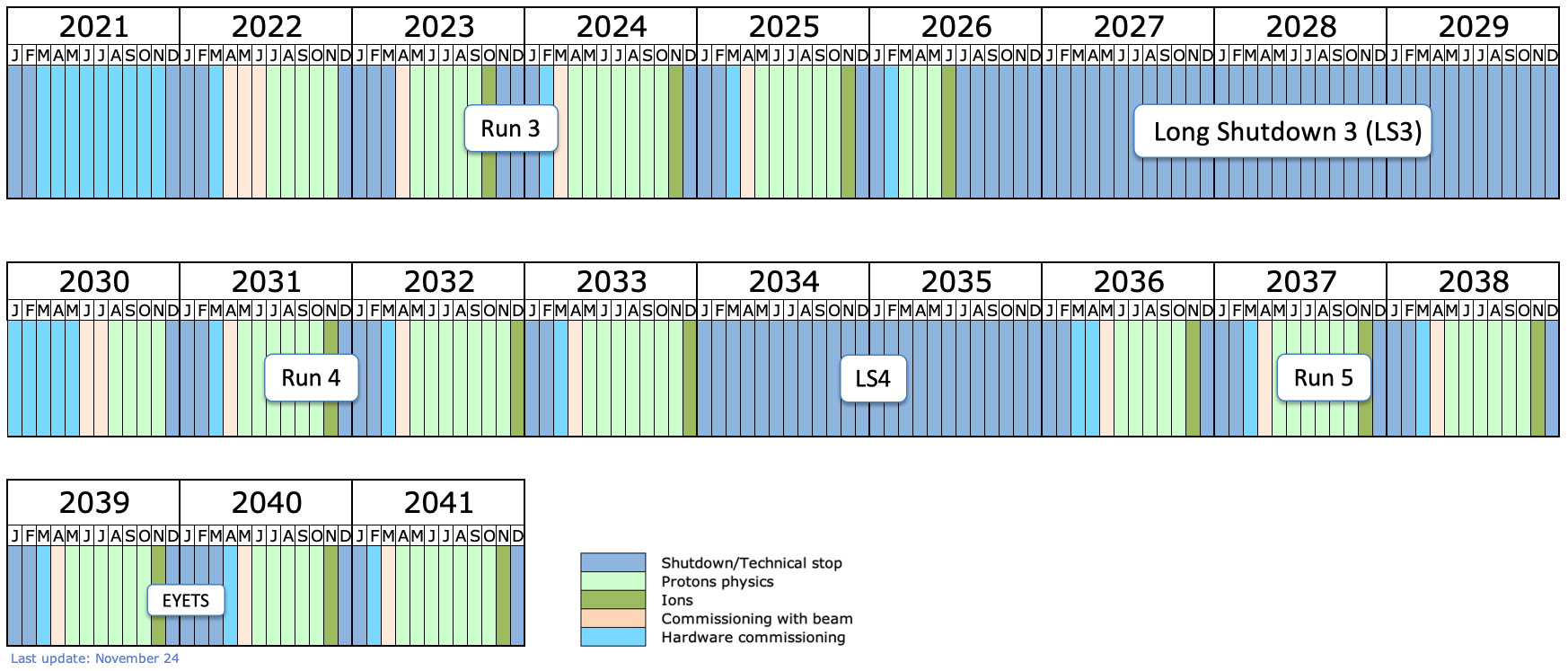}
\caption{The new LHC schedule approved by the CERN Research Board in September 2024 and revised in November 2024.\label{schedule-new}}
\end{figure}

The V4.1 HL-LHC operation schedule, approved by the CERN Research Board in September 2024 and revised in November 2024, foresees a later start of the HL-LHC operation, now planned for July 1st 2030, and one less long shutdown period (see Fig.~\ref{schedule-new}). At this stage in time, it is probably no longer realistic to assume an LHeC installation during LS4, only 10 years in the future. Without another long shutdown in the planning of the HL-LHC operation, it is only feasible to assume an LHeC installation after the end of the HL-LHC operation, after 2041. The ERL civil engineering work for the tunnel and the machine installation could still be carried out in parallel to the HL-LHC Run5 period. However, the LHeC installation would have to come after the end of the HL-LHC operation. In this scenario, there is no need anymore to limit the LHeC performance to a parasitic operation configuration and can instead assume for an operation after 2041 an entirely dedicated running period. Presuming nonetheless 1 year for the full commissioning of the LHeC setup with reduced performance, one would then expect to reach the $1$ ab$^{-1}$
after 6 years of operation. Assuming a 2 year long shutdown after 2041 for the installation of the LHeC infrastructure and detector upgrade / replacement planning for an additional 1 year long shutdown after 3 years of operation, one could complete the LHeC operation by 2050 (operation from 2044 to 2046 and from 2048 to 2050). The goals for $eA$ collisions could achieved by integrating $eA$ Runs in this program.

The energy consumption of the LHeC can be estimated as follows. The LHC total power consumption consists of 120\,MW for the machine plus 22\,MW for the experiments, resulting in 142\,MW. For the HL-LHC, the total power consumption further adds 10\,MW for experiments' upgrade plus 8\,MW for the machine cryogenics upgrade, resulting in 160\,MW.
The LHeC ERL has an energy consumption of 100\,MW to 130\,MW depending on the arc size of the ERL, 1/3$^{rd}$ or 1/5$^{th}$ - we assume 100\,MW in the following, corresponding to the 1/3$^{rd}$ configuration. The LHeC energy needs in dedicated operation mode can then be estimated as follows: one subtracts from the nominal 160\,MW of the HL-LHC operation 32\,MW for the cooling of the ATLAS and CMS detectors [22\,MW from the nominal LHC plus 10\,MW from the experiments' upgrade], 8\,MW from the HL-LHC machine upgrade and 10\,MW from the economy mode of cryogenics for operation with one proton beam only, resulting in saving of 50\,MW for the LHC machine when comparing it to the HL-LHC operation.
The energy consumption for the new detector infrastructure for the LHeC is assumed to be 10\,MW. Thus, we get total savings of 40\,MW in dedicated running mode wrt. nominal HL-LHC operation.
Therefore, the energy consumption of the LHeC dedicated mode operation consists of 160\,MW for HL-LHC consumption minus 50\,MW savings for not using machine and detector upgrades and no main experiments plus 10\,MW for the LHeC detector plus 100\,MW for the ERL machine, resulting in 220\,MW. This means an added power of 60\,MW when comparing to HL-LHC and 75\,MW additional power requirement when comparing to nominal LHC operation.

\begin{figure}
\centering
\includegraphics[width=1.0\textwidth]{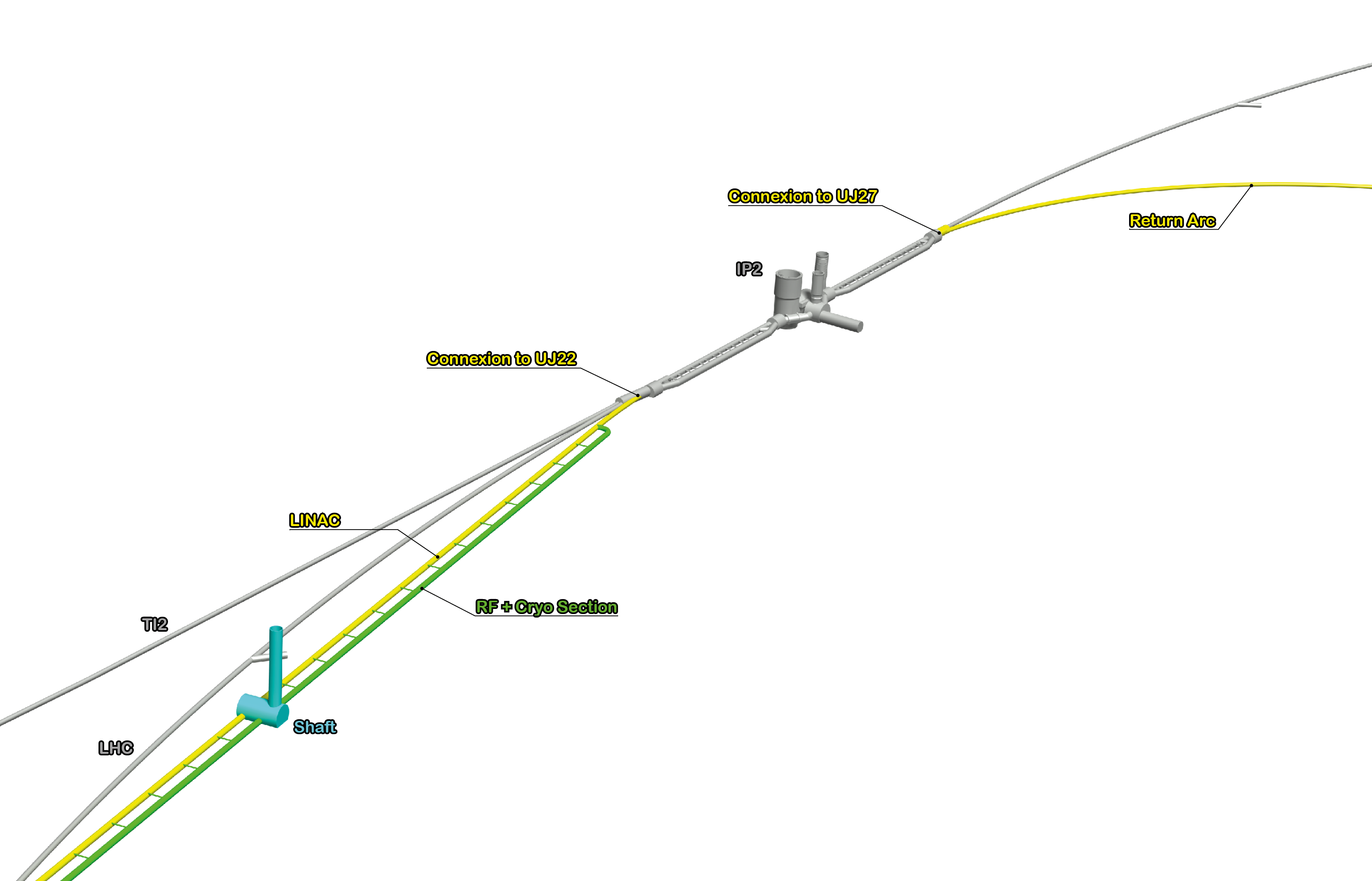}
\caption{Connection of the new LHeC installation with the existing LHC tunnel at Point2 of the LHC tunnel.\label{LHC-Tunnel-Intersection}}
\end{figure}

For 200 days of operation, 60\,MW yields 288\,GWh. Assuming an electricity price of 0.1\,EUR/kWh and an exchange rate of 1.06\,EUR/CHF, the LHeC implies additional annual operation cost of
27\,MCHF
with respect to the HL-LHC cost (or a total annual operation cost of
100\,MCHF for the HL-LHC+LHeC operation).
We consider the person power requirements for operating the LHeC with only one experiment and proton beam and the ERL facility comparable to the needs of the HL-LHC with two proton beams and four experimental insertions.

The baseline CDR configuration chose a beam energy of 50 GeV and a total ERL circumference of $1/5^{th}$ of the LHC circumference. However, several variations of this configuration were studied that could still be interesting alternatives. For example, choosing a larger circumference of the ERL of $1/3^{rd}$ of the LHC circumference opens to door of higher attainable beam energies [e.g. 60 GeV compared to the baseline 50 GeV] and reduces the power requirements arising from synchrotron radiation losses in the arcs [potential savings of ca. 30\,MW for the baseline beam parameters and 50 GeV operation], albeit these savings need to be balanced against the increased civil engineering cost, estimated at around 100\,MCHF~\cite{LHeC-cost}. Additional arguments for choosing the ${1/3}^{rd}$ ERL circumference as baseline, are related to the beam size preservation. 
We note that keeping the electron energy at 50\,GeV while increasing the arc length of the racetrack to a total of 1/3$^{rd}$ of the LHC decreases the power consumption by 30\,MW compared to the 1/5$^{th}$ option which, following the numbers done previously, diminishes the electricity costs in six years by 
$\sim 80$\,MCHF.

The LHeC could function as an attractive bridge project in case the 
foreseen operation of a new flagship project at CERN, e.g., FCC-ee or FCC-hh, will come after the 2040ies. The LHeC would serve as a technology driver and 'pre-series' production phase for the large scale SRF contracts required for the FCC-ee. Such a smaller scale initial contract could help in obtaining optimum industrial contracts for the FCC-ee, similar to the pre-series magnet production for the LHC construction. Furthermore, the LHeC ERL could be operated in recirculating linac mode as the injector into the FCC-ee, either in a reduced configuration with the current FCC-ee injection energy baseline of 20 GeV, which would require only one circulation loop in the LHeC, or at higher energy using several re-circulations in the LHeC for direct top-up injection into the FCC-ee.

\begin{figure}
\centering
\includegraphics[width=0.8\textwidth]{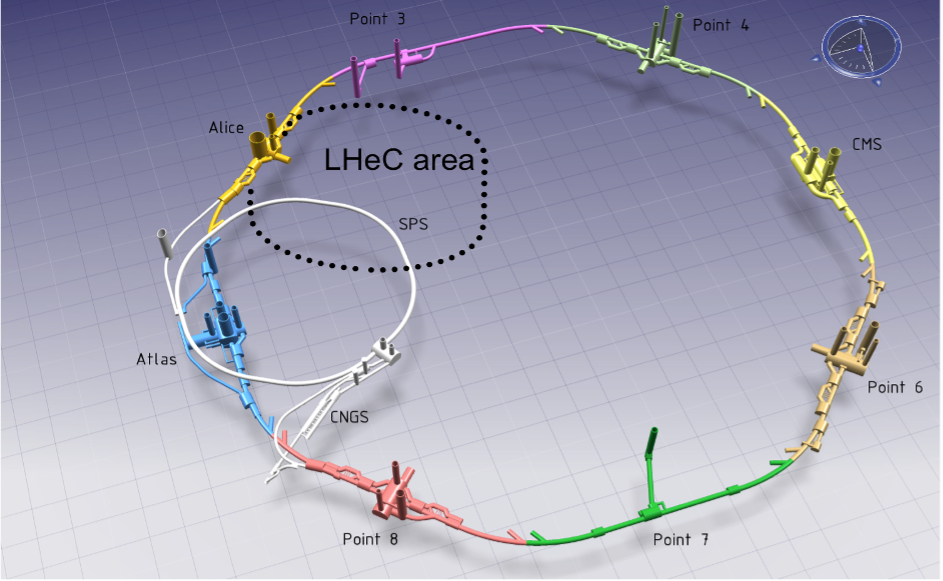}
\caption{Schematic layout of the LHeC tunnel within the existing LHC and SPS accelerator tunnels.\label{LHeC-Layout}}
\end{figure}

The underground structures proposed for LHeC in the proposed design with a 1/3$^{th}$ LHC circumference require a tunnel approximately 9 km long of 5.5 m diameter, including two LINACs. Parallel to the main LINAC tunnels, at 10 m distance apart, are the RF galleries, each 1070 m long. Waveguides of 1 m diameter are connecting the RF galleries and LHeC main tunnel (as depicted in Fig.~\ref{LHC-Tunnel-Intersection}). The civil engineering work involved for the 50 GeV ring at Point 2 (shown in Fig.~\ref{LHeC-Layout} with a black dotted line) can mostly take place during HL-LHC operation. The sections of tunnel which will have to be constructed outside of HL-LHC operation will be those in the closest proximity to Point 2. These sections which will have to wait until after HL-LHC operation to be constructed will likely be the beam dump and the connections with the LHC, such as the connecting tunnel to UJ22.
As a first pass estimate, assuming all enabling works and other construction can be completed before these connecting sections are added, an extra year should be sufficient to connect the LHeC to Point 2 of LHC. While IP2 was the option explored in the CDR, a reoptimization can be considered in function of transfer lines to the FCC-ee, the potential reuse of existing HL-LHC detectors and the location of the Forward Physics Facility~\cite{Anchordoqui:2021ghd}.

\begin{figure}
\centering
\includegraphics[width=0.8\textwidth]{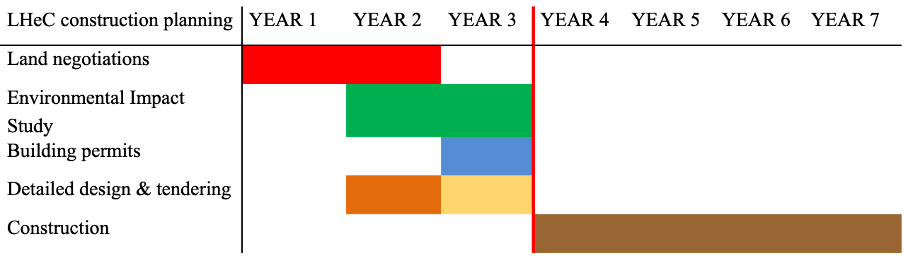}
\caption{Timeline for planning the construction and installation phases for the LHeC.\label{LHeC-Planning}}
\end{figure}

The table depicted in Fig.~\ref{LHeC-Planning} shows the estimated construction planning schedule from a time T=0 when land negotiations can begin. As can be seen from this, the total construction is expected to take 4 years, including the year-long connection to HL-LHC. Ahead of the construction activities, it has been estimated that three years of permits, engineering studies and tendering will be needed. Because the proposed locations of the LHeC shafts/surface facilities are either on, or close to existing CERN land, the environmental impact should be minimal.
To facilitate connection to the LHC, any equipment which may be sensitive to dust, vibration or water should be cleared from the tunnels immediately adjacent to the works. Additionally, if any of the access to transport and construction materials is in proximity with magnets etc., this could slow down construction and present logistical difficulties. 
The embodied carbon impact of concrete is mostly due to the amount of Portland cement that it contains, which accounts for 75-90\% of the overall embodied carbon impact of concrete. Portland cement is the most common type of cement used globally. As such, there are several avenues that can be considered for reducing carbon footprint of concrete as a material; this includes, but is not limited to:
\begin{itemize}
\item 
Partially replacing Portland cement (CEMI) with Supplementary Cementitious Materials (such as fly ash, GGBS, limestone powder, calcined clay and others).
\item 
Totally replacing Portland cement with “Portland cement-free” materials (such as alkali-activated materials / geopolymers).
\item 
Carbon sequestering in concrete (such as carbon negative aggregates and carbon injection in concrete). 
\end{itemize}

Electricity could be generated in solar (or wind) farms located in sunnier (or windier) areas in the member states such as in Spain, Portugal, Southern Italy, Greece, etc., and be fed into the European electricity grid and taken out at CERN. Currently (2022), for example,
- Netherlands’ average installed solar power per person increased to 1044\,W/capita (6\,GW), Germany with 816\,W/capita (68\,GW), Denmark with 675\,W/capita (4\,GW), and Belgium with 663\,W/capita (8\,GW). See also~\cite{solarpowereurope}
- Germany has the highest wind power installed capacity in Europe at more than 65\,GW, followed by Spain and the United Kingdom with roughly 30\,GW each.
In central Europe 1\,kW of peak solar power installed generates 1\,MWh per year. Other methods of lowering the power requirement should be pursued such as those mentioned in Sec.~\ref{sec:feasibility}, e.g., R\&D on improving the efficiency of klystrons, pushing the performance of RF cavities and further use of superconducting technology for some main magnet systems.

\section{PERLE: A Multi-Turn, High Power Energy Recovery Linac (ERL) Facility}
\label{sec:perle}

ERL is one of the five ESPP identified keys areas for an intensification of R\&D~\cite{Adolphsen:2022ibf,ERLwp}. Major benefits arise from operation in ERL mode:
(i) reduced RF power requirements, with the RF power needed being significantly lower as it only needs to excite the cavity and compensate for minor losses, thus leading to unprecedented intrinsic efficiency compared to conventional linacs (factor of reduction $>$15 for PERLE); (ii) minimized radioprotection, through dumping the beam at low energy, below the activation threshold, which minimizes the beam dump, thus limiting the environmental impact of the accelerator throughout its life cycle, and; (iii) excellent beam quality for experiments, as experiments can benefit from a continuously injected beam, with specifications being defined by the source, because the used beam is sent to the dump after recovering its energy.

Energy recovery has been successfully demonstrated at moderate beam power. Yet it was never experimentally proven in the multi-MW regime. The PERLE demonstrator here described aims to investigate the feasibility of the concept at high beam current and for multiple recirculation in the same linac. It will provide a unique platform to test energy recovery efficiency at high beam power, study collective effects, and investigate possible limitations. The energy recovery will be measured. 

\subsection{Experimental Facility and Specifications}

PERLE (Powerful ERL for Experiments,~\cite{Angal-Kalinin:2017iup,PERLEwp,Kaabi:2023cvi}) is a novel multi-turn, ERL facility currently under construction at IJCLab in Orsay, France. Its primary objective is to demonstrate high-current (20 mA), continuous wave (CW), multi-pass operation using superconducting cavities at 801.58 MHz.
PERLE will serve as a hub for validating and exploring a wide range of accelerator phenomena in an unprecedented operational power regime, contributing to the development of ERL technology for future high-energy and high-intensity machines. It is specifically designed to validate choices and provide experience with operation for an ERL envisioned in the design of the LHeC and the FCC-eh. In addition, PERLE will provide a high-energy ($>$100 keV) X-ray source through inverse Compton process thanks to a high finesse Fabry-Perot optical cavity. Moreover, PERLE could host unique experiments in nuclear physics, being the first ERL to study electron-nucleus interactions with radioactive nuclei.

The implementation will occur in two phases: Phase 1 involving a single-turn operation reaching an energy of 89 MeV and; Phase 2 consisting of three turns to achieve a 5 MW beam at 250 MeV.
Future plans include the possibility of reaching 10 MW and a final energy of 500 MeV by adding an additional cryomodule.

An international collaboration has formed around the project, which currently includes CERN, JLab, STFC, the University of Liverpool, Cornell University, An-Najah University, ESS-Bilbao, and IJCLab, along with LPSC from CNRS. Collaboration with BINP-Novosibirsk has been suspended, and IFJ-PAN Krakow and ESS-Lund are in the process of joining the collaboration. IJCLab is leading the collaborative efforts toward the realization of the project.

\subsection{PERLE Accelerator Complex}

\begin{figure}[ht]
\centering
\includegraphics[width=0.8\textwidth]{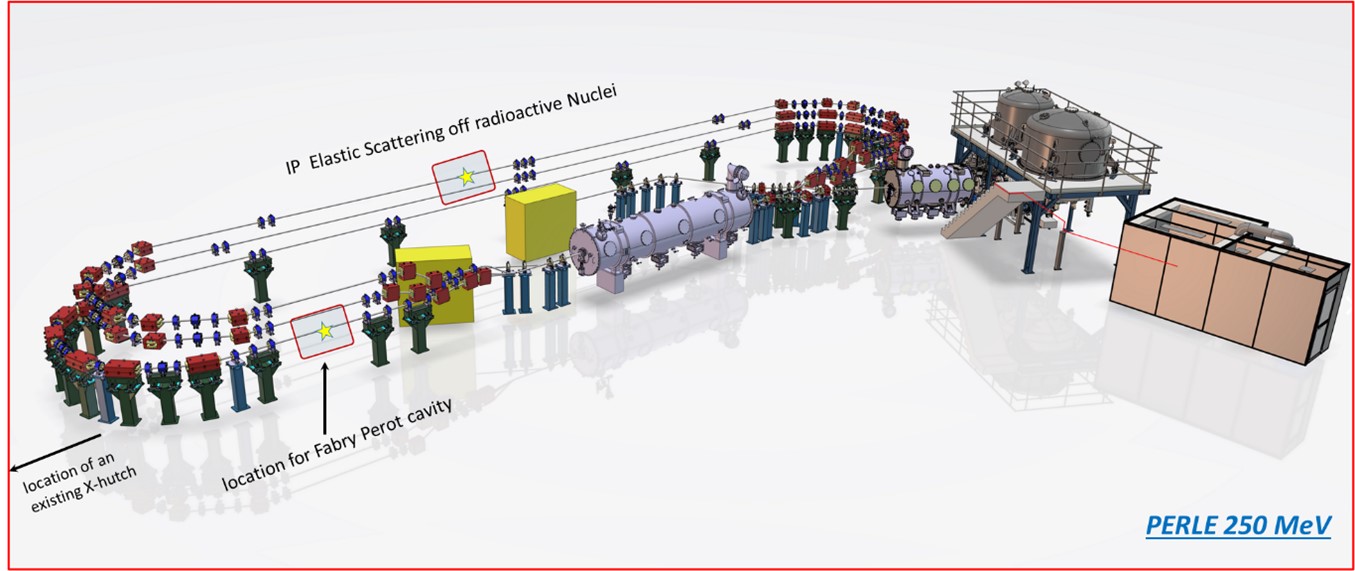}
\caption{Layout of the PERLE facility in the PHASE 2 configuration, featuring a cryomodule that houses four 5-cell superconducting (SC) cavities operating at 801.58 MHz, designed to achieve 250 MeV in three passes.}
\label{fig:layout}
\end{figure}

The PERLE accelerator complex (see Fig.~\ref{fig:layout}) is organized in a racetrack configuration, initially featuring one cryomodule. The electron beam is produced by a DC photogun operating at 350 kV. In order to reach the ambitious design value of 500 pC per bunch, multi-alkali photocathodes, having high quantum efficiency (QE), will be used and optically pumped by a green laser. High bunch charge will limit the lifetime of the photocathode: a dedicated photocathode preparation facility is directly connected to the photogun under vacuum for a fast photocathode exchange to maximize the accelerator availability.

The bunch length is adjusted by a warm buncher cavity and accelerated to 7 MeV by an SRF booster linac (4 single-cell cavities at 801.58 MHz). The beam is then injected into the ERL loop via a merging section and accelerated by an SRF linac. The cryomodule houses four 5-cell cavities operating at the same frecuency. Beam is circulated in the loop by a vertical stack of three recirculating arcs on each side as well as spreaders and recombiners, including matching sections.
The spreaders are positioned directly after each linac to separate beams of different energies and route them to their respective arcs. Conversely, the recombiners serve to merge beams of varying energies into a single trajectory. The beam can be used to generate X-rays with an optical Fabry-Perot cavity at the output of the cryomodule or for elastic scattering off radioactive nuclei in the straight section.

The path length of each arc is designed to be an integer multiple of the RF wavelength, except for the highest energy pass, arc 6, which is extended by half of the RF wavelength. This means that the beam will gain energy incrementally by 3 successive accelerations through the linac. Then the highest energy beam will be decelerated by traveling through the linac on its deceleration phase, 3 times incrementally down to its injection energy.

The cryomodule provides an energy change of up to 82 MeV to the high-average-current electron beam (20 mA). Consequently, over three turns, an energy increase of 246 MeV is achieved. When adding the initial injection energy of 7 MeV, the total energy reaches 253 MeV.

The main beam parameters of the PERLE facility are summarized in Tab.~\ref{tab:parameters}.
\begin{table}[ht]
\centering
\small{
\begin{tabular}{|l|l|}
\hline
\textbf{Parameters}                               & \textbf{Values} \\ \hline
Injection Energy                                   & 7 MeV           \\ \hline
Electron beam energy PHASE 1 (single turn)        & 89 MeV          \\ \hline
Electron beam energy PHASE 2 (three turns)        & 250 MeV         \\ \hline
Normalized Emittance \(\gamma \epsilon_{x,y}\)    & 6 mm.mrad       \\ \hline
Average beam current                               & 20 mA          \\ \hline
Bunch charge                                      & 500 pC          \\ \hline
Bunch length                                      & 3 mm            \\ \hline
Bunch spacing in injector                         & 25 ns           \\ \hline
RF frequency                                      & 802 MHz         \\ \hline
Duty factor                                       & CW              \\ \hline
\end{tabular}
}
\caption{Design machine parameters for PERLE.}
\label{tab:parameters}
\end{table}

Concerning the ERL operation, as the beam is decelerated in the cavity, it excites the cavity and transfers its power (thus energy) back to the accelerating structure. Thanks to the fast RF regulation system of the cavity, the power feeding the cavity is tuned down: the energy recovery linac thus allows to accelerate continuously injected bunches with minimal RF power into the cavity. After use and full deceleration of the beam, the remaining bunches are directed to a dump at its injection energy.

\subsection{Project Status}
\label{sec:perlestatus}

\begin{figure}[ht]
\centering
\includegraphics[width=0.8\textwidth]{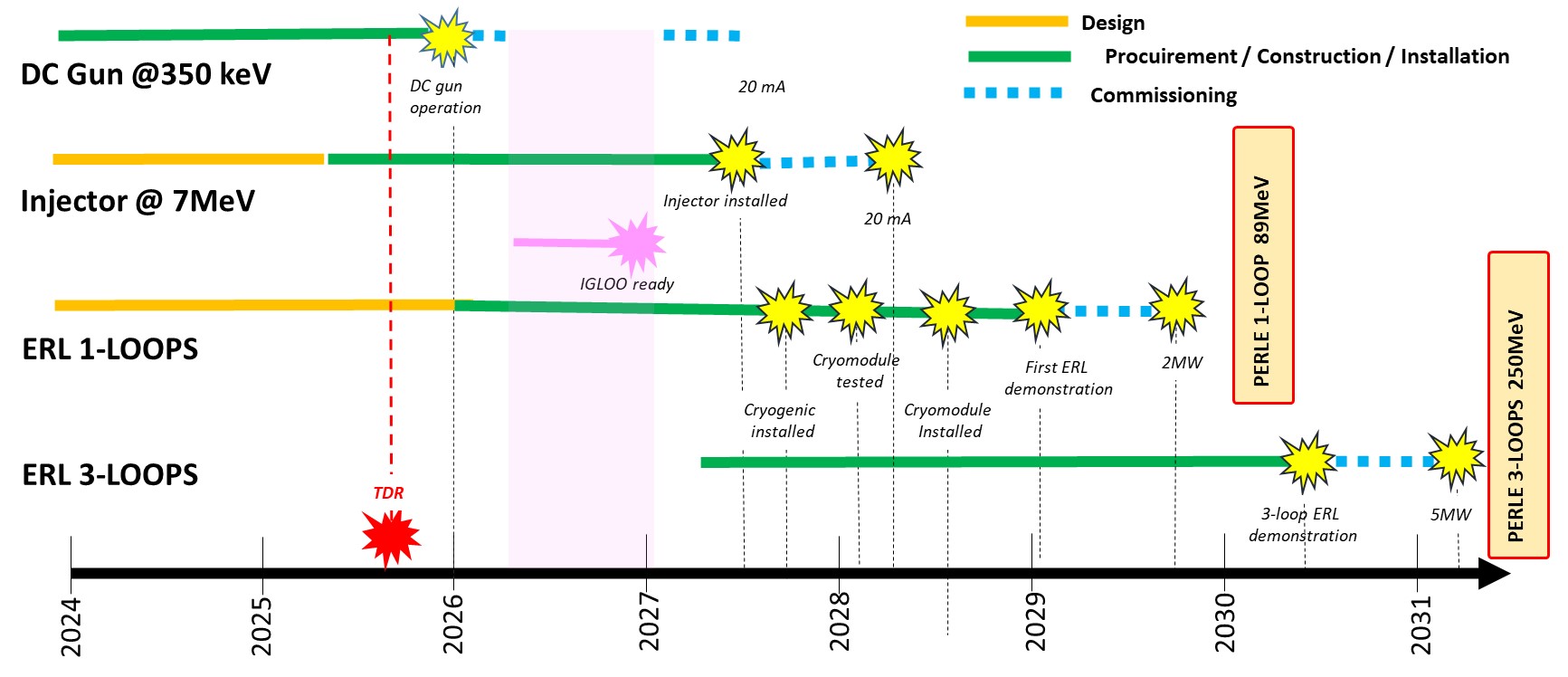}
\caption{Simplified version of the timeline for the realization of PERLE (Phase 1, Phase 2).}
\label{fig:planning-PERLE}
\end{figure}

The site has been selected for the project, specifically the building named IGLOO, which currently houses the ThomX machine. Infrastructure work and the installation of the first part of the injector (the DC gun) have already started. Large part of the funding has been secured for the development of the injector, and additional funding or in-kind contributions are being pursued for the completion of the first turn. Ongoing discussions aim to consolidate the budget for the three-turn phase. A sustainability working group is studying the efficiency of the accelerator and assess the carbon footprint of the PERLE machine. A detailed timeline, aligned with the current funding and available person power, has been established, and a simplified version is presented in Fig.~\ref{fig:planning-PERLE}.

Therefore, PERLE will test the performance of  high-power multiturn ERL by the early 2030ies. This is the timescale at which one would start the implementation phase of the LHeC.

\section{\NoCaseChange{LHeC} technology enabling a Higgs factory}
\label{sec:technology}

\subsection{LHeC investments improve the FCC-ee program}

A clear synergy with the FCC-ee injector design exists with the option to consider a recirculating LINAC (RLI) reusing the LHeC ERL facility,
with a first stage at 20 GeV, corresponding also to the latest top energy of the FCC-ee pre-injector, based on a warm full energy linac. 
The objective of a future study will be to produce coherent set 
of parameters for the RLI to replace LINAC pre-injectors and Booster, including positron production. 
The design could be further extended to cover the full FCC-ee energy range (i.e., replacing the FCC-ee booster). Note that this is an attractive option,
as the FCC-ee injector parameters with respect to the required  particle flux are largely within the capabilities of the LHeC ERL.
Finally, one could review the idea of a TLEP option based on ERL ($Z$, $W$, $H$-factory in the LHC tunnel).

There is quite some overlapping with the FCC-ee with respect to the development of the RF system technologies, in particular SRF 
pre-series production (including High-Temperature SRF at 4.5 K), while building at CERN expertise and industrial validation. 
In addition, the mentioned development of high-efficiency klystrons currently pursued at CERN can impact present and future studies and projects, 
including the HL-LHC, the LHeC, linear colliders and the FCC.

As sustainability and cost optimization is key for future accelerators, a study on the impact of the cryogenic system towards reduced power and cost
will be necessary but also Civil Engineering optimization based on CO$_2$ emission. 

Thereby, the FCC-ee would profit from the investment for the LHeC with respect to cost and resources, in particular for the injector development and common 
key technologies, and at the same time offering to the physics users a complementary  program. 
Additionally, technological risk will be reduced by prototyping and testing of certain FCC-ee accelerator technologies.
The schedule for LHeC operation is bridging the HL-LHC towards the FCC-ee.
Training opportunities  for building ERL expertise (in collaboration with  PERLE and bERLinPRO) and gaining operational experience are added values of an initial LHeC investment towards accelerator research and technology of the future.

\subsection{LHeC as a Detector Development Stepping-Stone towards Higgs factories} 
\label{detector:steppingstone}

The baseline design for the LHeC detector 
(Sec.~\ref{sec:feasibility}) and~\cite{LHeC:2020van})
is based on technology choices that are well established; 
either already deployed at the LHC or
expected to be so in the HL-LHC phase. A solution
to LHeC instrumentation thus exists that 
could be built already. On the other hand, the
project also offers opportunities as a testing ground for new ideas and technologies that may be strategically important 
for future 
flagship facilities such as a 
lepton-based Higgs factory or the next energy-frontier 
hadron collider. Decision points on the best technology
choices for such future facilities will be informed by
the work of the DRD collaborations through 
which European detector R\&D is now organised~\cite{detectorroadmap}.  
The picture may also evolve based on experience with
detector components at the HL-LHC or at the EIC~\cite{AbdulKhalek:2021gbh}. In the 
latter case, new ideas with direct relevance to DIS 
experiments are currently being developed into full
designs, with construction scheduled to begin from
2026, supported by a large and growing community
of instrumentation experts, many of whom
may find a natural home at the LHeC in the longer term. 
Changes to the baseline design of the LHeC
are therefore to be expected, 
and upgrades can be considered. 

In terms of functionality and performance requirements, 
the instrumentation needs of the LHeC overlap substantially
with those of a future electron-positron flagship facility.
The physics drivers of both begin with 
the precision study of Higgs decays 
in a low pile-up environment, with
decays to wide-ranging
final states including $b$ and $c$ quark modes. 

The strong emphasis on high resolution
tracking and secondary vertexing required by future
Higgs factories
(FCC-ee targets $\sigma(p_T)/p_T \sim 10^{-3}$ at 
$p_T = 50 \ {\rm GeV}$) 
has clear technical 
synergies with the needs of the LHeC. 
Both benefit from the relatively modest 
expected fluences compared with the HL-LHC, enabling the exploitation
of less radiation-hard technologies. 
A MAPS-based silicon solution
has already been identified as the obvious choice in the baseline
LHeC design and is likely to remain so, offering low material
budget, high spatial resolution and cost-effectiveness. However, 
the design may yet evolve in the light of experience 
with the ITS3
upgrade to ALICE for the HL-LHC~\cite{ITS3}.
Meanwhile,
an ITS3-derived solution is also being 
developed for the ePIC
detector at the EIC~\cite{ePIC}, sharing in common with the LHeC a layout
that mixes barrel layers with multiple
disks to maximise pseudorapidity
coverage.
The integration of precision timing layers for pile-up suppression 
in the HL-LHC general-purpose detectors through the use of
LGAD detectors is taken forward 
as a low momentum
particle identification tool in ePIC. Similar ideas
may be explored for the LHeC and ultimately for Higgs factories.

The need for precision electromagnetic calorimetry
at Higgs Factories (a few $\% / \surd E$)
is mirrored in DIS experiments where they play
a crucial role in the pivotal task of 
scattered electron reconstruction. 
The use of silicon detectors in forward and backward calorimetry has become more common recently, as is notably being 
deployed in CMS\,\cite{CMS-HGCAL-TDR}. Further 
forward, the 
silicon pixel-based ALICE FOCAL~\cite{CERN-LHCC-2024-004} 
will come on-line on the HL-LHC timescale and encourages 
the use of similar solutions even in beam-line instrumentation, as has been considered for the zero degree calorimeter at ePIC and 
in the baseline LHeC design. 
The current LHeC design already 
widely deploys silicon
in sampling calorimetry for the end-cap regions, 
offering a clear testing ground to Higgs factories. 
At ePIC, the
deployment of silicon layers extends to the barrel 
region, where
4 astropix layers are interleaved with a basic lead-scintillating
fibre design to provide precision pointing 
capabilities in the 
electromagnetic calorimetry.
Whereas the current LHeC 
barrel electromagnetic
calorimeter baseline design, see Sec.~\ref{sec:detector}, is an evolution of the liquid
argon technology in use by ATLAS, there is scope to adjust, 
particularly if there is a 
motivation from future field-wide needs.

The current hadron calorimeter design for the LHeC is 
relatively modest compared with the FCC-ee aims of 
$30 \% / \surd E$. On the other hand, the 
FCC-ee physics driver
of precision reconstruction and separation of $W$, $Z$ and 
Higgs decays to hadronic jets is also highly relevant to the
LHeC physics program. The LHeC therefore provides an 
opportunity to test novel ideas such as dual readout 
calorimetry, and also to test and optimise approaches
to particle flow algorithms that match high granularity
calorimetry with precision tracking.

Particle ID requirements,
for example for $\pi^0 / \gamma$ and  
for $K/\pi$ separation are also fundamental to Higgs factory
physics. Although not a major consideration in the LHeC 
baseline design, the EIC
strongly emphasises these aspects, and their introduction 
at LHeC would open up new physics opportunities
in the area of semi-inclusive DIS, for example probing the 
role of strange quarks or the transverse momentum dependence in proton and nuclear structure.
The widespread use of Cerenkov-based detectors for 
particle ID at high momentum is already a testing
ground for the future and could potentially be 
further explored at the LHeC if there is space to integrate
it into the design without compromising other key
functionalities. 

Whilst the FCC-ee target for luminosity monitoring at the $10^{-4}$ level is not readily achievable in a DIS environment, the
common need for a high quality near-beam electromagnetic 
response in  
a high rate environment for both
$e^+ e^-$ Bhabha scattering and the Bethe-Heitler
process in the $ep$ case is clear. 
Elsewhere in the detector, similar technologies 
(RPCs, $\mu$-RWELL,\ldots) are under consideration 
for muon detectors at the LHeC and Higgs factories, while there
are also common needs in terms of data acquisition and
handling.

\section{The \NoCaseChange{LHeC} Cost and Resource Estimates}
\label{sec:cost}
A 2018 study looked at cost estimates for the LHeC accelerator part and studied different configurations for the implementation~\cite{LHeC-cost}. The SRF system was identified as the main cost driver for the LHeC accelerator. In addition to savings in SRF installation, reducing the LHeC beam energy also reduces the synchrotron radiation power loss and, therefore, allows the use of smaller return arcs and thus a smaller overall accelerator footprint. 
The study showed that a 30\,GeV LHeC version with a total circumference of $1/5^{th}$ of that of the LHC can be constructed for just below 1 BCHF. Increasing the SRF installation and the overall accelerator footprint increases the performance reach of the LHeC but also implies an increase in the accelerator cost. The study showed that a 60\,GeV beam energy version of the LHeC with a total ERL circumference of $1/3^{rd}$ of the LHC circumference will cost about 1.76\,BCHF, with saving of about 0.13\,BCHF if the energy is decreased to 50\,GeV. Therefore, the cost of the baseline version of the LHeC accelerator would be around 1.6\,BCHF (in the 2018 cost model).

Given a comparable overall material budget as the HL-LHC (total budget at completion of about 1.1 BCHF), it seems reasonable to also assume comparable person power needs for the implementation of the accelerator part. The HL-LHC person power needs were estimated for the project approval during the June 2016 Council session at a total number of 1600 Person Years of integrated person power over ca. 10 years from project approval to start of operation~\cite{HL-LHC-Council}. The person power needs were later updated in 2024 to a total need of approximately 2300 Person Years of CERN staff~\cite{Bruning:2024zai} plus person power from the international collaborations. We therefore assume for the LHeC a total person power need of ca. 2500 Person Years for the accelerator implementation.

A preliminary estimate of the cost for the central part of the baseline detector, the central tracker, calorimetry, and the muon system, amounts to roughly 360 MCHF, based on recent cost evaluations for LHC detector upgrades. The cost for magnet, beam pipe, common DAQ and trigger, mechanical structure and infrastructure is not included. The largest fraction, around 270 MCHF, comes from calorimetry. Optimising the sampling fraction and granularity would lead to significant cost reductions. 

As indicated in Sec.\ref{sec:implementation}, we consider the person power
requirements for operating the LHeC with only one experiment and proton beam and the ERL facility comparable to
the needs of the HL-LHC with two proton beams and four experimental insertions. The detector would require human resources for construction and operation similar to that of an LHC general-purpose detector.

\section{Summary and Outlook}
\label{sec:outlook}

In the present manuscript, which served as accompanying document to the LHeC submission to the 2026 Update of the ESPP~\cite{LHeCSub}, we  have developed the case for the LHeC as a bridge project between the HL-LHC and the next flagship at CERN. We have reviewed the physics reach that DIS alone offers, the implications of such physics on hadron colliders, and the complementarity with other existing projects. We have discussed the technical feasibility, the implementation plan, the status of the development of the ERL required to achieve the highest luminosities, the technical synergies with Higgs factories, and the financial and person power cost estimates.

In the Recommendations by the European Strategy Group~\cite{ESGRec}, where the FCC-ee appears as Plan A and its descoped version as Plan B, the LHeC is listed as one of the alternative options in case the FCC-ee would not be feasible.
 In the coming years the LHeC plan will be further developed with particular focus on the technology readiness of the ERL-based electron accelerator and the exploration of the physics impact. Key to this plan is the further development of the ERL technology at PERLE. While the LHeC alone cannot constitute a project for the very long term future of CERN, the plan to be developed will show its feasibility as a bridge between major colliders, with a strong physics case on its own and large implications, both on the physics and on the technology, for future projects in particle physics.

\acknowledgments

The research of NA was supported by European Research Council project ERC-2018-ADG-835105 YoctoLHC, by Xunta de Galicia (CIGUS Network of Research Centres), by European Union ERDF, and by the Spanish Research State Agency under project PID2023-152762NB-I00 supported by MCIU /AEI /10.13039/501100011033 / FEDER, EU, and  is part of the project CEX2023-001318-M financed by MCIN/AEI/10.13039/501100011033.
The research of LF and KP has been supported by the Polish National Agency for Academic Exchange under grant number BPN/PPO/2021/1/00011. LF has also been supported by the NCN grants 2022/01/1/ST2/00022 and 2023/49/B/ST2/03273, and by the CERN COAS programme. HK appreciates financial support from NAWA under grant number BPN/ULM/2023/1/00160, 
as well as from the IDUB program at the AGH University.
TL, HM, HP and MT are supported by the Research Council of Finland, the Centre of Excellence in Quark Matter (projects 346324 and 346326) and project 338263. HM is supported by the European Research Council (ERC, grant agreements ERC-2023-COG-101123801 GlueSatLight). FIO is supported by the U.S. Department of Energy grant No. DE-SC0010129. AMS is supported by the U.S. Department of Energy grant No. DE-SC-0002145. AMS and FIO are supported within the framework of the Saturated Glue (SURGE) Topical Theory Collaboration.
CS acknowledges support by the Deutsche Forschungsgemeinschaft (DFG, German Research Foundation) under Germany‘s Excellence Strategy – EXC 2121 “Quantum Universe” – 390833306. 
BM is grateful to the South African Department of
Science, Technology and Innovation for the support through the SA-CERN program.

\appendix
\section{Some examples of the physics reach with lower integrated luminosities}
\label{sec:lowlumi}

The nominal integrated luminosity planned for the LHeC is 1\,ab$^{-1}$. A natural question is to which extent the physics program becomes limited if such integrated luminosity cannot be achieved. While a full answer can only be given by repeating all studies versus integrated luminosity, here we provide some specific examples.

The effect on PDF determination of decreasing the integrated luminosity to 50\,fb$^{-1}$ (i.e., a decrease of a factor twenty) is shown in Fig.~\ref{fig:pplumi} and discussed in detail in Sec. 3.3.7 in~\cite{LHeC:2020van}.  The conclusion is that the vast majority of the improvements with respect to the respective determination using only HERA data comes from the first 50\,fb$^{-1}$, with larger luminosities resulting in modest further improvements.
The effect of the same factor-20 decrease on the strong coupling constant is illustrated in Fig. 4.5 in~\cite{LHeC:2020van}, where is is shown that for DIS-only fits the uncertainty turns out to be a factor two larger, while the effect for DIS+jet fits is very small. Other QCD physics topics, like small $x$ or diffraction, are not statistically limited and, therefore, are not affected by such a decrease in luminosity.

\begin{figure}[h]
\begin{center}
\includegraphics[width=0.49\textwidth]{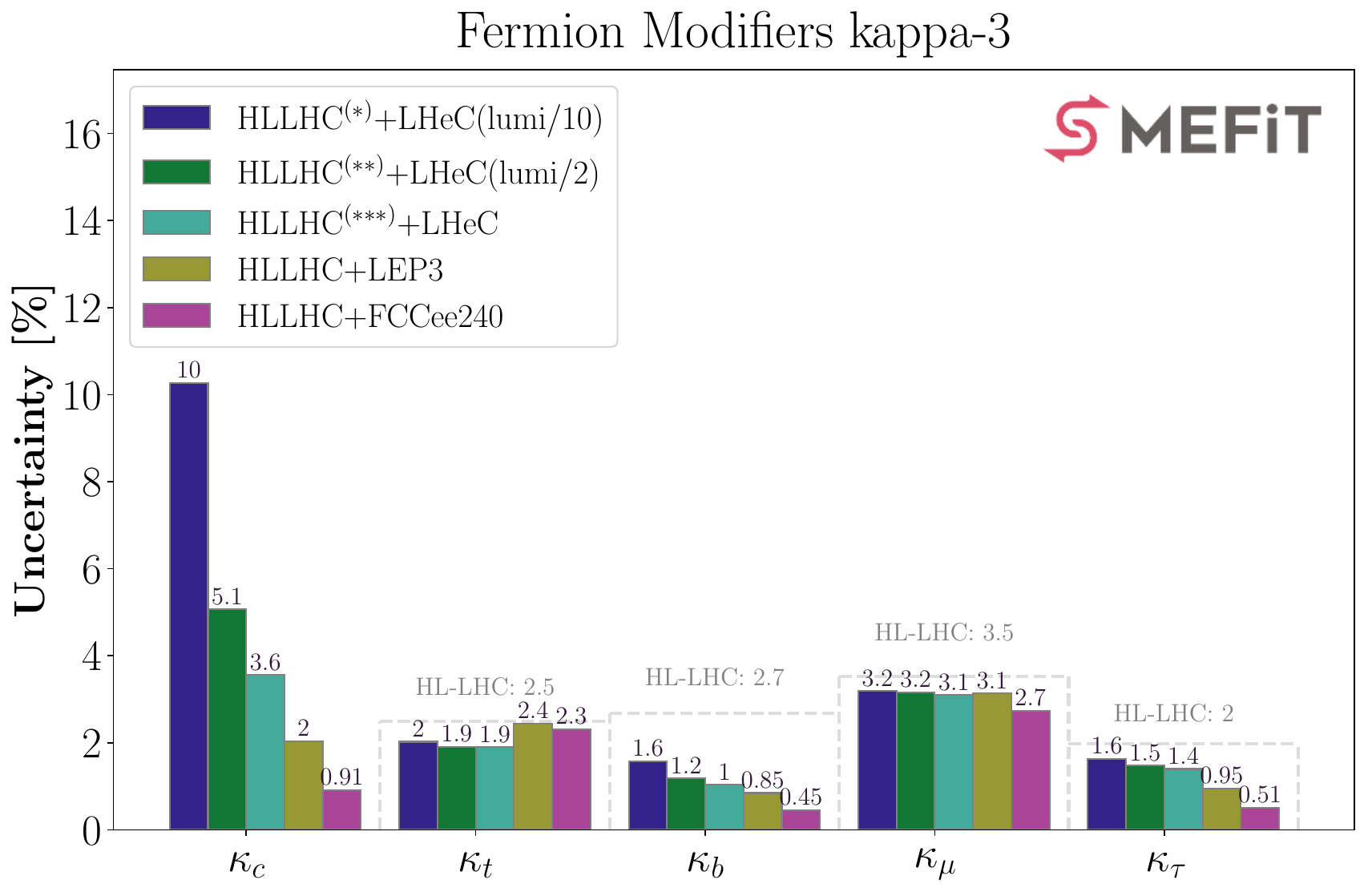}
\includegraphics[width=0.49\textwidth]{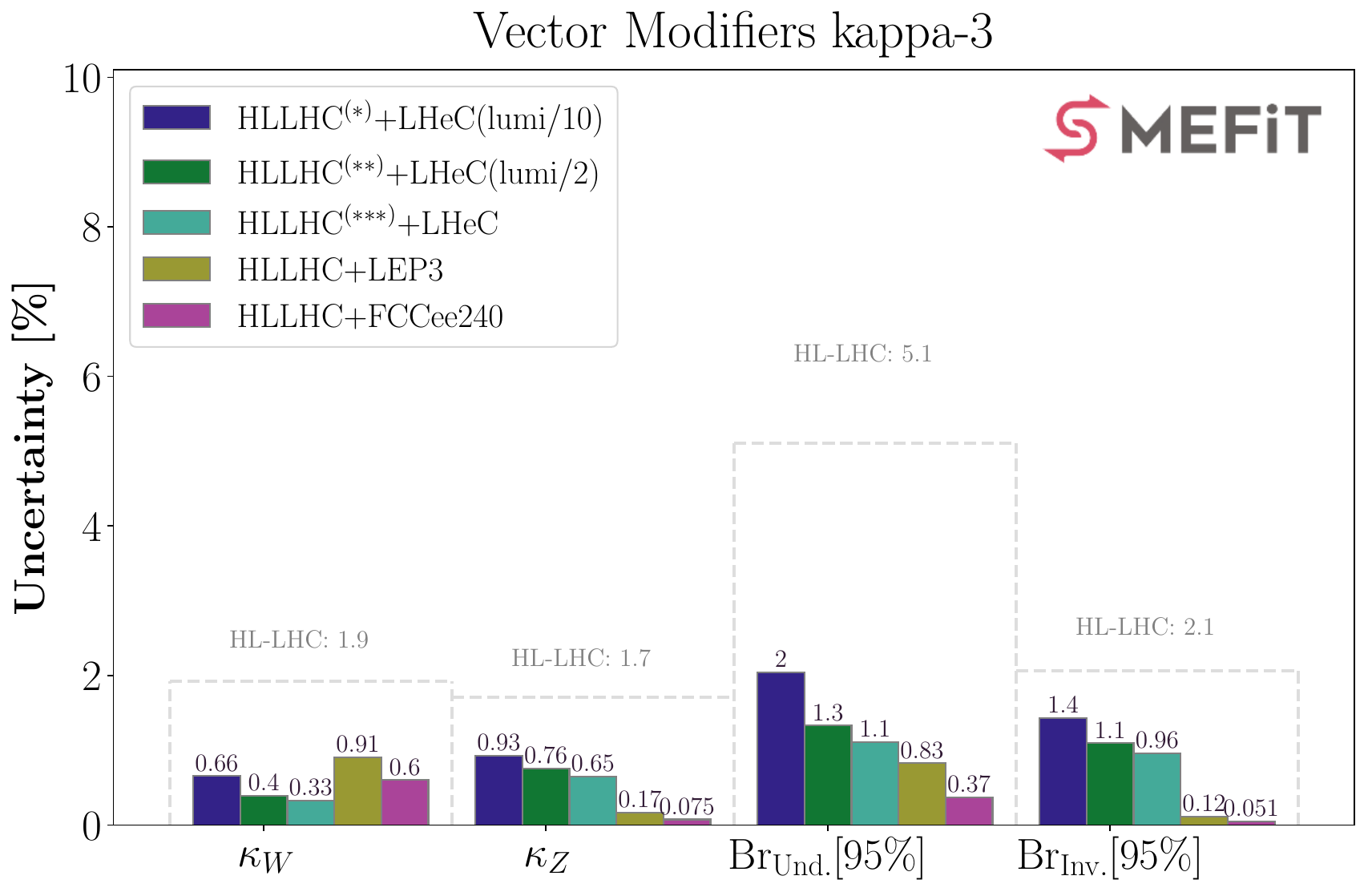}
\includegraphics[width=0.49\textwidth]{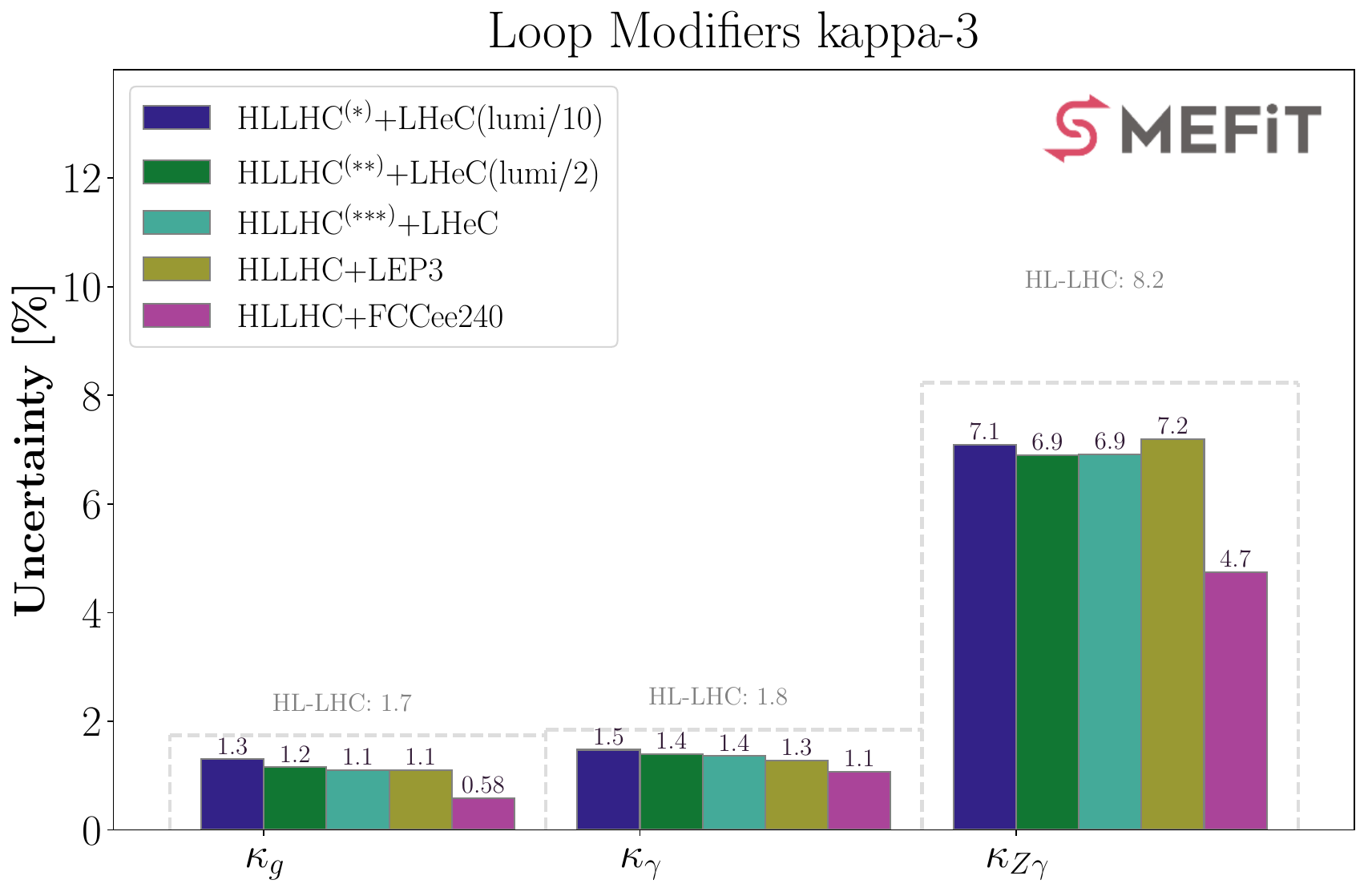}
\end{center}

\vskip -0.6cm
\caption{Id. to  Fig.~\ref{fig:smefit2} but showing the combinations HL-LHC+LEP3 and HL-LHC+FCC-ee (240 GeV) in  absence of LHeC improvements together with HL-LHC+LHeC for different LHeC integrated luminosities. The asterisks in HL-LHC denote that the considered integrated luminosity is also taken into account for the improvements that more precise PDFs and $\alpha_s$ bring to the HL-LHC measurements. The dashed lines indicate the uncertainties achievable at the HL-LHC. Results have been obtained with the {\sc\small SMEFiT} framework~\cite{Celada:2024mcf,terHoeve:2023pvs,Giani:2023gfq,Ethier:2021bye,Armadillo:2026mvp}.
}
\label{fig:smefit3}
\end{figure}

The effect of having smaller integrated luminosities, 0.5 and 0.1\,ab$^{-1}$, on the Higgs couplings is shown in Fig.~\ref{fig:smefit3}, 
the reduction in HL-LHC uncertainties follows the same procedure detailed in the main text, but dividing the PDF and $\alpha_s$ related uncertainties by a factor 2.8 (2) in the case of one half (one tenth) LHeC luminosity scenario, instead of the factor 3 used for the main LHeC proposal. A decrease of a factor ten in luminosity has a modest effect on the uncertainties -- smaller than 50\% -- for all couplings except for $c$, while a decrease of a factor two has a minor effect.

Concerning top physics, Fig.~\ref{fig_Wtx_couplings} (left) shows the uncertainties in the determination of anomalous top couplings on which integrated luminosities beyond 100\,fb$^{-1}$ have very little effect. Fig.~\ref{fig_Wtx_couplings} (right) demonstrates that for $|V_{ts}|$ current limits are already improved with 100\,fb$^{-1}$ but values close to the SM fit require larger integrated luminosities.

Finally, some very preliminary considerations on BSM physics with 100\,fb$^{-1}$ for ALPs, HNs and dark photons~\cite{JorgenVenice} indicate that sensitivity exists with such an integrated luminosity. The achievable limits are complementary to those at the HL-LHC and, in some cases, the difference between 1/10th and full LHeC integrated luminosity is smaller than the corresponding difference between the LHeC and the FCC-eh.

\bibliography{biblio}

\end{document}